\documentclass[fleqn,usenatbib]{mnras}

\usepackage{newtxtext,newtxmath}
\usepackage[T1]{fontenc}
\usepackage{ae,aecompl}

\usepackage{color}
\usepackage{xcolor}
\usepackage{ulem}
\usepackage{float}

\usepackage{tikz,xcolor,hyperref}

\definecolor{lime}{HTML}{A6CE39}
\DeclareRobustCommand{\orcidicon}{
	\begin{tikzpicture}
	\draw[lime, fill=lime] (0,0) 
	circle [radius=0.16] 
	node[white] {{\fontfamily{qag}\selectfont \tiny ID}};
	\draw[white, fill=white] (-0.0625,0.095) 
	circle [radius=0.007];
	\end{tikzpicture}
	\hspace{-2mm}
}

\foreach \x in {A, ..., Z}{\expandafter\xdef\csname orcid\x\endcsname{\noexpand\href{https://orcid.org/\csname orcidauthor\x\endcsname}
			{\noexpand\orcidicon}}
}

\newcommand{\te}{$T_{\rm e}$}
\newcommand{\elecd}{$n_{\rm e}$}
\newcommand{\hb}{H$\beta$}
\newcommand{\ha}{H$\alpha$}
\newcommand{\fci}{[C\,{\sc i}]}

\newcommand{\foi}{[O\,{\sc i}]}
\newcommand{\foii}{[O\,{\sc ii}]}
\newcommand{\foiii}{[O\,{\sc iii}]}
\newcommand{\fsii}{[S\,{\sc ii}]}
\newcommand{\fsiii}{[S\,{\sc iii}]}
\newcommand{\fnitroi}{[N\,{\sc i}]}
\newcommand{\fnii}{[N\,{\sc ii}]}
\newcommand{\fariii}{[Ar\,{\sc iii}]}
\newcommand{\fariv}{[Ar\,{\sc iv}]}
\newcommand{\farv}{[Ar\,{\sc v}]}
\newcommand{\fcliii}{[Cl\,{\sc iii}]}
\newcommand{\fcliv}{[Cl\,{\sc iv}]}
\newcommand{\oiii}{O\,{\sc iii}}
\newcommand{\nii}{N\,{\sc ii}}

\newcommand{\oi}{O\,{\sc i}}
\newcommand{\oii}{O\,{\sc ii}}
\newcommand{\cii}{C\,{\sc ii}}
\newcommand{\ciii}{C\,{\sc iii}}

\newcommand{\hi}{H\,{\sc i}}
\newcommand{\hii}{H\,{\sc ii}}
\newcommand{\hei}{He\,{\sc i}}
\newcommand{\heii}{He\,{\sc ii}}

\newcommand{\useaa}{} 
\newcommand{\ionic}[1]{$\,${\sc #1}}
\newcommand{\forb}[2]{[{#1}\ionic{#2}]}

\newcommand{\forbr}[4]{\forb{#1}{#2}$\,\lambda$#3/$\lambda$#4\useaa}

\newcommand{\perm}[2]{#1\ionic{#2}}

\newcommand{\permr}[4]{\perm{#1}{#2}$\,\lambda\,$#3/$\lambda\,$#4\useaa}

\usepackage{graphicx}	
\usepackage{amsmath}	

\usepackage{amssymb}	
\usepackage{longtable}
\usepackage{lscape}
\usepackage{bm}

\title[MUSE spectroscopy of planetary nebulae with high adf]{MUSE spectroscopy of planetary nebulae with high abundance discrepancies}

\author[J. Garc\'{\i}a-Rojas et al.]{
J. Garc\'{\i}a-Rojas,$^{1,2\orcidA{}}$\thanks{E-mail: jogarcia@iac.es},
C. Morisset$^{3\orcidB{}}$,
D. Jones,$^{1,2\orcidC{}}$,
R. Wesson$^{4\orcidD{}}$, 
H.~M.~J. Boffin$^{5\orcidE{}}$,
\newauthor
H. Monteiro$^{6\orcidF{}}$,
R.~L~M. Corradi$^{1,2,7\orcidG{}}$,
P. Rodr\'{\i}guez-Gil$^{1,2\orcidH{}}$
\\
$^{1}$Instituto de Astrof\'isica de Canarias, E-38205 La Laguna, Tenerife, Spain\\
$^{2}$Departamento de Astrof\'isica, Universidad de La Laguna, E-38206 La Laguna, Tenerife, Spain\\
$^{3}$Instituto de Astronom\'ia (IA), Universidad Nacional Aut\'onoma de M\'exico, Apdo. postal 106, C.P. 22800 Ensenada, Baja California, M\'exico\\
$^{4}$Department of Physics and Astronomy, University College London, Gower St, London WC1E 6BT, UK\\
$^{5}$European Southern Observatory, Karl-Schwarzschild-Str. 2, 85738 Garching bei M\"unchen, Germany\\
$^{6}$Instituto de F\'{\i}sica e Qu\'{\i}mica, Universidade Federal de Itajub\'a, Av. BPS 1303-Pinheirinho, 37500-903, Itajub\'a, Brazil \\
$^{7}$Gran Telescopio CANARIAS S.A., c/ Cuesta de San Jos\'e s/n, Bre\~na Baja, E-38712 Santa Cruz de Tenerife, Spain\\
}

\date{Accepted XXX. Received YYY; in original form ZZZ}

\pubyear{2015}

\begin{document}
\label{firstpage}
\pagerange{\pageref{firstpage}--\pageref{lastpage}}
\maketitle

\begin{abstract}
We present MUSE deep integral-field unit spectroscopy of three planetary nebulae (PNe) with high abundance discrepancy factors (ADF $>$ 20): NGC\,6778, M\,1--42 and Hf\,2--2. We have constructed flux maps for more than 40 emission lines, and use them  to build extinction, electron temperature ({\te}), electron density ({\elecd}), and ionic abundances maps of a number of ionic species. The effects of the contribution of recombination to the auroral {\fnii} and {\foii} lines on {\te} and the abundance maps of low-ionization species are evaluated using recombination diagnostics. As a result, low {\te} values and a downward gradient of {\te} are found toward the inner zones of each PN. Spatially, this nearly coincides with the increase of abundances of heavy elements measured using recombination lines in the inner regions of PNe, and strongly supports the presence of two distinct gas phases: a cold and metal-rich and a warm one with ``normal'' metal content. We have simultaneously constructed, for the first time, the ADF maps of O$^+$ and O$^{2+}$ and found that they centrally peak for all three PNe under study. We show that the main issue when trying to compute realistic abundances from either ORLs or CELs is to estimate the relative contribution of each gas component to the {\hi} emission, and we present a method to evaluate it. It is also found that, for the studied high-ADF PNe, the amount of oxygen in the cold and warm regions is of the same order. 
\end{abstract}




\section{Introduction}
\label{sec:intro} 

The abundance discrepancy problem is one of the major unresolved questions in nebular astrophysics, having being around for more than seventy years since \cite{wyse42}. In photoionised nebulae---both {\hii} regions and planetary nebulae (PNe)---optical recombination lines (ORLs) systematically yield larger chemical abundance values than collisionally excited lines (CELs). Solving this problem has obvious implications for the measurement of the chemical content of nearby and distant galaxies, because this is most often done using emission from their ionised interstellar medium from using CELs only.
The discrepancy is generally parameterised in terms of the abundance discrepancy factor (ADF), which for a given ion is defined as the ratio between ionic chemical abundances derived from ORLs and CELs. The origin of this discrepancy has been the subject of strong debate in the past decades and several scenarios have been put forward \citep[see e.g.][for details]{garciarojasetal19, garciarojas20}. 

\citet{garciarojasesteban07} proposed that the source of the abundance discrepancies may be different in {\hii} regions compared to PNe. In addition, detailed studies of {\hii} regions showed that high-velocity gas flows \citep{mesadelgadoetal09a,mesadelgadoetal09b} or the presence of high-density clumps, such as protoplanetary discs \citep[proplyds,][]{tsamisetal11, mesadelgadoetal12}, may have a significant impact upon the abundance determinations using CELs. Therefore, the hypothesis that several sources act to produce the observed discrepancies has gained strength. 

In the case of PNe, several authors have proposed that the abundance discrepancy can be explained by the presence of two different components of gas  \citep[e.g]{liuetal00, liuetal01, tsamispequignot05, liuetal06, yuanetal11}: a hot ($\sim$10\,000\,K) gas with standard metallicity where the CELs can be efficiently excited, and a much cooler ($\sim$1\,000\,K) H-poor gas component with a highly enhanced content of heavy elements and almost no CEL emission, in which the bulk of heavy-element ORLs is produced.

\citet{liuetal06} speculated that the large ADF found in the PN Hf\,2--2 may be the result of the common envelope evolution of its known close binary central star. 
%
Indeed, \citet{corradietal15}, \citet{jonesetal16} and \citet{wessonetal18} showed that the extremely large ADFs observed in some PNe are associated with the presence of close binary central stars. 

Although the presence of two gas phases is not predicted by standard mass loss theories, several explanations have been propounded. Some of them are naturally linked to binarity, such as the hypothesis that the ORL-emitting gas is a low temperature, metal-enhanced phase likely produced by multiple episodes of mass loss and even fallback and reprocessing of the ejected material. This is supported by the significantly low mass of the ionised gas measured in PNe\footnote{Although it should be noted that all PNe, not just those with extreme ADFs and/or close binary central stars, suffer from the so-called ``missing mass problem'' \citep[see][for a full discussion]{boffinjones19}.} \citep{liuetal06, corradietal15}. Other explanations involve the presence of planetary debris that survived the whole evolution of the central stars, or the tidal destruction, accretion and ejection of Jupiter-like planets \citep[see the discussion in][]{corradietal15}. However, at this point it is not possible to favour any scenario owing to the lack of sufficient observational constraints.

To overcome this, several spectroscopic studies have been carried out, finding that in many PNe the ORL-emitting gas tends to concentrate at the central parts of the nebula. This is mainly found in PNe with close binary central stars and high ADFs \citep[e.g.][]{corradietal15, garciarojasetal16, jonesetal16, wessonetal18}. However, this behaviour has also been seen in PNe with low-to-moderate ADFs and no indication of binarity \citep[e.g.][]{garnettdinerstein01, liuetal01} and PNe with relatively high ADFs but no known close binary central star \citep[e.g. M\,1--42;][]{climent16, garciarojasetal17}.
\citet{garciarojasetal16,garciarojasetal17} obtained the first direct images of PNe (NGC\,6778 and Abell\,46) in the light of the {\oii} ORLs using the tunable filters at the 10.4-m GTC telescope, finding  that the {\oii} $\lambda\lambda$4649+50 ORL and the [{\oiii}] $\lambda$5007 CEL emission did not coincide spatially, with the ORLs originating closer to the nebula's core.

In this work, we aim to exploit the high spatial resolution capabilities of the MUSE 2D spectrograph to further test these findings in three high-ADF PNe, namely NGC\,6778, Hf\,2--2, and M\,1--42. The former two have confirmed post-common envelope binary central stars.

Since the advent of integral field unit (IFU) spectroscopy, many authors have used this technique to study the physical properties of PNe, probing their three-dimensional structure  \citep{monteiroetal13, danehkaretal13, danehkaretal14}, or specific nebular components such as low ionization structures \citep{danehkaretal16} or halos \citep{monrealiberoetal05}. An extensive IFU survey of southern PNe using the 2.3m ANU telescope at Siding Spring Observatory  \citep{alietal16, basurahetal16, alidopita17,  dopitaetal17, dopitaetal18, alidopita19} has provided valuable observational constraints on the physical structure of PNe. However, only a handful of these studies have addressed in some way the abundance discrepancy problem \citep{tsamisetal08, alidopita19}. Deeper studies of PNe  can be performed using MUSE at the 8.2-m Very Large Telescope (VLT), as done by \citet{walshetal16, walshetal18} and \citet{monrealiberowalsh20}. These authors used the nominal mode (MUSE--NOAO--N), which covers the wavelength range $480-930$\,nm, and hence does not include the recombination lines of the multiplet 1 of {\oii} at $\lambda \simeq 465$\,nm. In the present study, we instead use the MUSE extended mode (WFM--NOAO--E), that  covers  the  wavelength range $460-930$\,nm allowing the measure of these critical ORL lines.
For a review  on the capabilities of MUSE for PNe studies see \citet{walshmonrealibero20}.

The structure of the paper is as follows: in Section~\ref{sec:obs} we describe the observations and reduction processes. In Section~\ref{sec:lines} we present the methods followed to construct the emission line maps and describe how uncertainties in the line fluxes are propagated. In Section~\ref{sec:extinction} the extinction maps are computed and discussed. In Section~\ref{sec:phys_cond} we focus on the analysis of physical conditions using several line diagnostics. In Section~\ref{sec:ionic} we give details on the ionic chemical abundance determinations and the ionic abundance discrepancy factor. We also discuss the chemical analysis and determination of the ionization correction factors in the integrated spectra of each object. Finally, in Sections~\ref{sec:discuss} and \ref{sec:conclu} results and conclusions are presented.

\section{Observations and data reduction}
\label{sec:obs} 

\begin{table}
\caption{Log of the MUSE observations.}\label{Tab:Obs}
\begin{tabular}{ccccc}
\hline
UT Start & n & Exp &  Airm. & Seeing \\
& & (s) &  & ($"$) \\\hline
Target: M\,1--42 & \multicolumn{4}{c}{Mode:  WFM--NOAO--E}\\ \hline
 2016-07-07 00:49:13.114 &1/9 & 30.0 & 1.32  & 0.82 \\
 2016-07-07 00:51:36.457 &2/9 & 60.0 & 1.31  & 0.99 \\
 2016-07-07 00:54:28.527 &3/9 &450.0 & 1.298 & 0.9 \\
 2016-07-07 01:03:18.209 &4/9$^{\rm a}$ &180.0 & 1.264 & 0.84 \\
 2016-07-07 01:08:13.358 &5/9 &450.0 & 1.246 & 0.85 \\
 2016-07-07 01:17:40.152 &6/9 &450.0 & 1.214 & 0.77 \\
 2016-07-07 01:26:31.242 &7/9$^{\rm a}$ &180.0 & 1.187 & 0.84 \\
 2016-07-07 01:31:30.920 &8/9 &450.0 & 1.173 & 0.85 \\
 2016-07-07 01:40:57.004 &9/9 &450.0 & 1.149 & 0.87 \\ \hline
Target: NGC\,6778 & \multicolumn{4}{c}{Mode:  WFM--NOAO--E}\\ \hline
 2016-07-07 03:56:54.262&1/9&30.0&1.127&1.17\\
 2016-07-07 03:59:04.483&2/9&120.0&1.124&0.89\\
 2016-07-07 04:02:46.600&3/9&450.0&1.12&0.88\\
 2016-07-07 04:11:44.399&4/9$^{\rm a}$&180.0&1.11&0.78\\
 2016-07-07 04:16:35.484&5/9&450.0&1.105&0.82\\
 2016-07-07 04:25:46.574&6/9&450.0&1.098&0.86\\
 2016-07-07 04:34:44.604&7/9$^{\rm a}$&180.0&1.093&0.99\\
 2016-07-07 04:39:34.461&8/9&450.0&1.09&1.37\\
 2016-07-07 04:48:45.581&9/9&450.0&1.088&1.19\\ \hline
Target: Hf 2--2 & \multicolumn{4}{c}{Mode:  WFM--NOAO--E}\\ \hline
 2016-07-07 05:05:07.675 &1/9 & 30.0& 1.024&0.88\\
 2016-07-07 05:07:36.225&2/9&120.0&1.026&1.2\\
 2016-07-07 05:11:34.044&3/9&1800.0&1.029&0.98\\
 2016-07-07 05:43:05.639&4/9$^{\rm a}$&600.0&1.066&1.13\\
 2016-07-07 05:55:17.824&5/9&1800.0&1.088&0.86\\
 2016-07-07 06:27:18.454&6/9&1800.0&1.157&0.75\\
 2016-07-07 06:58:49.237&7/9$^{\rm a}$&300.0&1.251&0.63\\
 2016-07-07 07:05:57.674&8/9&1800.0&1.281&0.74\\
 2016-07-07 07:37:59.288&9/9&1800.0&1.431&0.78 \\
\hline
\end{tabular}
\begin{description}
\item $^{\rm a}$ {Sky frames taken far enough away from the object to ensure that there is no nebular contamination.}
\end{description}
\end{table}

\begin{figure}
\centering
\includegraphics[width=0.83\columnwidth]{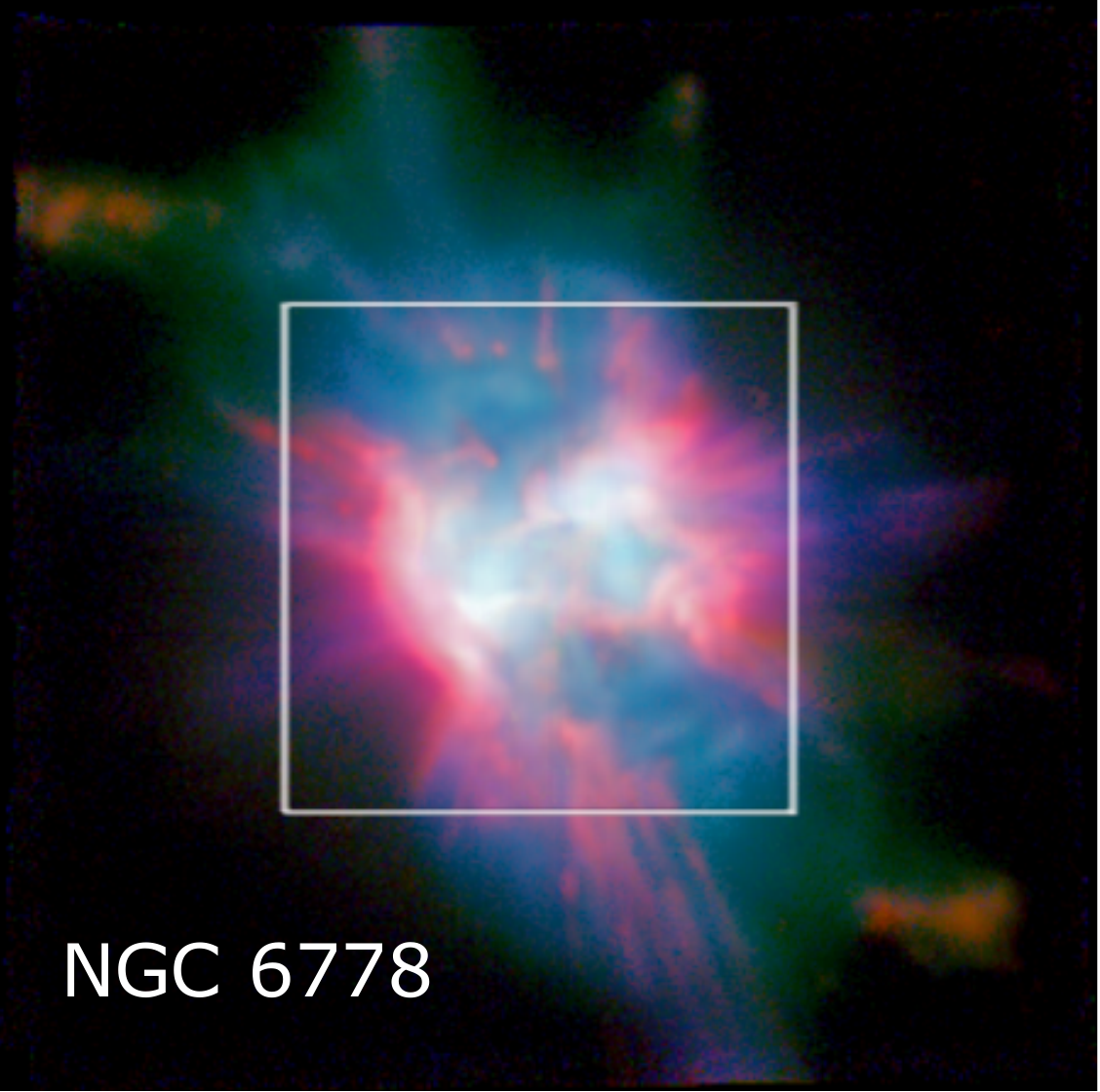}
\includegraphics[width=0.83\columnwidth]{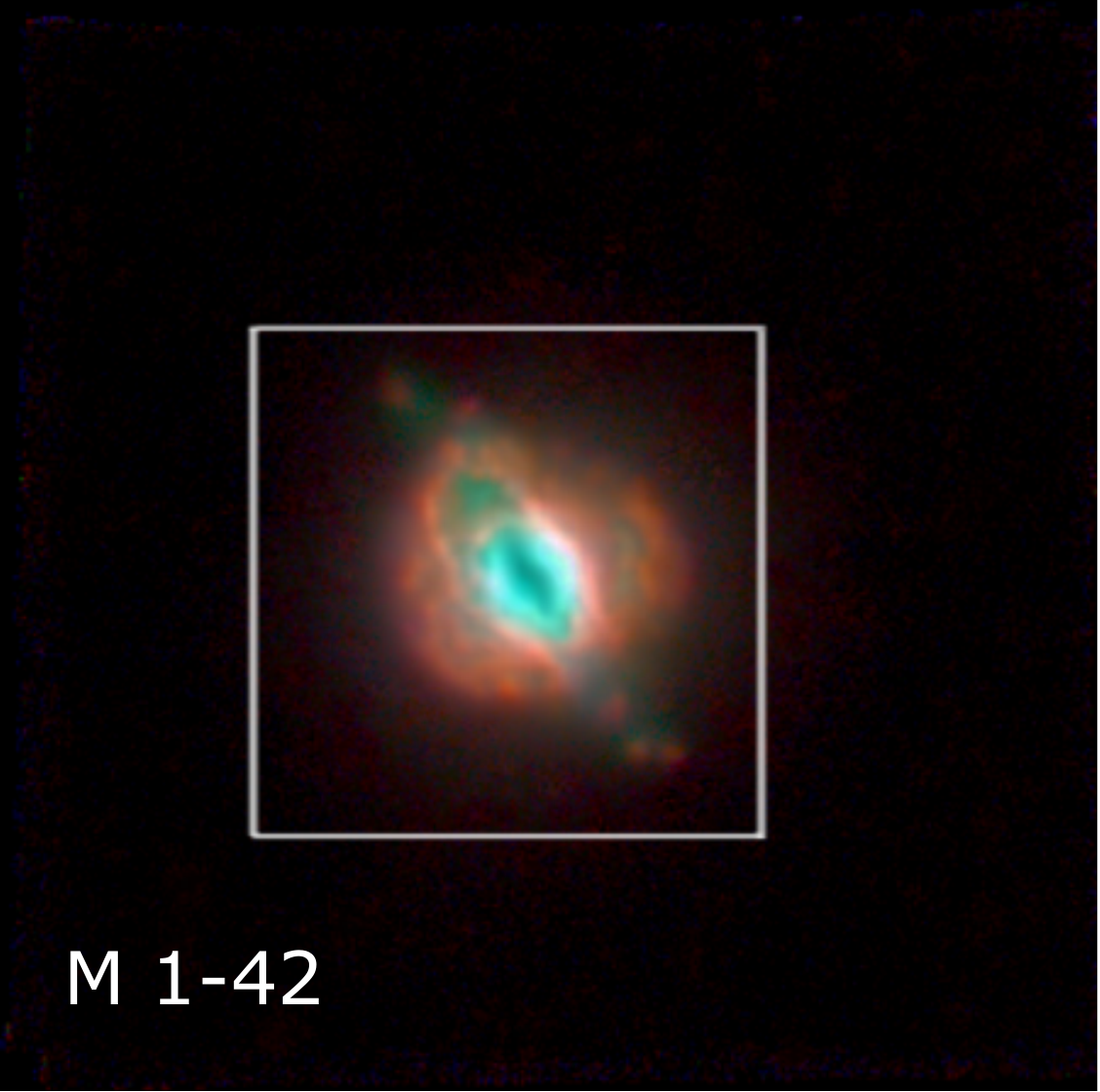}
\includegraphics[width=0.83\columnwidth]{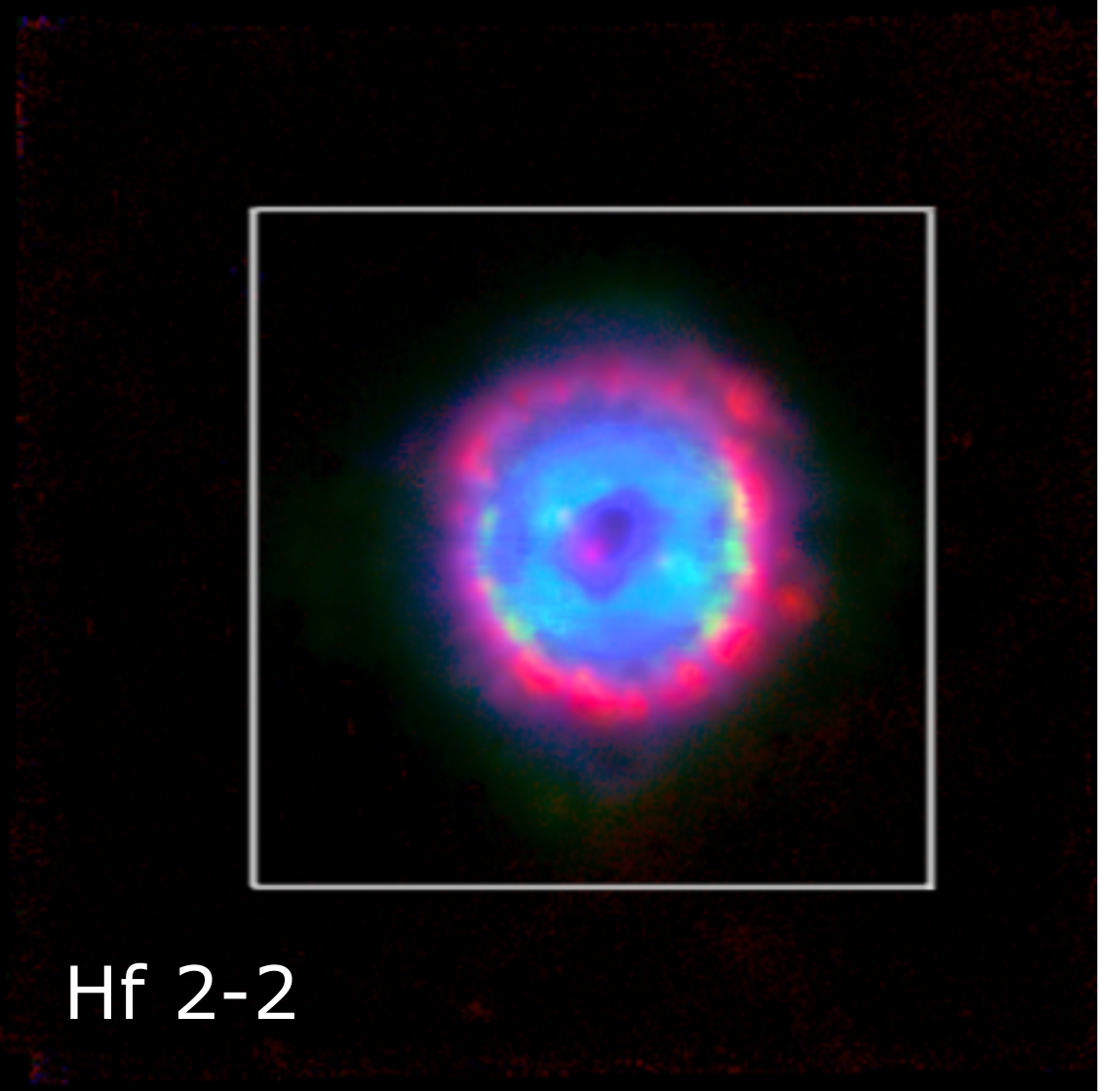}
\caption{Composite RGB images of the MUSE field of view for NGC\,6778 (top panel), M\,1--42 (middle panel) and Hf\,2--2 (bottom panel). {\foiii} $\lambda$5007 emission is shown in blue, H$\alpha$ in green and an average of {\fnii} $\lambda$6583 and {\fsii} $\lambda$6731 in red. The white squares represent the trimmed area of the data cubes that we study in this paper: 30\,arcsec$^2$ ($150 \times 150$ spaxels) for NGC\,6778 and M\,1--42 and 40\,arcsec$^2$ ($200 \times 200$ spaxels) for Hf\,2--2. In all panels, north is up and east is to the left.
\label{fig:extraction}}
\end{figure}

Our targets were observed with the Multi Unit Spectroscopic Explorer (MUSE) integral-field spectrograph  \citep{baconetal10} on the Very Large Telescope (VLT), in seeing-limited mode, on the night of 6 to 7 July 2016. The log of the observations is provided in Table~\ref{Tab:Obs}. The sky conditions were rather good with a seeing of around 1 arcsec and under thin clouds. 
The instrument was used in its 
Wide Field Mode with the natural seeing (WFM--NOAO) configuration. This provides a nearly contiguous 1\,arcmin$^2$ field of view with  0.2-arcsec spatial sampling. 
In all cases, we used the extended mode of MUSE (WFM--NOAO--E), that covers the wavelength range $460-930$\,nm with an effective spectral resolution that increases from $R=1609$ at the bluest wavelengths to $R=3506$ at the reddest wavelengths.
The data are available at the ESO Science Archive under Prog.\ ID 097.D--0241(A). 
For each object, observations were obtained with dithering and a rotation of 90\degr{} between each exposure to remove artefacts during data processing.  Given that the objects are extended we obtained two sky frames on each sequence (see Table~\ref{Tab:Obs}) to perform an adequate sky subtraction. A few short exposures were also taken to analyse the strongest emission lines.
The data reduction was done with {\tt esorex}, using the dedicated ESO pipeline \citep{weilbacheretal14,weilbacheretal20}. For each individual frame the pipeline performs bias subtraction, flat-fielding and slice-tracing, wavelength calibration, geometric corrections, illumination correction using twilight sky flats, sky subtraction (telluric absorption/emission lines and continuum) making use of the sky frames obtained between science exposures. In the final steps of the reduction process, differential atmospheric correction and flux calibration were performed.

In the nominal wavelength range of MUSE,  second-order contamination in the red is suppressed by a blue cut-off filter that also avoids ghosts in that region. This is not the case with the extended wavelength range that we use, which extends down to 465\,nm, in order to have access to some important recombination lines. Thus, strong second-order contamination may appear redwards of 790\,nm, that reaches levels of $\simeq 12$ per cent at wavelengths redder than 850\,nm.
For the reddest of our lines, we can assume that this contamination is not significant as there is little continuum in our nebulae and the chance that one emission line in the blue superimposes on one in the red at exactly twice its wavelength is minimal. 

In this work we are mainly interested in the spatial distribution of the emission lines and the physical and chemical properties in the inner zones of PNe. Therefore, we trimmed the original data cubes to the central 30\,arcsec ($150 \times 150$ spaxels) for NGC\,6778 and M\,1--42 and the central 40\,arcsec ($200 \times 200$ spaxels) for Hf\,2--2. In Fig.~\ref{fig:extraction}, we show RGB composite images of the three PNe along with a box indicating the extracted area for each data cube. The images clearly show that the trimmed sections fully cover the emission of M\,1--42 and Hf\,2--2. However, for NGC\,6778 we focus on the bright central regions, neglecting the high-velocity northern and southern jets and part of the southern knot \citep[see fig.~3 of][for the nomenclature]{guerreromiranda12}.  These outer regions of NGC\,6778 are not analysed in this work as the faint ORLs (critical to study the ADF) are not detected there.

We checked for possible saturation of the brightest lines in preliminary emission maps extracted using {\sc qfitsview} \citep{ott12}. 
The {\foiii} $\lambda$5007 and $\lambda$4959 lines originate from the same energy level, so their intensity ratio only depends on the transition probabilities (A-values) ratio, being the most accepted theoretical value 2.98 \citep{storeyzeippen00}. We computed the {\foiii} $\lambda$5007/$\lambda$4959 intensity ratio maps and concluded that in NGC\,6778 and M\,1--42 this ratio departs significantly from the theoretical value in most exposures. This indicates that the {\foiii} $\lambda$5007 line is saturated in a large number of spaxels in the field of view, and hence we decided to measure only the {\foiii} $\lambda$4959 line in these PNe.
Similarly, we found for the same two objects that the {\fnii} $\lambda$6583 line is also saturated in a number of spaxels, as the observed {\fnii} $\lambda$6583/$\lambda$6548 line ratio departs significantly from the theoretical value of 2.94 \citep[see][]{frosefischertachiev04}, so we neglected this line in any calculations for  NGC\,6778 and M\,1--42. Finally, H$\alpha$ is also slightly saturated in some spaxels of M\,1--42. In Section~\ref{sec:extinction} we discuss the actions taken to address this issue. In Hf\,2--2, we found no signs of line saturation.

\section{Emission line measurements}
\label{sec:lines} 

To obtain the emission line maps with their uncertainties, we used a {\sc python} code that combines tools from the {\sc astropy} \citep{astropy:2013, astropy:2018} and {\sc pyspeckit} \citep[an extensible spectroscopic analysis toolkit for astronomy;][]{ginsburgmirocha11} packages. The code reads in the MUSE data cubes and performs individual Gaussian fits to previously selected  emission lines in each spaxel. The user supplies initial guesses for the line centre, amplitude and width and the fit is then performed in a specified region of the spectra extracted around the line centre. The fits are done using the Levenberg-Marquardt least-squares routine \citep[{\sc mptif};][]{markwardt2009}. We adopted the variance estimates of the data cube derived in the reduction process as the data uncertainties, which were treated as weights in the fits. The code outputs maps of the emission line flux, central wavelength, width and continuum as well as their respective uncertainties. We quadratically added an extra 5 per cent to the uncertainty in the line fluxes to account for systematic sources of error.

Continuum maps in the blue and red zones off the Paschen jump were also constructed using the automated line-fitting algorithm {\sc alfa} \citep{wesson16}, which provides continuum measurements at 8100 and 8400\,\AA.

\subsection{Emission line maps}
\label{sec:maps} 

\begin{figure*}
\includegraphics[scale = 0.45]{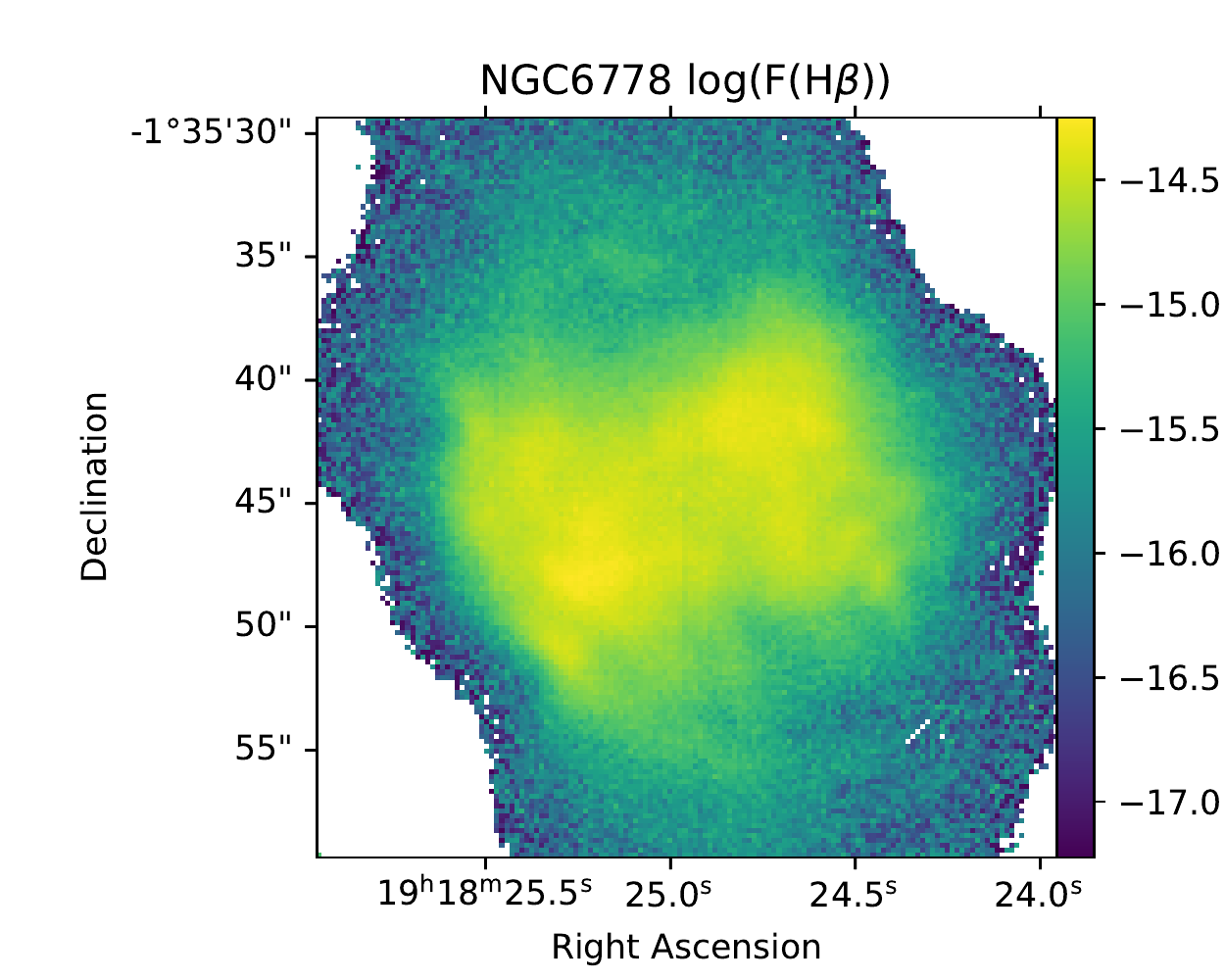}
\includegraphics[scale = 0.45]{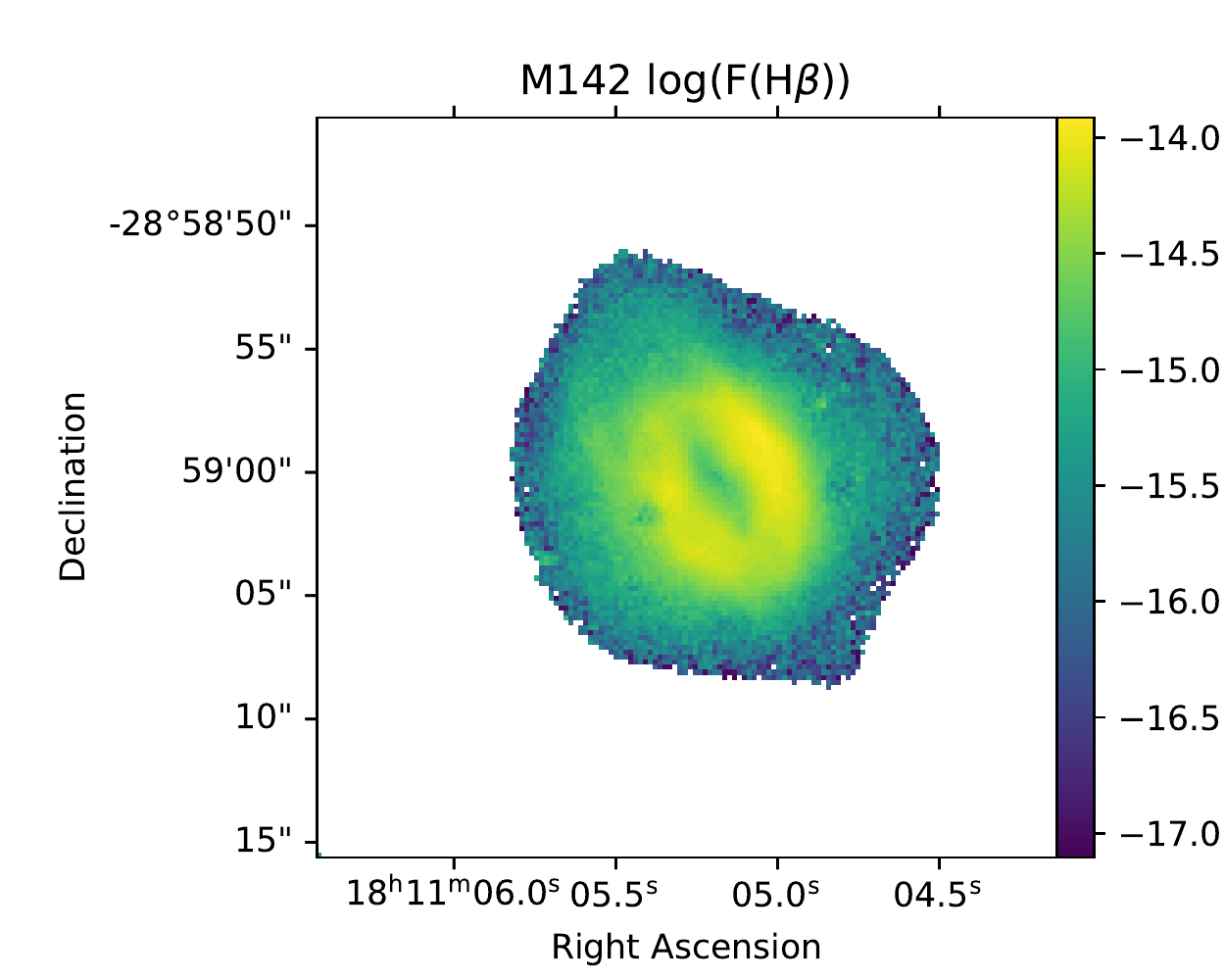}
\includegraphics[scale = 0.45]{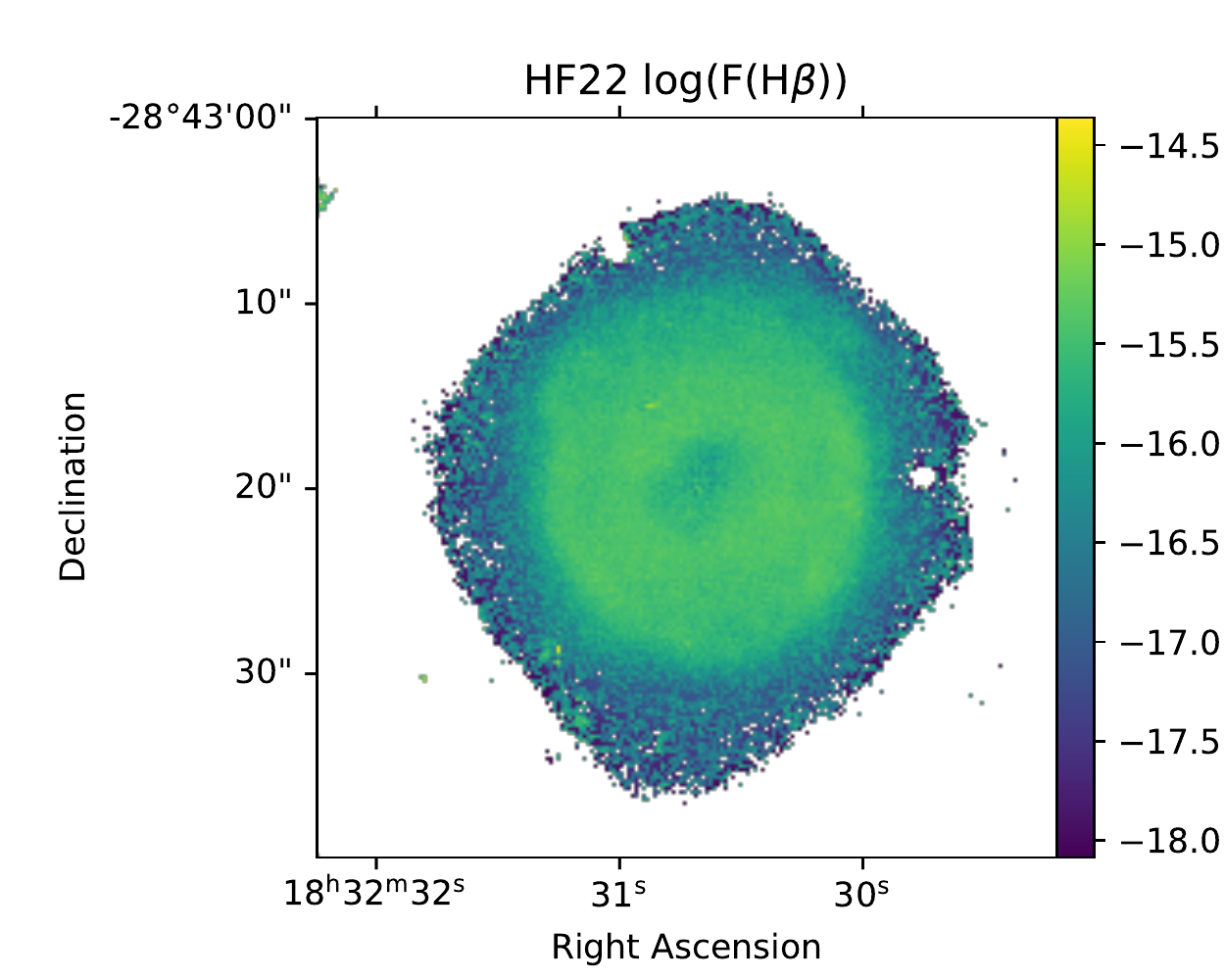}
\caption{H$\beta$ emission line maps of our three PNe on logarithmic scale. \label{fig:hbeta}}
\end{figure*}

For each PN, we constructed flux maps of more than 40 emission lines. We masked out the spaxels with very low signal-to-noise ratio: only spaxels with $F({\rm H}\beta) > 0.005 \times F({\rm H}\beta)_{\rm max}$ in the unreddened $F({\rm H}\beta)$ map were considered. This criterion does not affect the main body of each nebula and only masks spaxels where the measured quantities are likely unreliable.

In Figs.~\ref{fig:maps_ngc6778}, \ref{fig:maps_m142} and~\ref{fig:maps_hf22} we present the unreddened flux maps of some of the most relevant emission lines in NGC~6778, M\,1--42 and Hf\,2--2, respectively. We show the emission maps by increasing element mass and increasing ion ionization stage. Maps are displayed on logarithmic scale to enhance the fainter emitting structures of each line. For each nebula, the ionization structure is apparent, with the higher ionization stages emitted more internally than the lower ionization ones. A remarkable exception to this behaviour in found for the CNO recombination lines, which reveal that the emission is produced in a volume more internal than that expected for the CEL emission of the corresponding ions (see Section~\ref{sec:orls_cels}). The three PNe exhibit moderate excitation, with centrally peaked {\heii} emission. However, only M\,1--42 shows clear emission of the high-excitation {\farv} line in its central parts, while NGC\,6778 reveals a quite faint emission of this line, which is completely absent in Hf\,2--2. 
The flux distribution of the main gas shells in each nebula can be inferred from the H$\beta$ emission line maps on logarithmic scale presented in Fig.~\ref{fig:hbeta}, for which we have adopted the mask defined above. 

\subsubsection{Morphology from the emission line maps}

The morphology and kinematics of NGC\,6778 have been extensively studied by \citet{guerreromiranda12}. These authors found that its main nebular shell consists of a disrupted equatorial ring and kinematically disturbed bipolar lobes. The highly distorted and fragmented equatorial ring, aligned close to the east-west direction, can be easily seen in the {\foi}, {\fnitroi}, {\fsii} and {\fnii} emission line maps in Fig.~\ref{fig:maps_ngc6778}. On the other hand, the {\hi}, {\hei} and  mid-ionization ions like {\foiii}, {\fsiii} and {\fariii} show a much smoother spatial distribution than the low-ionization ones.

M\,1--42 was imaged by \citet{schwarzetal92} using narrow-band H$\alpha$ and {\foiii} $\lambda$5007 filters. It exhibits an elliptical morphology, with a slightly more extended angular diameter in H$\alpha$ than in {\foiii}, and a ring-like structure with a faint lobe to the north. More recently, \citet{guerreroetal13} analysed deep H$\alpha$ HST images and catalogued M\,1--42 as an elliptical PN with ansae/bipolar lobes. The presence of such ansae is evident in Fig.~\ref{fig:extraction}, where it is clear that they are relatively bright in low-ionization emission lines. However, they are quite faint in H$\beta$ and, therefore, have been excluded from our analysis following the criteria applied to build our maps (see above). 

\begin{figure}
\centering
\includegraphics[width=0.5\columnwidth]{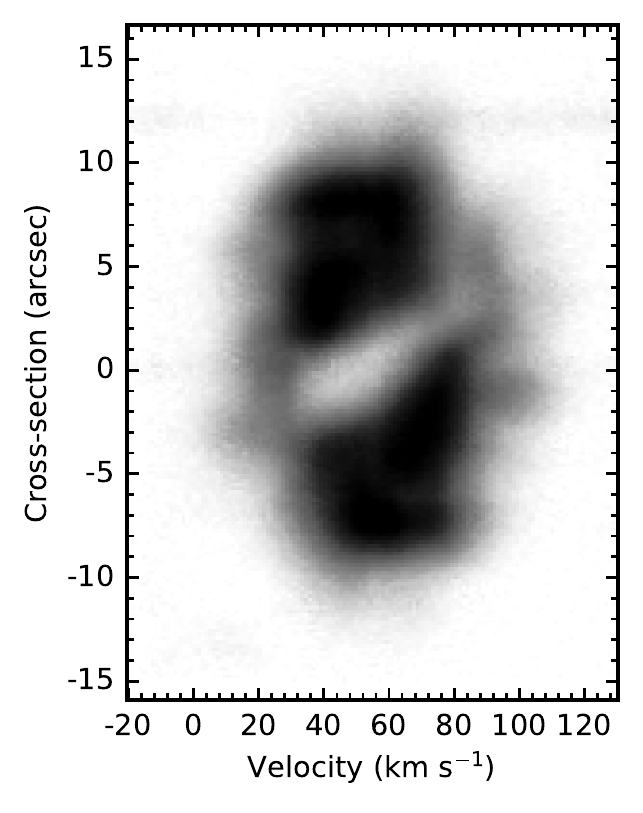}
\caption{H$\alpha$ position-velocity diagram from VLT--UVES long-slit data crossing the central star of Hf\,2--2 at a position angle of 0\degr{} (i.e. north is up). The observed profile is consistent with a bipolar morphology observed almost pole-on similar to that of Sp\,1 \citep{jones12}.
\label{fig:hf22_uves}}
\end{figure}

Hf\,2--2 was also observed by \citet{schwarzetal92}. This PN displays a highly symmetric morphology, with an inner disc-like shell very bright in {\foiii} and an outer limb-brightened shell with an angular diameter of $\simeq 20$\,arcsec. The inner disc (shell) is brighter in {\foiii} and has a central cavity, while the outer limb-brightened shell is bright in H$\alpha$ and low-ionization species, such as {\foi}, {\foii}, {\fnii} and {\fsii}, where the emission of the shell is shown to be disrupted in several bright knots.
This could be consistent with the PN being bipolar and observed nearly pole-on---such a nebular inclination would be consistent with the orbital inclination of the central binary star \citep{hillwig16}. Nevertheless, a detailed spatio-kinematical model is needed to confirm this hypothesis, but archival VLT--UVES high-resolution spatially-resolved spectroscopy (Fig.~\ref{fig:hf22_uves}) obtained in long-slit mode \citep[see][for a full description of this setup and the data reduction]{tyndall12} is consistent with a low inclination (i.e. pole-on) bipolar morphology similar to that of Sp\,1 \citep{jones12}. 

\subsubsection{Emission line maps of heavy element RLs}
\label{sec:orls_cels}

Only a handful of works have shown the spatial distribution of faint recombination lines in PNe, such as {\oi}, {\oii} and {\cii} \citep[see e.g.][]{tsamisetal08, alidopita19}. To our knowledge, this is the first time this has been done in high-ADF PNe using IFU data. As we point out in Section~\ref{sec:intro}, the emission of these lines is more centrally concentrated than the CEL emission of the same ions in PNe with high ADFs. This is clearly illustrated in Fig.~\ref{fig:maps_ngc6778}, where the RL emission of the {\oi} triplet $\lambda\lambda$7771+74+75 is produced in more central parts than the {\foii} $\lambda\lambda$7329+30 CEL emission. The same applies to the RL emission of {\oii} $\lambda\lambda$ 4649+50 RLs and the {\foiii} $\lambda$4959 CEL. Moreover, the spatial distribution of the {\cii} $\lambda$6462 and {\nii} $\lambda$5679 RLs resembles that of the {\oii} $\lambda\lambda$4649+50 RLs, although the C$^{2+}$, N$^{2+}$ and O$^{2+}$ ions do not share exactly the same ionization potential range (see Section~\ref{sec:ionic_orls} for a more detailed discussion on this). 

\subsection{Using Monte Carlo simulations to propagate uncertainties}
\label{sec:montecarlo}

We used a Monte Carlo (MC) based method  to assess the uncertainties in the physical and chemical spatial properties of the objects derived from the pipeline. After obtaining the extracted emission line maps (see Section~\ref{sec:lines}), we create 150 artificial ``fake" maps for each emission line.  Line intensities in each spaxel are generated using random values obtained from a normal distribution with a mean value equal to the observed intensity and a standard deviation given by its uncertainty. This produces data cubes of size $150 \times 150 \times 151$ or $200 \times 200 \times 151$, depending on the object, for each emission line (and continuum value). These synthetic data cubes are then used in each step of the pipeline to compute and apply the reddening corrections, to determine the physical parameters (electron temperatures and densities), to determine the ionic abundances corresponding to each emission line and the adopted ionic abundances and to determine the ionic ADFs. For each of these parameters and each spaxel of the object image, we then obtain a distribution of 151 values that allow to determine the ``original" value obtained from the observed line intensity, the ``median" value obtained from the distribution and the ``standard deviation" of these values.

The line intensities obtained from the integrated spectra (see Section~\ref{sec:collapsed}) follow the same process: we actually added the integrated value of each emission line to the pixel (0,0) of the corresponding map, ensuring that the integrated value follows exactly the same steps in the analysis pipeline than the individual spaxels of the maps.  

\section{Extinction maps}
\label{sec:extinction}

\begin{figure}
\includegraphics[scale = 0.6, trim={2cm 0cm 0cm 0cm}, clip]{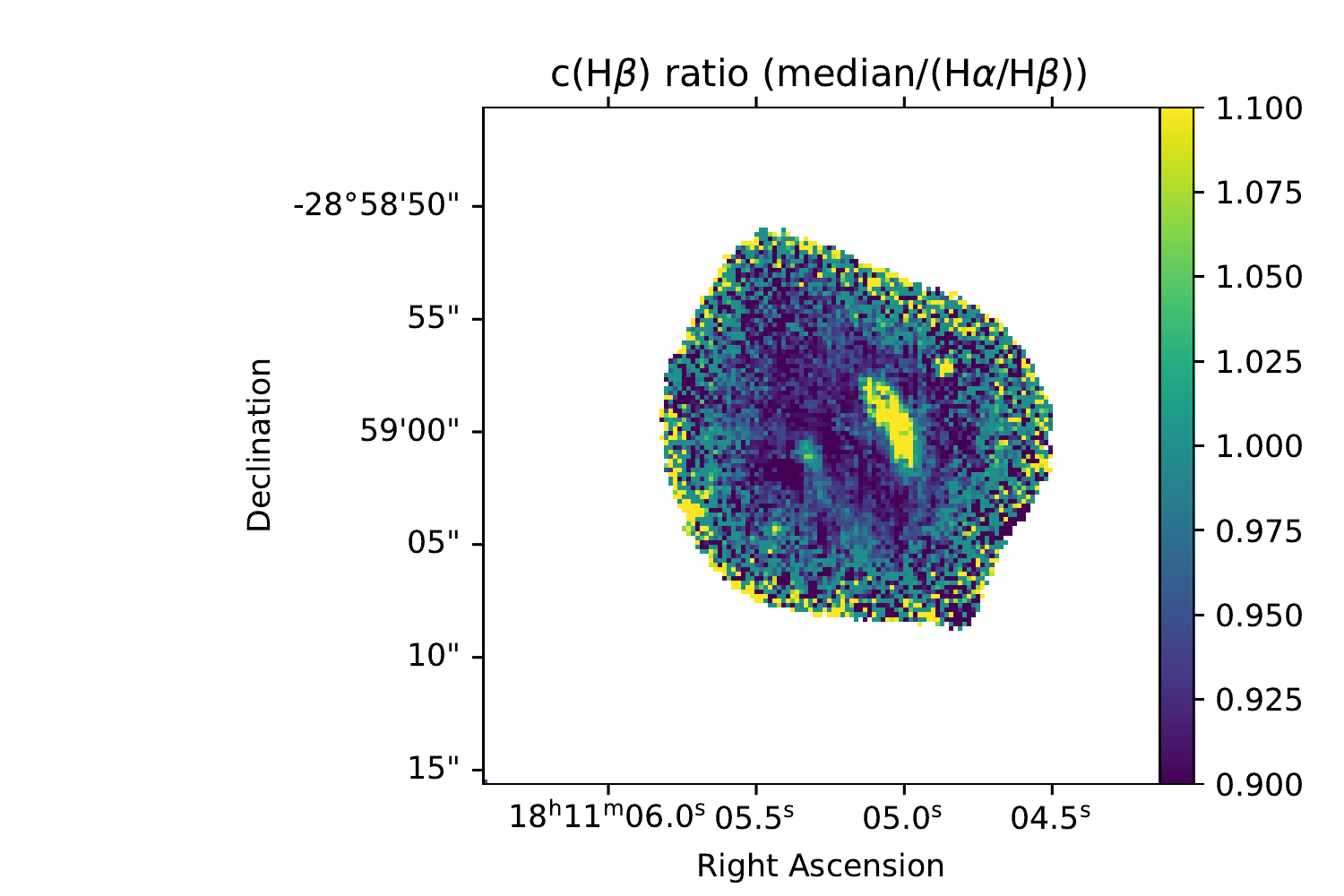}
\caption{The $c({\rm H}\beta)$ ratio obtained from the median of five {\hi} line ratios and from only the H$\alpha$/H$\beta$ in M\,1--42 (see text for details).
\label{fig:comp_chb_M142}}
\end{figure}

\begin{figure*}
\includegraphics[scale = 0.45]{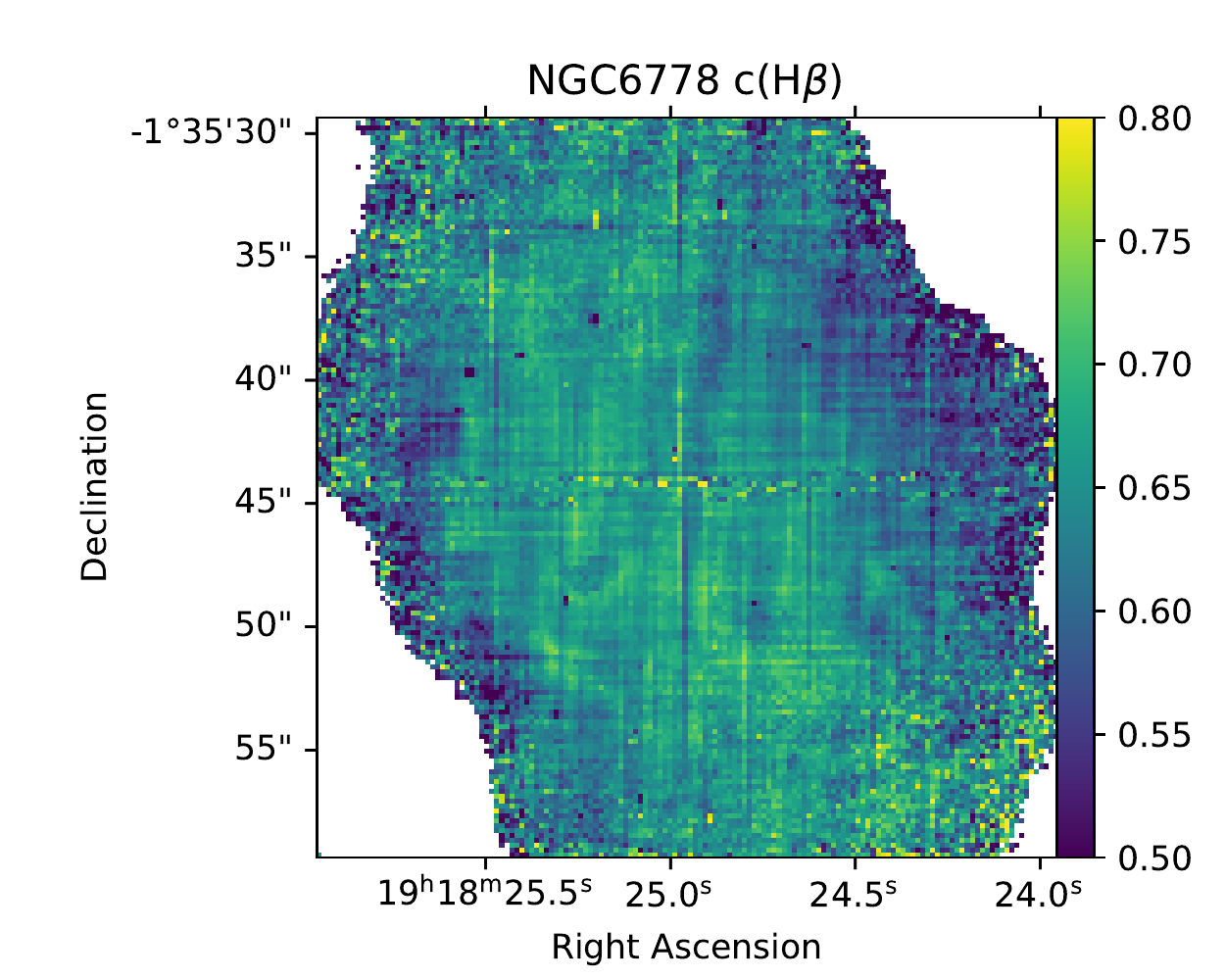}
\includegraphics[scale = 0.45]{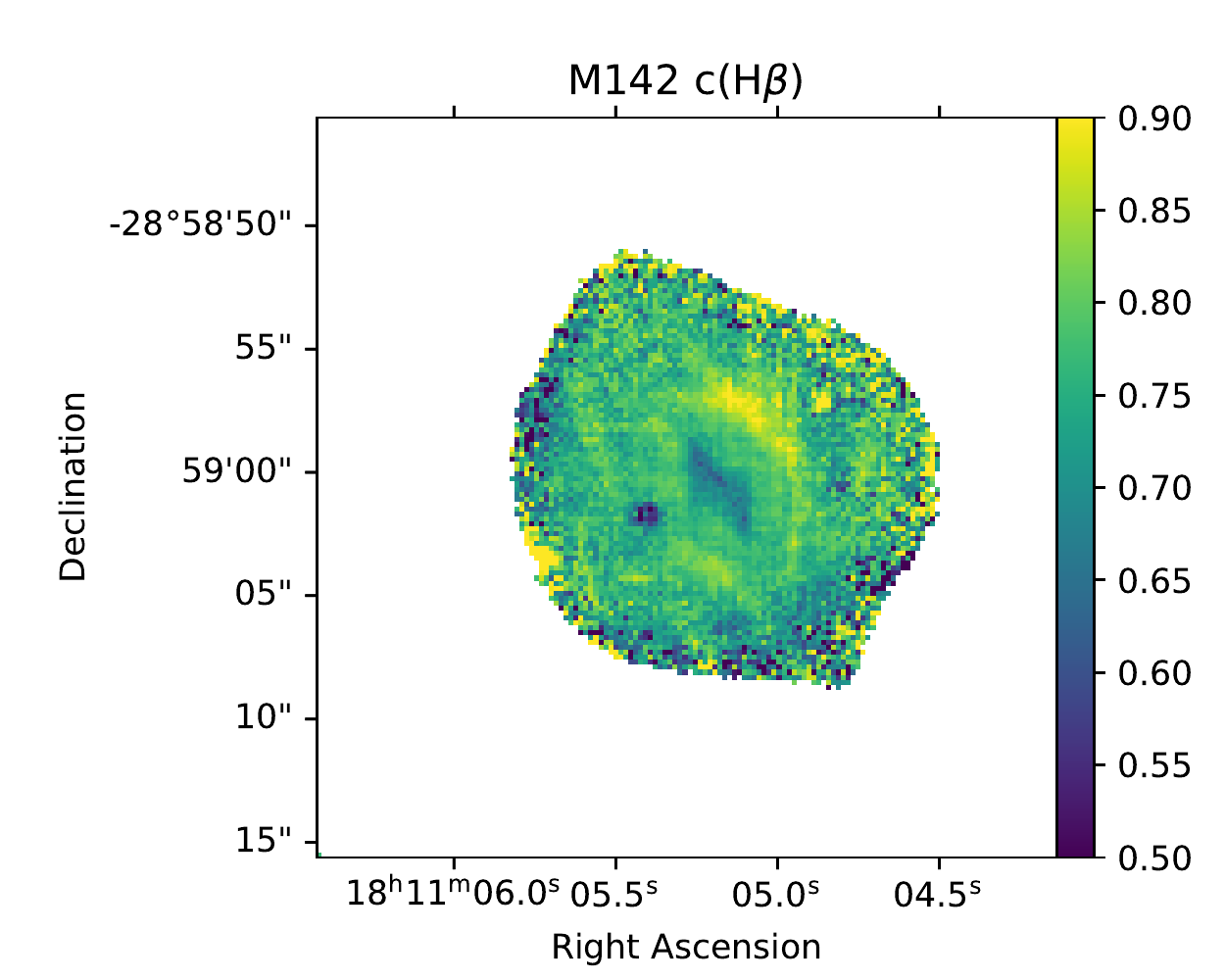}
\includegraphics[scale = 0.45]{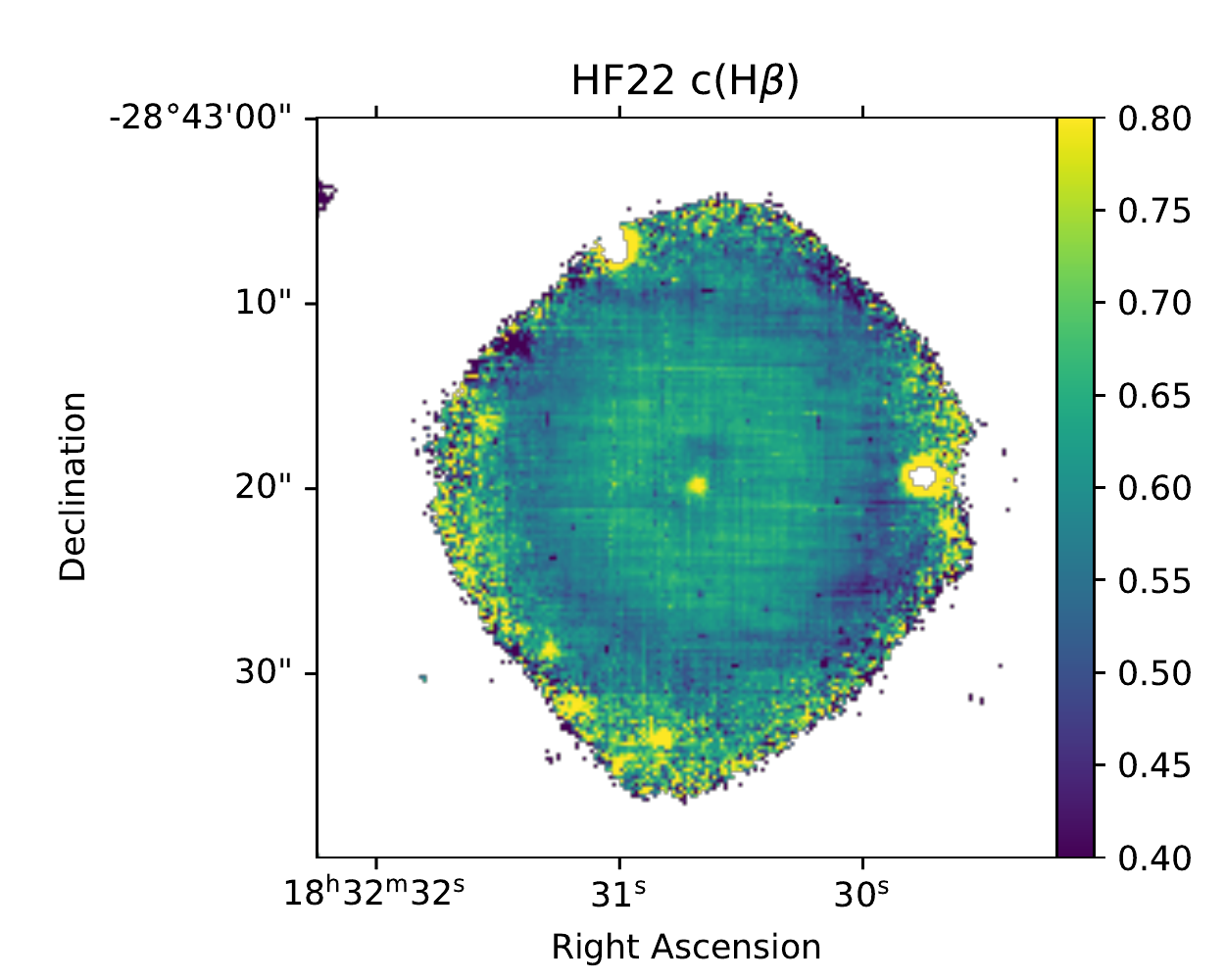}
\caption{Extinction maps, given by $c$(H$\beta$) computed from the H$\alpha$/H$\beta$ ratio for the three planetary nebulae of our sample
\label{fig:chb}}
\end{figure*}

The extinction coefficient, $c$(H$\beta$), maps were constructed using {\sc pyneb} \citep{luridianaetal15} by comparing the observed and theoretical H$\alpha$/H$\beta$ line ratios and adopting the Galactic extinction law by \citet{fitzpatrick99} and the theoretical {\hi} line ratios from \citet{storeyhummer95} for $T_{\rm e}=10\,000$\,K and $n_{\rm e}=1\,000$\,cm$^{-3}$. Notice that we did not try to correct the {\hi} emission lines for the adjacent {\heii} faint emissions, because at the relatively moderate excitation of our three PNe this contribution was always below 1 per cent, i.e. within the uncertainties. In M\,1--42, we detected that H$\alpha$ is slightly saturated in a small region located few arcseconds to the northwest of the central star, which provides an apparent lower extinction and also produces artifacts in the adjacent pixels. For this nebula we tried to construct the extinction map using a shorter exposure of 60\,s. However, the extracted extinction map had large uncertainties in the outer parts of the nebula, which propagated to the computed physical conditions and chemical abundances. Therefore, we decided to adopt a different strategy for the computation of the extinction map for M\,1--42: we built additional extinction maps using Paschen (P9 to P12) over H$\beta$ line ratios and computed the median $c({\rm H}\beta)$ map considering all the available maps. This strategy diluted the saturation effect in the inner parts of the nebula and conserved an adequate determination of the extinction coefficient in the outer parts. This is shown in Fig.~\ref{fig:comp_chb_M142}, where we plot the ratio between the median $c({\rm H}\beta)$ map and the one computed using only H$\alpha$/H$\beta$; the ratio is almost constant and close to 1 in all the nebula with the exception of the saturated spaxels. 

The extinction maps for each PN are shown in Fig.~\ref{fig:chb}. In Table~\ref{Tab:chb} we show the median $c$(H$\beta$) obtained using different {\hi} line ratios for our three objects, as well as the median value and standard deviation in the integrated spectra. For NGC\,6778 and  Hf\,2--2, only the $c({\rm H}\beta)$ obtained from the H$\alpha$/H$\beta$ ratio is considered. Overall, the $c({\rm H}\beta)$ values obtained using different {\hi} lines are in a relatively good agreement for all the objects, with the exception of the value obtained with the P10/H$\beta$ ratios, which is systematically lower and may be indicative of telluric absorption affecting the flux measurements of the P10 line, especially in the case of Hf\,2--2, for which the signal-to-noise ratio of the lines is the lowest in our sample (and the exposure times the largest by far).

\begin{table}
\caption{Median extinction coefficient, $c({\rm H}\beta)$, computed from different line ratios with respect to H$\beta$.}
\label{Tab:chb}
\begin{center}
\resizebox{\columnwidth}{!}{%
\begin{tabular}{ccccccc}
\hline
Object & H$\alpha$ & P9  & P10 & P11 & P12 & {\bf Integrated}  \\
\hline
NGC\,6778 & 0.64  & 0.58  &  0.54  &  0.66  & 0.61 & \bm{$0.66\pm0.09$} \\
M\,1--42   & 0.80  &  0.75  &  0.71  &  0.78  & 0.80 & \bm{$0.77\pm0.04$}  \\
Hf\,2--2   & 0.60  &  0.46  &  0.27  &  0.51  & 0.48 & \bm{$0.61\pm0.08$} \\
\hline
\end{tabular}
}
\end{center}
\end{table}

In general, the maps do not exhibit large extinction variations for each nebula, although relatively small differences can be appreciated. In the case of NGC\,6778, it is apparent that the higher extinction values follow a region delimited by the curved jets, where H$\alpha$ dominates the emission of the nebula \citep[see the H$\alpha$/{\foiii} and {\fnii}/{\foiii} ratios in][]{guerreromiranda12}. For M\,1--42, the extinction is quite constant, with the exception of the central cavity, where the extinction is consistently lower, and a thin slab of gas northwest of the central cavity with somewhat higher extinction. Finally, Hf\,2--2 apparently shows a slightly higher extinction in the inner parts surrounded by a ring that delimites a lower extinction. 

\section{Mapping the physical conditions}
\label{sec:phys_cond}

The physical conditions were computed with {\sc pyneb} v1.1.15 \citep{luridianaetal15} using the atomic data set presented in Table~\ref{tab:atomic_data}. As our targets have the peculiarity of being strong emitters of heavy-element recombination lines, we have to be careful when computing physical conditions for the different plasma components. Thus, in this Section we describe the methodology followed to compute the physical conditions using classical CEL diagnostics, paying special attention to the potentially strong effect of the recombination contribution to the faint auroral lines. We finally describe how we compute {\te} using recombination line and continuum diagnostics.

\begin{table}
\caption{Atomic data sets used for the CELs and ORLs. \label{tab:atomic_data}}
\resizebox{\columnwidth}{!}{%
\begin{tabular}{lcc}
\hline
\multicolumn{3}{c}{CELs} \\
Ion & Transition Probabilities & Collision Strengths \\
\hline
O$^{0}$   &  \citet{wieseetal96} & \citet{bhatiaetal95}\\
O$^{+}$   &  \citet{wieseetal96} & \citet{kisieliusetal09}\\
O$^{2+}$  &  \citet{frosefischertachiev04} & \citet{storeysochi14}\\
          &  \citet{storeyzeippen00} & \\
N$^{+}$   &  \citet{frosefischertachiev04} & \citet{tayal11}\\
S$^{+}$   &  \citet{rynkunetal19} & \citet{tayal10}\\
S$^{2+}$  &  \citet{froesefischeretal06} & \citet{tayalgupta99}\\
Cl$^{2+}$ &  \citet{rynkunetal19} & \citet{butlerzeippen89}\\
Cl$^{3+}$ &  \citet{mendozazeippen82a} & \citet{butlerzeippen89}\\
 &  \citet{kaufmansugar86}  & \\
Ar$^{2+}$ &   \citet{munozburgosetal09}  & \citet{munozburgosetal09}\\
Ar$^{3+}$ &   \citet{rynkunetal19}  & \citet{ramsbottombell97}\\
Ar$^{4+}$ &   \citet{kaufmansugar86}  & \citet{galavisetal95}\\
 &  \citet{mendozazeippen82a}  & \\
\hline
\multicolumn{3}{c}{ORLs} \\
Ion & \multicolumn{2}{c}{Effective Recombination Coefficients}
 \\
\hline
H$^{+}$   & \multicolumn{2}{c}{\citet{storeyhummer95}} \\
He$^{+}$   & \multicolumn{2}{c}{\citet{porteretal12,porteretal13}} \\
He$^{2+}$   & \multicolumn{2}{c}{\citet{storeyhummer95}} \\
O$^{+}$   & \multicolumn{2}{c}{\citet{pequignotetal91}} \\
O$^{2+}$   & \multicolumn{2}{c}{\citet{storeyetal17}}\\
C$^{2+}$   & \multicolumn{2}{c}{\citet{daveyetal00}} \\
N$^{2+}$   & \multicolumn{2}{c}{\citet{fangetal11, fangetal13}}\\
\hline
\end{tabular}
}
\end{table}

\subsection{Electron temperatures and densities from CELs}
\label{sec:te_ne_cels}

Every 2D data set of a given emission line considered in this work contains $150^2$ or $200^2$ pixels, depending on the object. For each pixel, we consider an MC distribution of 150 ``fake" observations (see Section~\ref{sec:montecarlo}). This leads to between three and six million intensity values for each emission line. Computing the values of the electron temperature and density corresponding to the intersection of two diagnostic line ratios in the {\te}--{\elecd} plane for each one of these data points may be very CPU time consuming. To speed up the computation process, we took advantage of the {\sc pyneb} \verb!Diagnostic.getCrossTemDen! method that uses a Machine Learning (ML) algorithm (from v1.1.13 on). An artificial neural network (ANN) is trained with 900 pairs of diagnostic line ratios corresponding to the {\te}--{\elecd} pairs covering the range of expected values ($3\,000 \leq T_{\rm e}\,{\rm [K]} \leq 30\,000$ and $1 \leq \log\,(n_{\rm e}\,{\rm [cm^{-3}])} \leq 6$). This training phase is done only once and takes a few minutes. The ANN is then used to predict the {\te}--{\elecd} values corresponding to the millions of observed and MC line ratios in less than two seconds. The ANN is implemented directly in {\sc pyneb} using the {\sc ai4neb} facility (Morisset et al., in prep.) that, in the present case, interfaces the {\sc scikit-learn} package \citep{pedregosaetal11}. The ANN consists of three hidden layers of 10, 20 and 10 cells using \verb!tanh! activation. The solver used is \verb!lbfgs!. The input line ratios have been transformed into their logarithmic values and scaled using the \verb!StandardScaler! routine from \verb!sklearn!. More details on the {\sc ai4neb} facility will be provided in Morisset et al. (in prep.).

\begin{figure*}
\includegraphics[scale=0.45,trim={3cm 5cm 0 5cm}, clip]{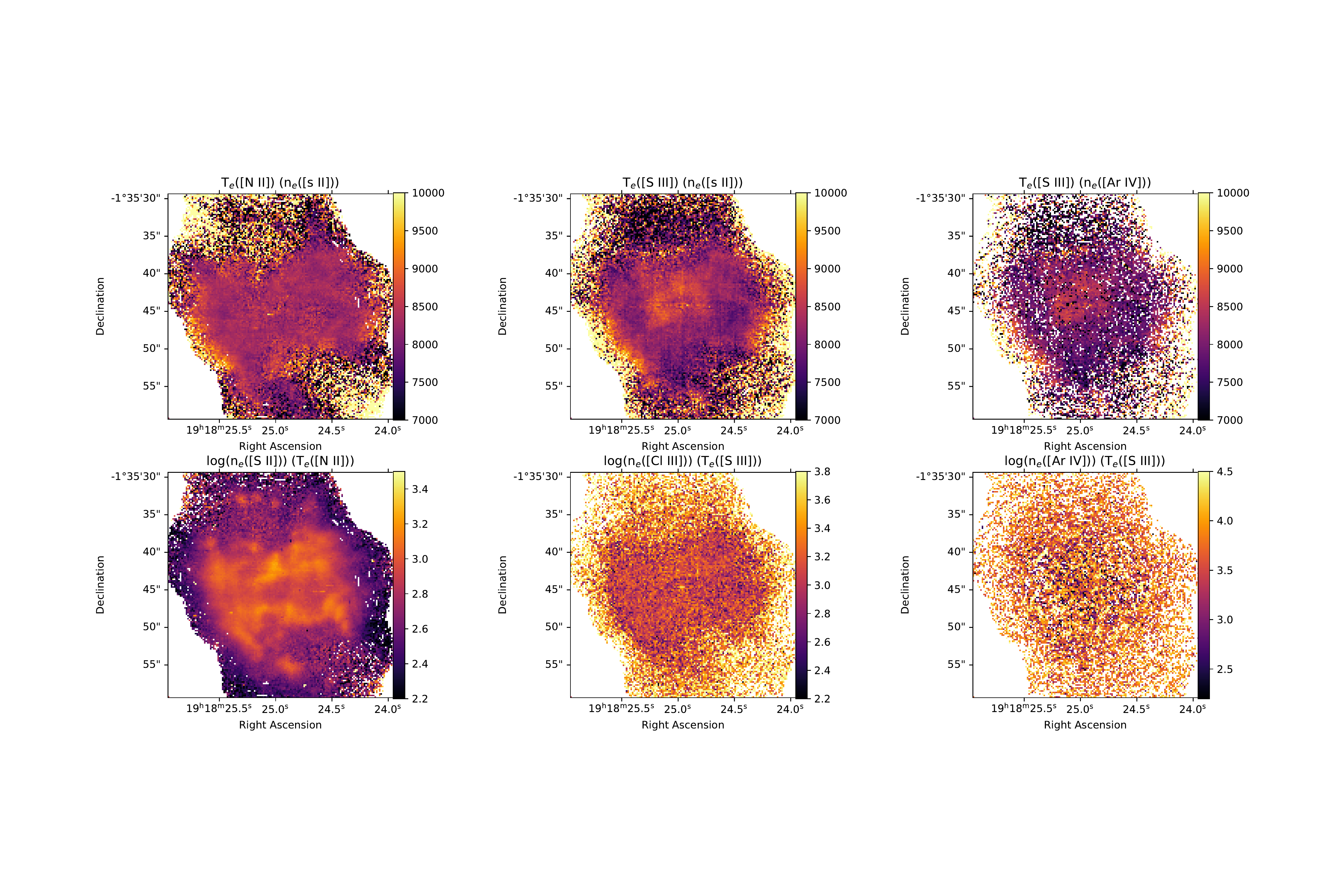}
\caption{Electron temperature, \te, and density, \elecd, maps obtained from the combination of different temperature and density diagnostics for NGC\,6778. The recombination contribution to {\fnii} $\lambda$5755 assuming $T_{\rm e}=4\,000$\,K has been considered. The same maps for M\,1--42 and Hf\,2--2 can be found in the online material.
\label{fig:NGC6778_TeNe_4000}}
\end{figure*}

We obtain the electron densities and temperatures for both each observed and MC spaxel, for the pairs of diagnostic line ratios described in Table~\ref{Tab:diags}. We adopted a classical two-zone scheme, i.e. we use the {\te} determined from \forbr{N}{ii}{5755}{6548} and \forbr{S}{iii}{6312}{9069} for the low ($\leq 17$\, eV) and high ($> 17$\,eV) ionization potential zones, respectively (see Section~\ref{sec:ionic}). It is important to emphasize that, although {\fsiii} $\lambda$9069 emission line can be affected by H$_2$O and O$_2$ telluric absorption bands, the MUSE pipeline effectively removes such signatures using dedicated sky exposures (see Section~\ref{sec:obs}). The electron density is determined using \forbr{S}{ii}{6731}{6716} in both zones. Generally, {\fcliii} and/or {\fariv} diagnostics are better suited for the high-ionization zone. However, the {\fcliii} density diagnostic leads to rather similar values to {\elecd}({\fsii}) in the main body of each PN , but a much higher {\elecd} in the external parts where the {\fcliii} lines emissivity significantly drops (see Figs.~\ref{fig:maps_ngc6778} to~\ref{fig:maps_hf22}). Electron densities based on {\fariv} are higher but the maps are quite noisy. This behaviour is due to the intrinsic difficulty in deblending {\hei} $\lambda$4713 from {\fariv} $\lambda$4711. Taking into account that in the density range below $10^4\,{\rm cm}^{-3}$ (which is the case for all the PNe studied here), the effect of {\elecd} on the absolute values derived for {\te}({\fsiii}) is small, we avoid the use of this {\elecd} diagnostic. In Figs.~\ref{fig:NGC6778_TeNe_4000},~\ref{fig:M142_TeNe_4000} and~\ref{fig:HF22_TeNe_4000} we show the obtained {\te} (upper panels) and {\elecd} (bottom panels) for NGC\,6778, M\,1--42 and Hf\,2--2, respectively. In these figures, we have corrected for the recombination contribution to the auroral {\fnii} $\lambda$5755 line assuming $T_{\rm e}=4\,000$\,K for the zone where recombination emission arises (see next Sections). Physical conditions diagnostic maps assuming $T_{\rm e}=1\,000$\,K, 8\,000\,K and no correction are available in the online material.

\begin{table}
\caption{Diagnostic line ratio pairs used to compute the physical parameters. The  {\te}--{\elecd} diagnostics used to determine ionic abundances are shown in bold font.}
\label{Tab:diags}
\begin{tabular}{llcl}
\hline
Ioniz. Zone & {\te} diagnostic &--& {\elecd} diagnostic  \\
\hline

Low & {\bf \forbr{N}{ii}{5755}{6548}} &--& {\bf \forbr{S}{ii}{6731}{6716}} \\
High & \forbr{S}{iii}{6312}{9069} &--& \forbr{Cl}{iii}{5538}{5518} \\
& {\bf \forbr{S}{iii}{6312}{9069}} &--& {\bf \forbr{S}{ii}{6731}{6716}} \\
& \forbr{S}{iii}{6312}{9069} &--& \forbr{Ar}{iv}{4740}{4711} \\
\hline
\end{tabular}
\end{table}

\subsection{Correction for recombination contribution}
\label{sec:rec_cont}

\citet{rubinetal86} pointed out that the low-lying metastable levels of several CELs, including the auroral {\fnii} $\lambda$5755 line and trans-auroral {\foii} $\lambda\lambda$7320+30 lines can be significantly populated by recombination processes. To illustrate how this can affect the emission maps, in Figs.~\ref{fig:maps_ngc6778} to~\ref{fig:maps_hf22} one can compare  the spatial distribution of {\fsii} $\lambda$6731 and {\foii} $\lambda\lambda$7329+30 (labelled as {\foii} 7330+). Although O$^+$ and S$^+$ have similar ionization potential ranges, the spatial distribution of their emission are remarkably different in {\foii} $\lambda\lambda$7330  and {\fsii} $\lambda\lambda$6731 lines (see Figs.~\ref{fig:maps_ngc6778} to~\ref{fig:maps_hf22}). As several authors have indicated \citep[see][and references therein]{wessonetal18}, second-row elements such as O and N can show a strong enhancement of their recombination lines. However, this does not seem the case for third-row elements (like S), so it is not expected that S lines can be affected by any recombination process. 

\begin{figure*}
\includegraphics[scale=0.45]{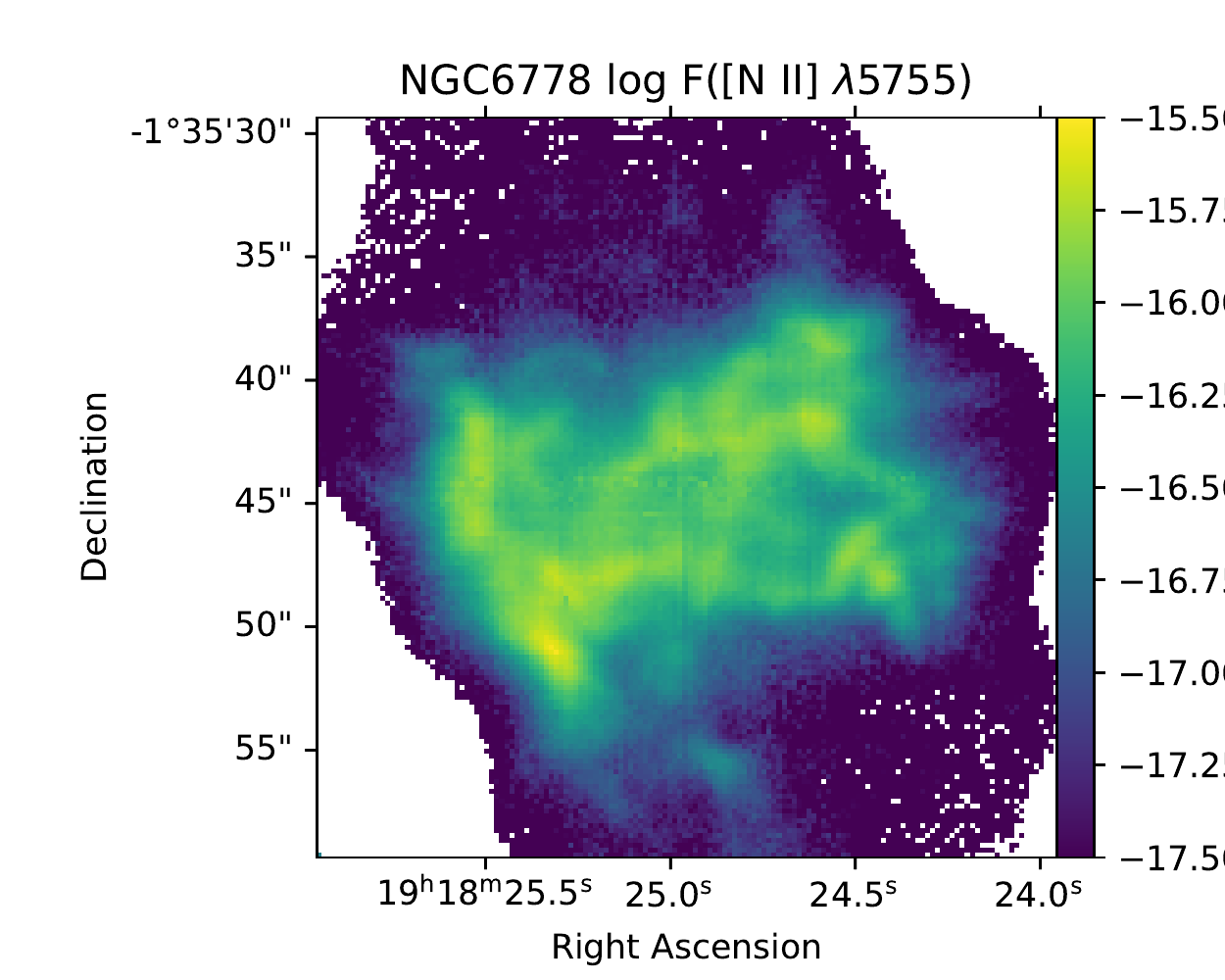}
\includegraphics[scale=0.45]{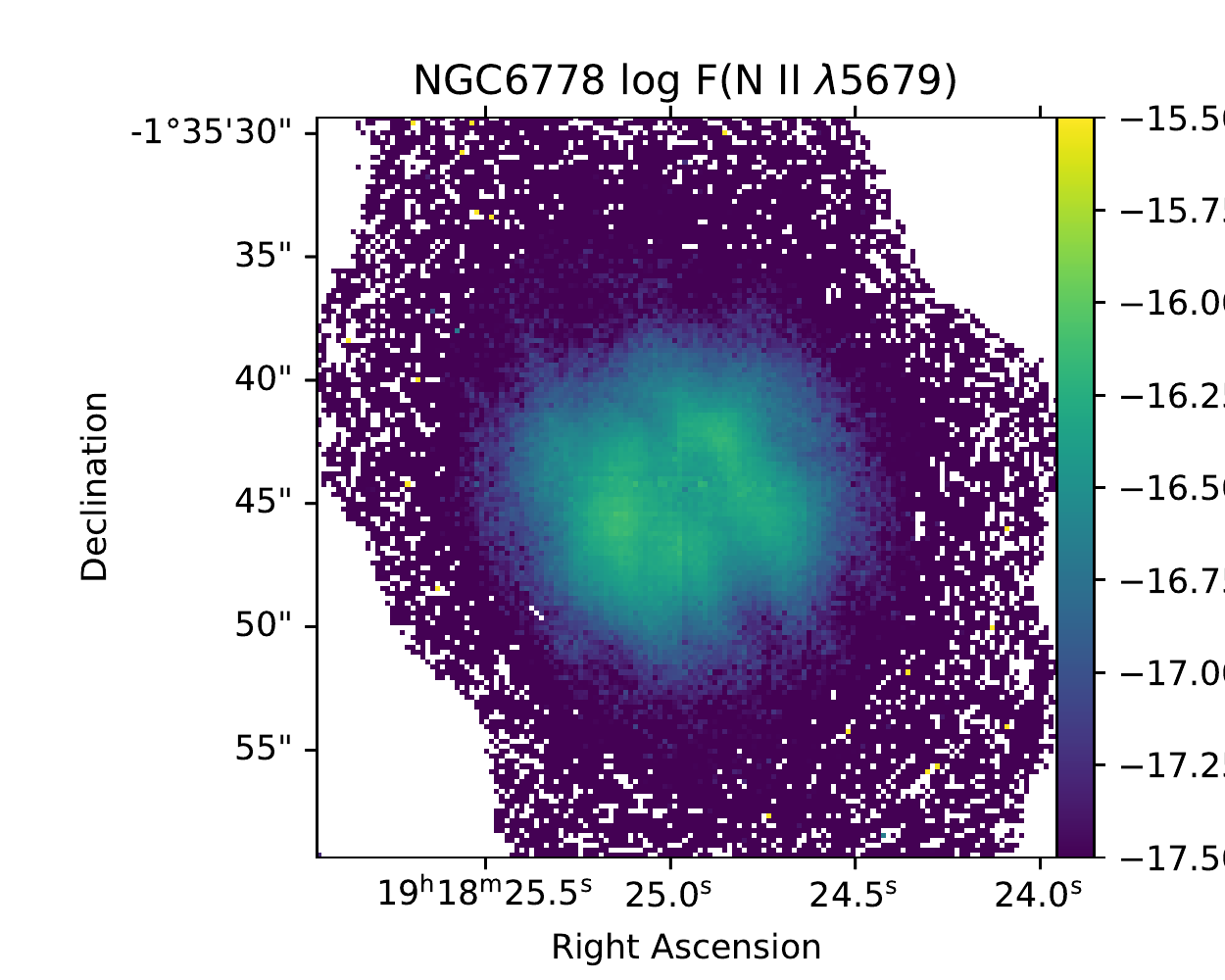}
\includegraphics[scale=0.45]{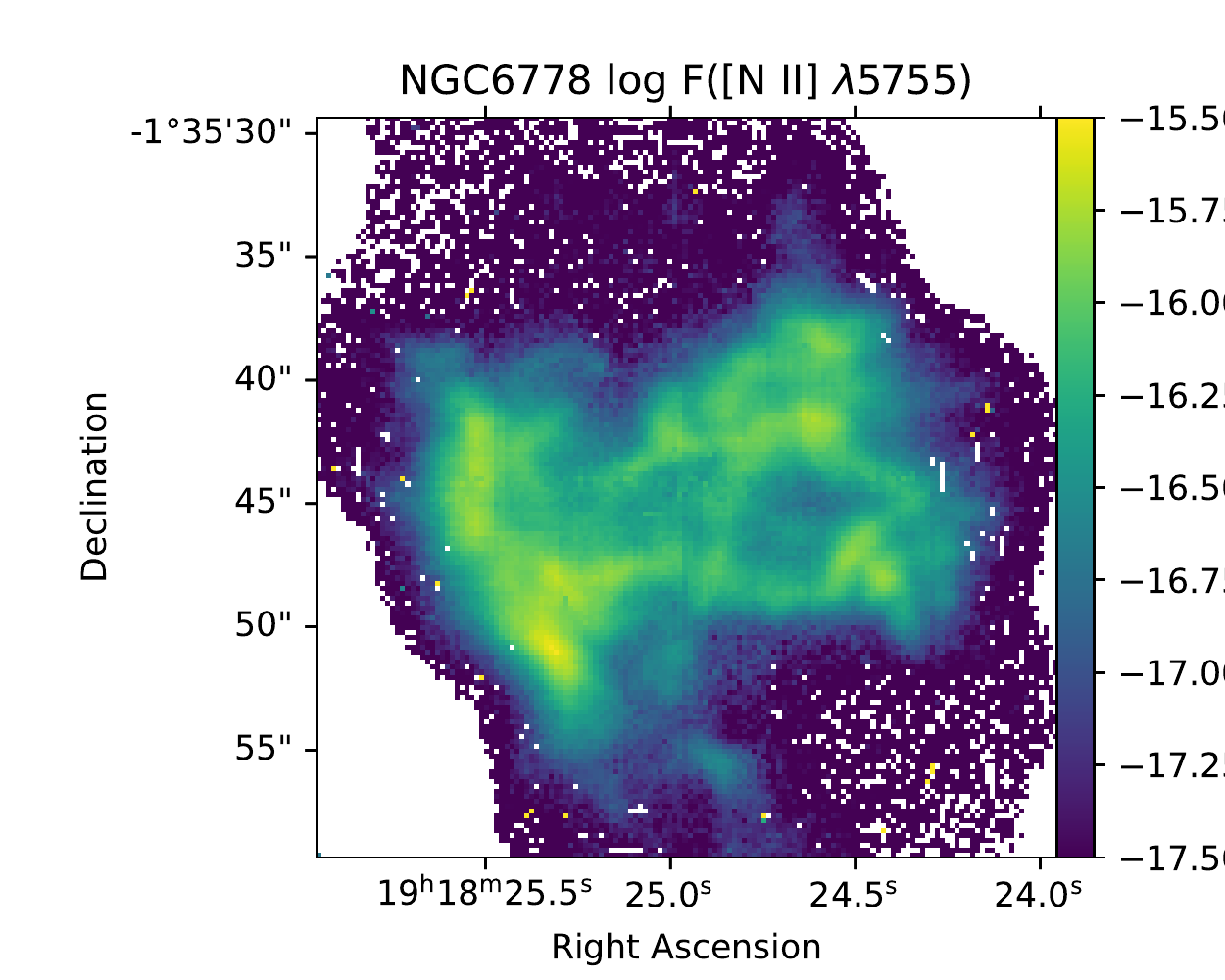}
\includegraphics[scale=0.45]{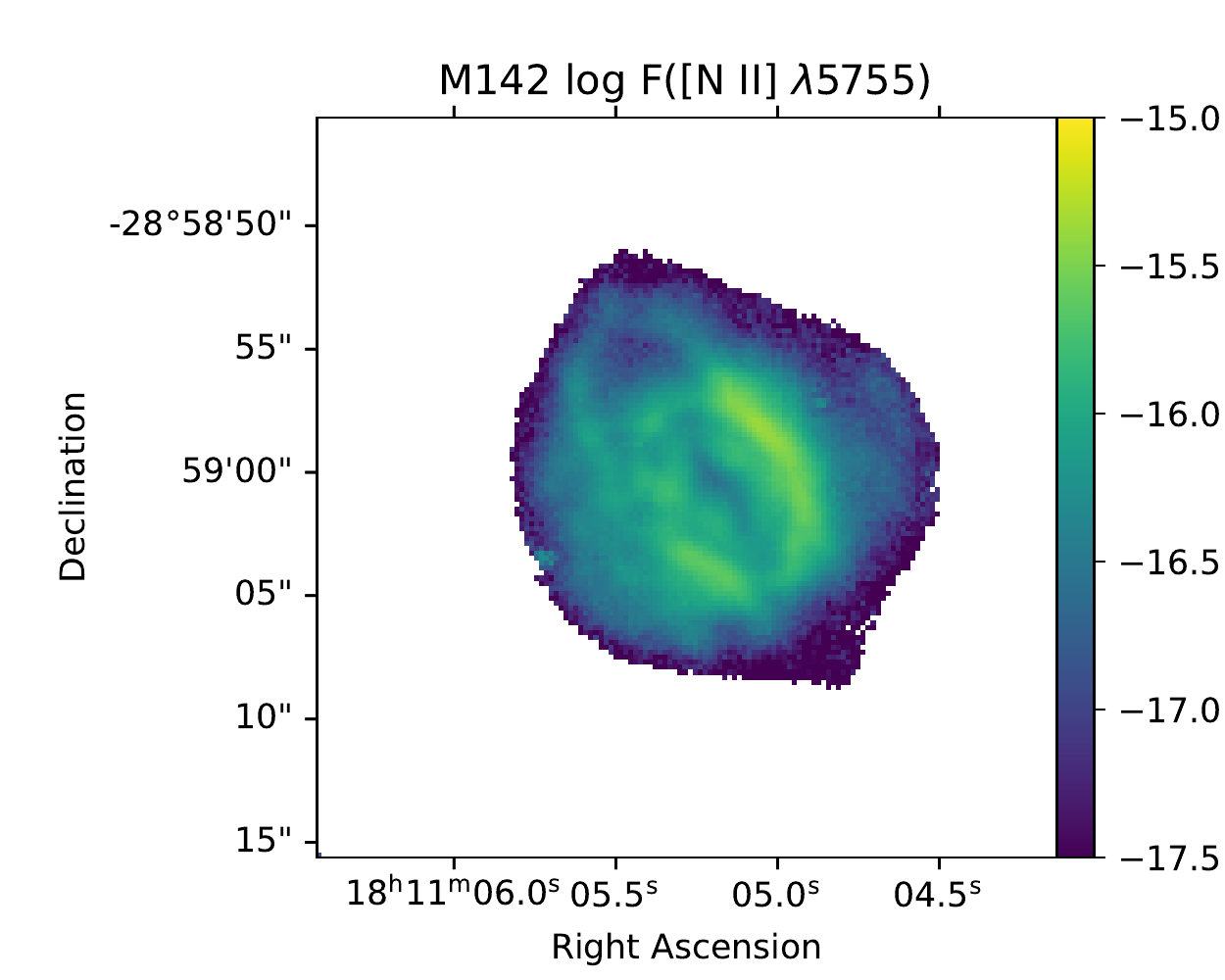}
\includegraphics[scale=0.45]{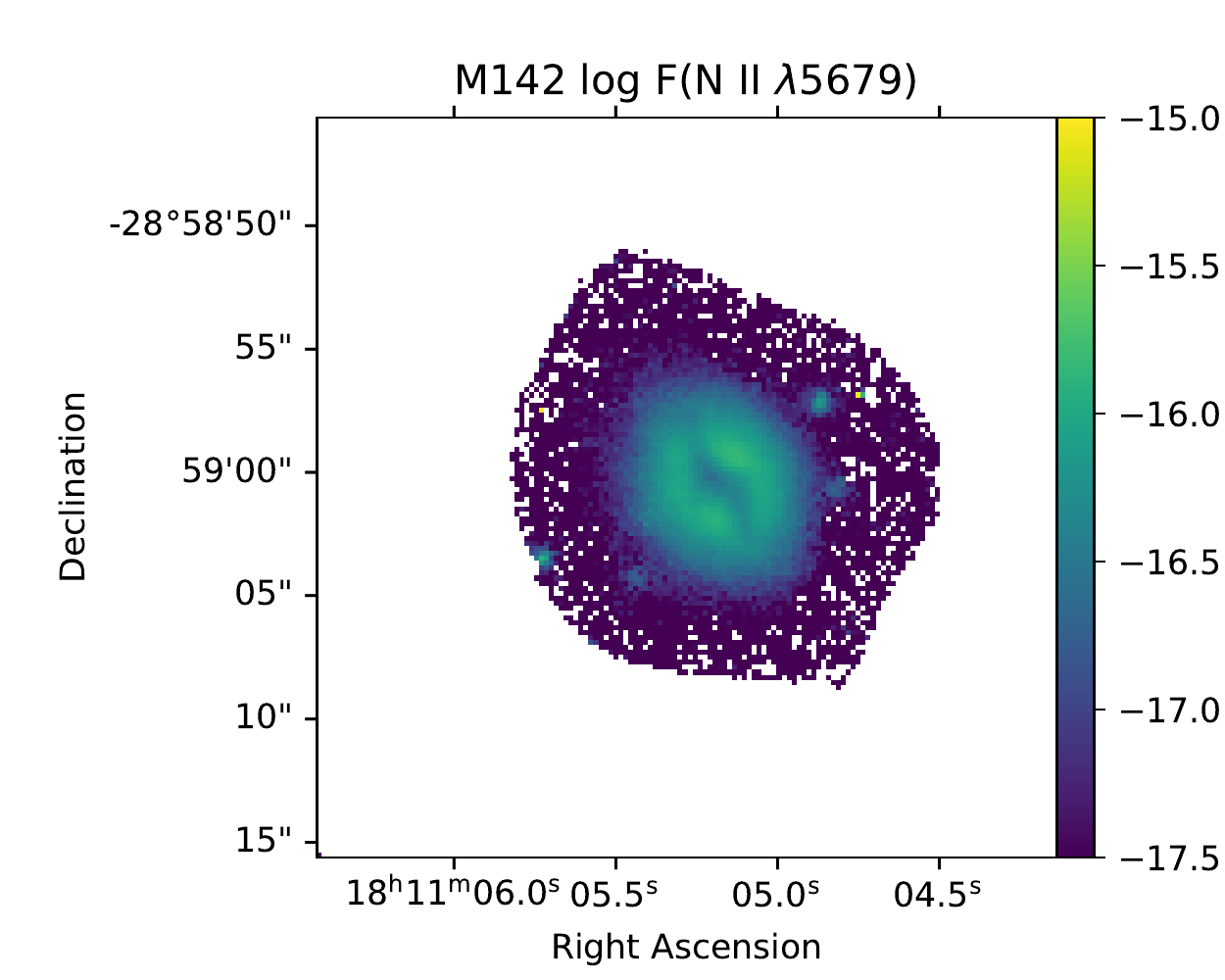}
\includegraphics[scale=0.45]{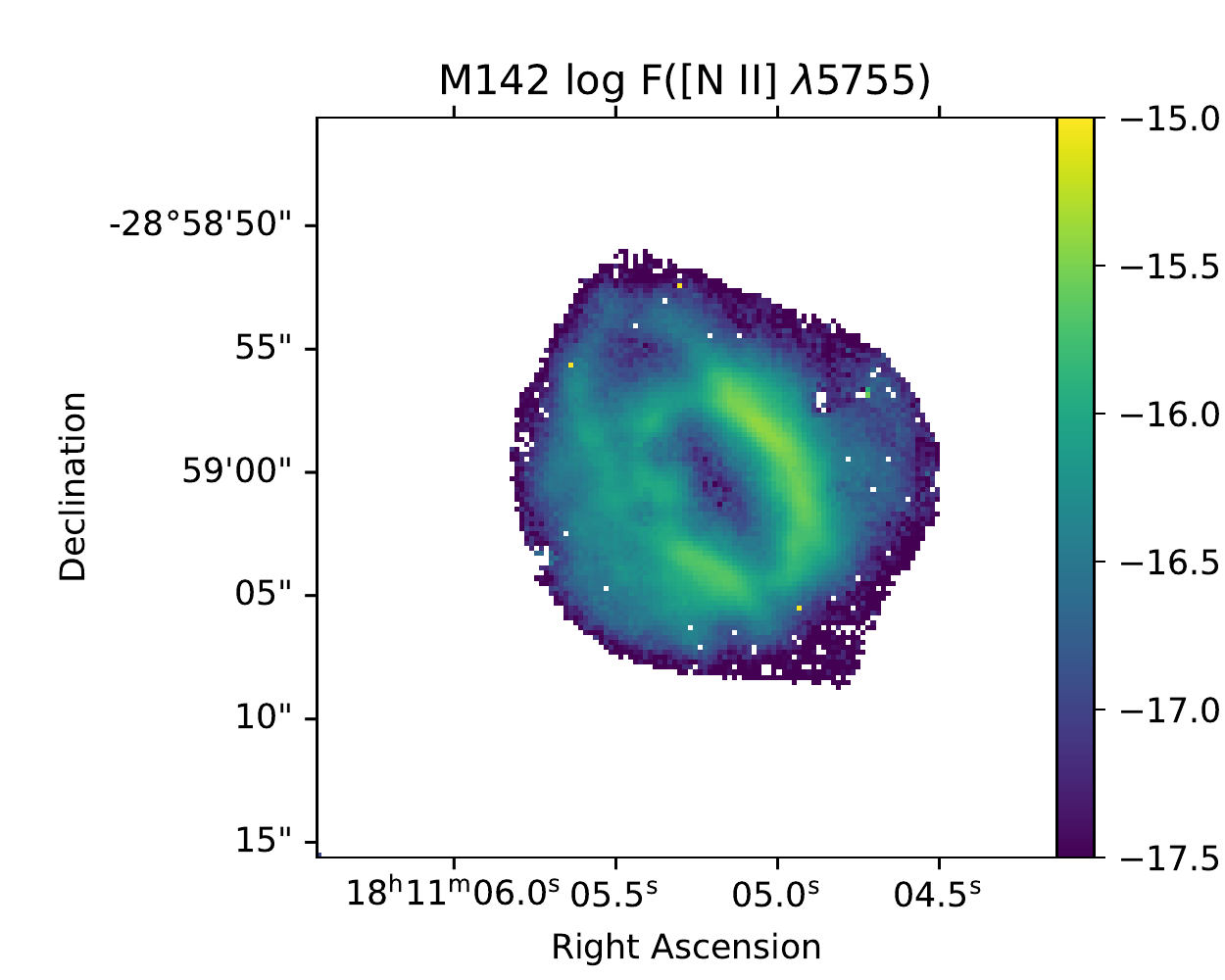}
\includegraphics[scale=0.45]{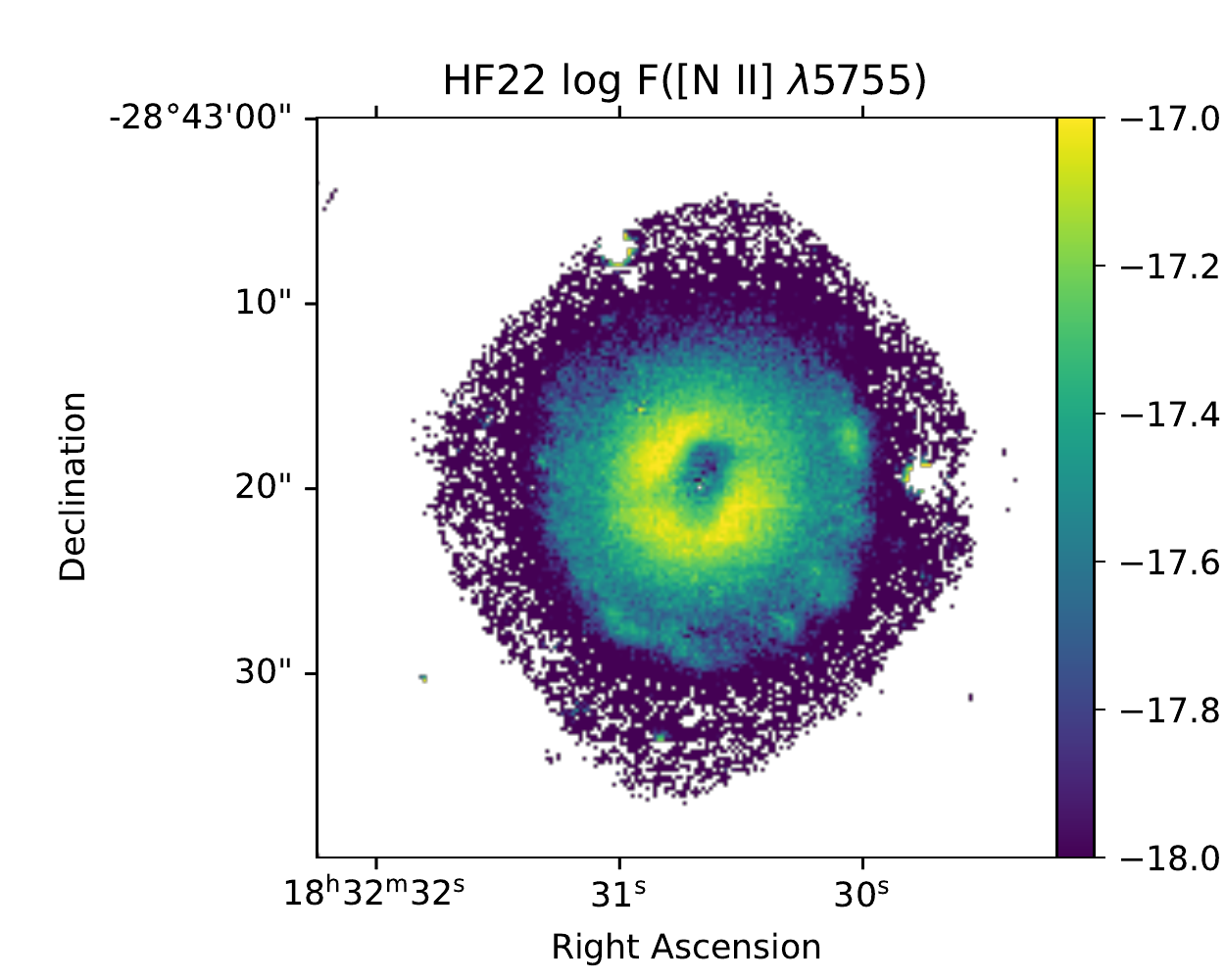}
\includegraphics[scale=0.45]{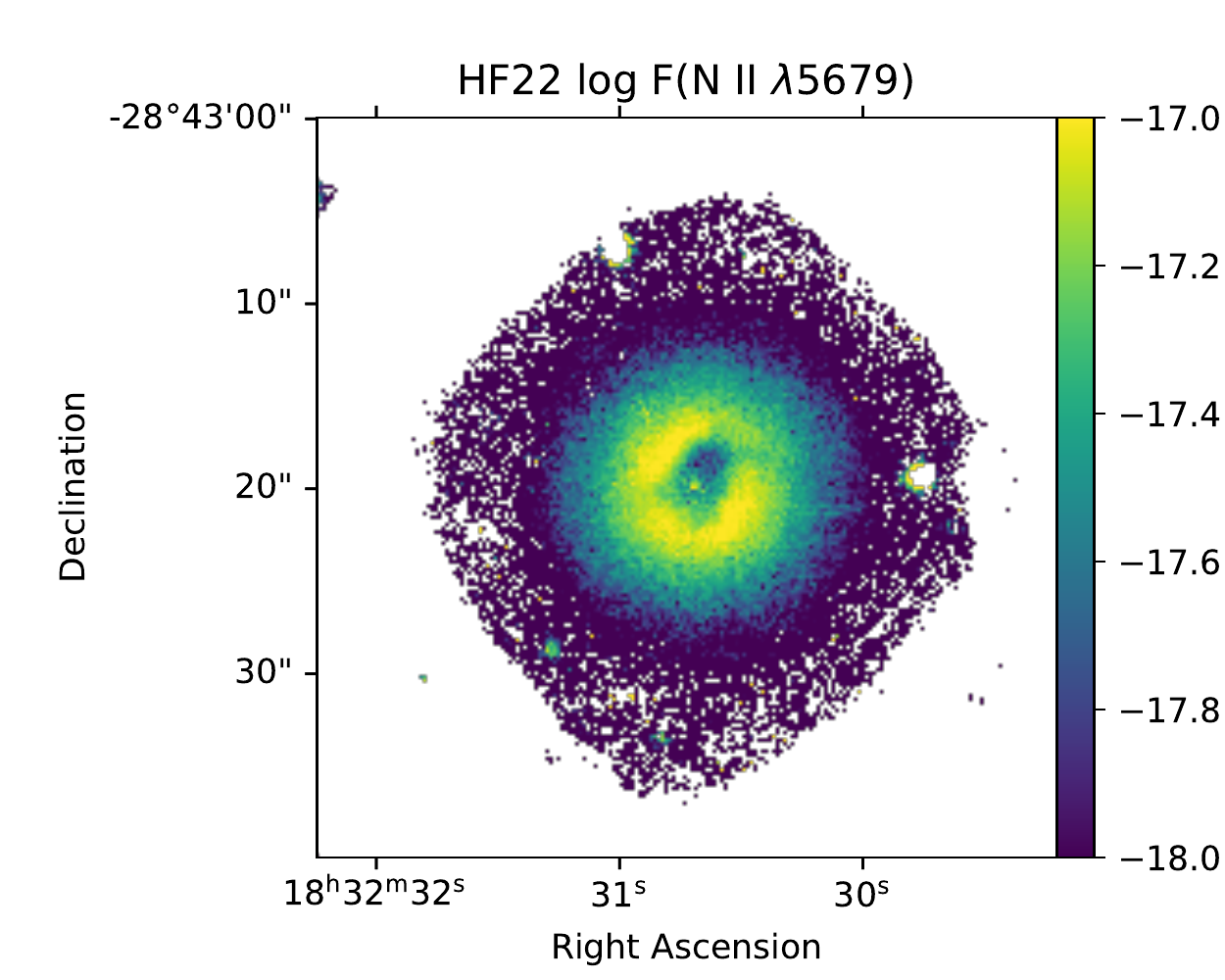}
\includegraphics[scale=0.45]{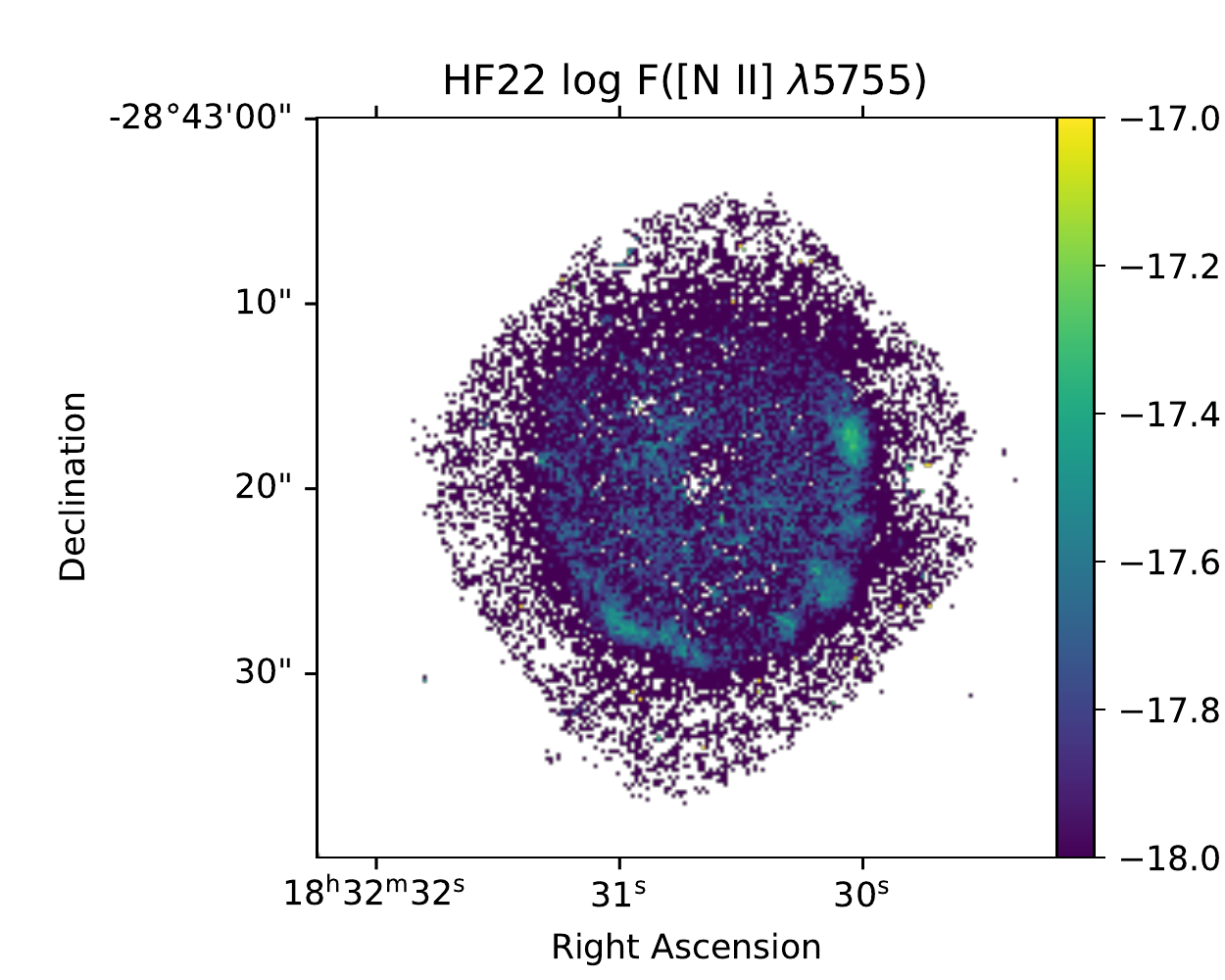}
\caption{Left panels: Spatial distribution of the auroral {\fnii} $\lambda$5755 emission line in our three PNe prior to applying the recombination contribution. Middle panels: Spatial distribution of the {\nii} $\lambda$5679 RL. Right panels: Same as left panels but after applying the recombination contribution correction, considering a constant {\te} for the recombination emission of 4\,000\,K. 
\label{fig:RecCont}}
\end{figure*}

\citet{liuetal00} proposed some recipes to compute the recombination contribution to the {\fnii} $\lambda$5755 and {\foii} $\lambda\lambda$7320+30 lines, which depends on the computed abundances of N$^{2+}$ and O$^{2+}$ from RLs and the assumed {\te}. However, these recipes are only valid over a limited {\te} range ($5\,000 \leq T_{\rm e}\,{\rm [K]} \leq 20\,000$) and it is reasonable to suspect that {\te} could be lower in high-ADF PNe. Additionally, \citeauthor{liuetal00}'s method requires a first guess of the N$^{2+}$ and O$^{2+}$ abundances and, therefore, would need an iterative process to properly apply the correction. For these reasons, we have decided to apply a more direct method to estimate this contribution. We used the radiative recombination coefficients for the metastable levels of {\fnii} and {\foii} calculated by \citet{pequignotetal91} and radiative recombination coefficients for {\nii} $\lambda$5679 \citep{fangetal11} and {\oii} $\lambda\lambda$4649+50 \citep{storeyetal17}, which are the brightest {\nii} and {\oii} ORLs in our spectra, and used their extinction corrected emission maps to estimate the contribution by simply applying the following: 

\begin{equation}
    I(5755)_{\rm corr} = I(5755) - \frac{j_{5755}(T_{\rm e}, n_{\rm e})}{j_{5679}(T_{\rm e}, n_{\rm e})}\times I(5679)
\end{equation}

\begin{equation}
    I(7320+30)_{\rm corr} = I(7320+30) - \frac{j_{7320+30}(T_{\rm e}, n_{\rm e})}{j_{4649+50}(T_{\rm e}, n_{\rm e})}\times I(4649+50)~,
\end{equation}

\noindent
where $j_{\nu}$({\te}, {\elecd}) are the recombination emissivities for each line. In these equations, {\elecd} was set to a constant value of $10^3$\,cm$^{-3}$ and {\te} was explored using three different values (1\,000, 4\,000 and 8\,000\,K). In Fig.~\ref{fig:RecCont} we present the {\fnii} $\lambda$5755 (left panels) and {\nii} $\lambda$5679 (central panels) emission maps, as well as the corrected {\fnii} $\lambda$5755 maps assuming $T_{\rm e}=4\,000$\,K for the recombination emission (right panels). It is evident that the effect is not negligible in any of the three PNe and can be dramatic in Hf\,2--2, where almost all the observed emission of the auroral {\fnii} line is due to recombination.

We have checked that the corrections obtained using our recipes are within 10 per cent of those derived from \citet{liuetal00}. As we pointed out before, the advantage of our method is that it does not rely on the ionic abundances of the recombining ions, and then can be applied at an earlier step in the analysis pipeline, before starting to compute the ionic abundances.

Possible effects of continuum fluorescence on the intensities of the {\cii}, {\nii}, {\oi} and {\oii} lines in the low-ionization PN IC\,418 were explored by \citet{escalanteetal12}. These authors found that continuum fluorescence significantly affects the intensity of the {\nii} $\lambda$5679 ORL (by $\simeq 70$ per cent) and {\oii} $\lambda$4649+50 ($\simeq 20-30$ per cent), leaving the {\cii} $\lambda\lambda$5342, 6462 and {\oi} $\lambda$7771+74+75 lines unaffected. Given that our objects are all highly excited PNe, we do not expect such a high contribution of fluorescence to be responsible for the excitation of {\nii} and {\oii} lines from the s, p and d states. However, as we will see in detail in Section~\ref{sec:ionic_orls}, the excitation in the cold gas is much lower than in the ``normal'' gas in all our targets, so we cannot completely rule out that fluorescence can contribute to some extent to the observed fluxes. Unfortunately, without detailed information from the UV spectra of the ionizing sources, we cannot evaluate the significance of this effect on the observed intensities. 

Last but not least, it is important to emphasize that in highly ionized nebulae the recombination contribution can also be important for the auroral {\foiii} $\lambda$4363 CEL \citep{liuetal00}. However, we are currently unable to check for this as {\foiii} $\lambda$4363 lies outside the wavelength range of our MUSE data. However, this effect should be carefully assessed when computing {\te}({\foiii}), because in some highly-excited PNe with high ADFs, the auroral {\foiii} $\lambda$4363 line could be strongly excited by recombination, making this {\te} diagnostic unreliable \citep[see][]{gomezllanosetal20}.

\subsubsection{Effect of the recombination contribution on the {\normalfont [N\,{\sc ii}]} electron temperature maps.}
\label{sec:rec_cont_te}

\begin{figure*}
\includegraphics[scale=0.33]{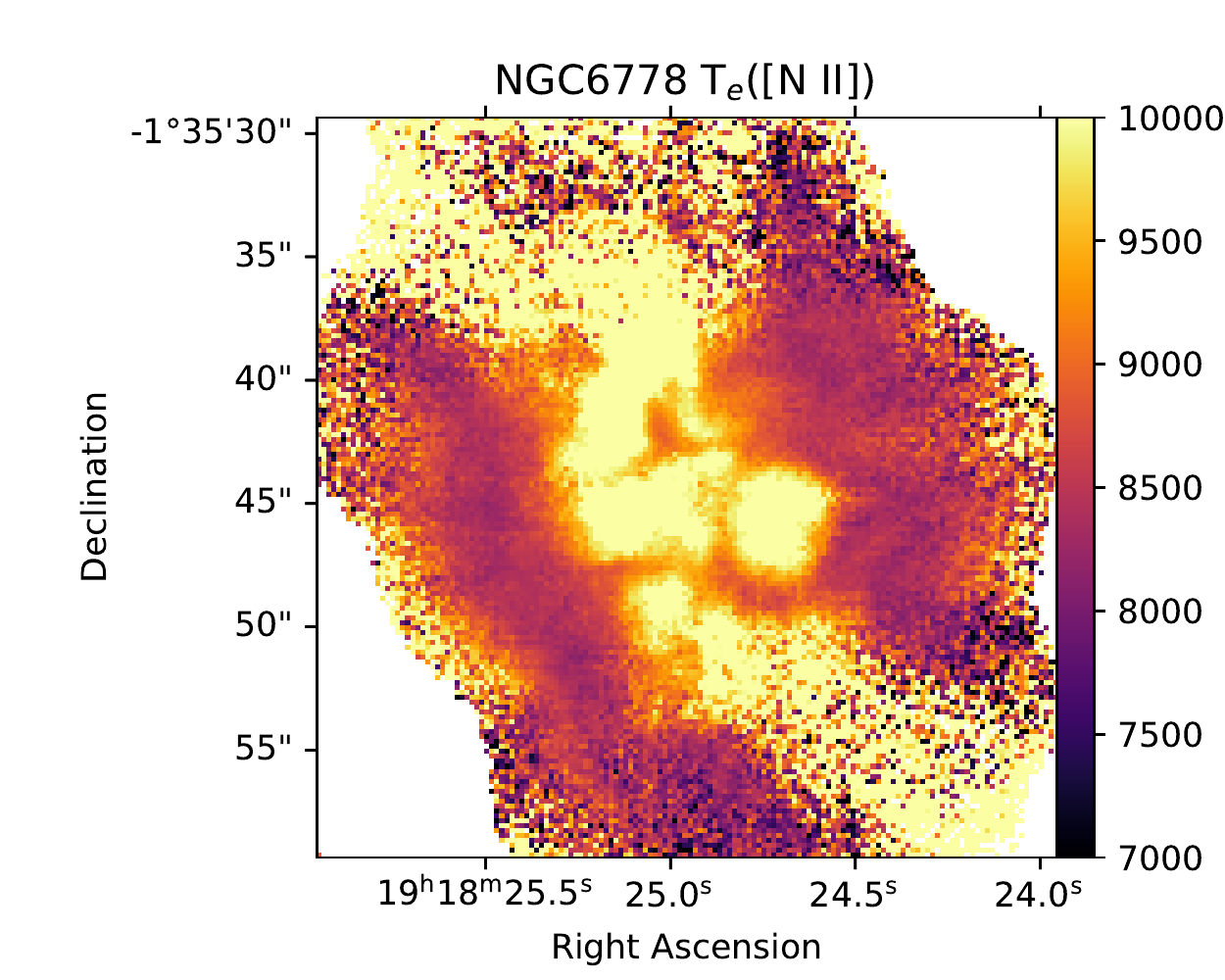}
\includegraphics[scale=0.33]{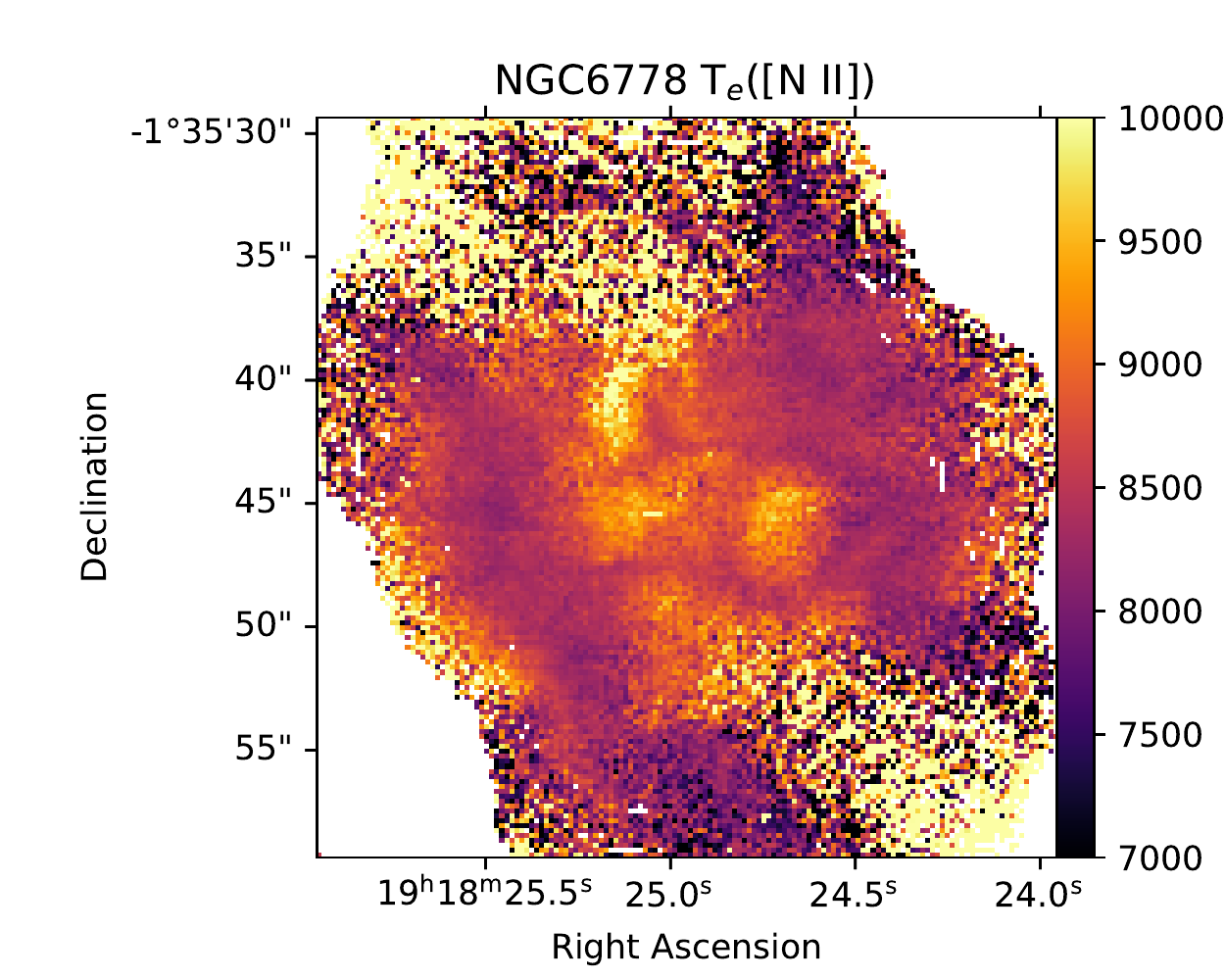}
\includegraphics[scale=0.33]{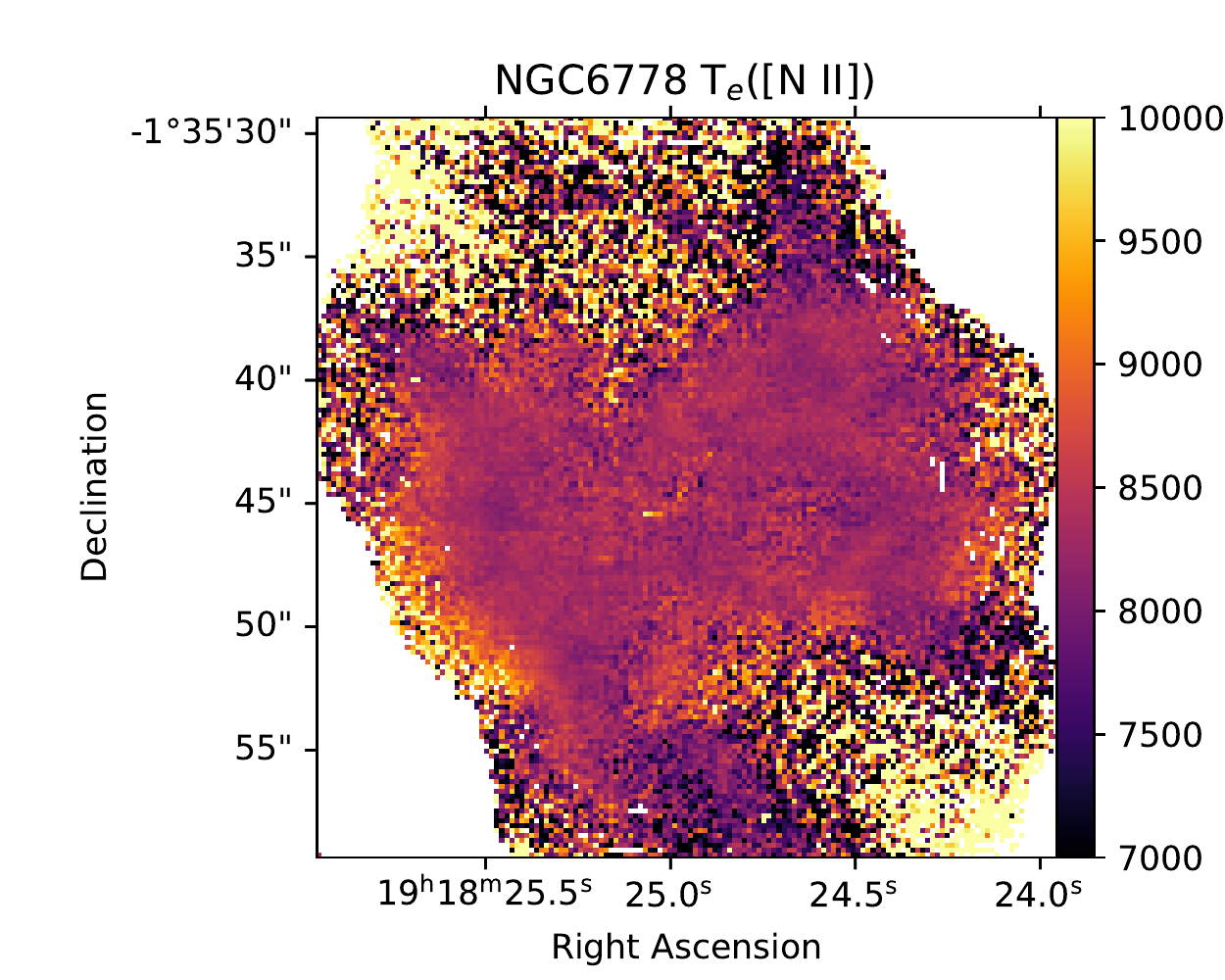}
\includegraphics[scale=0.33]{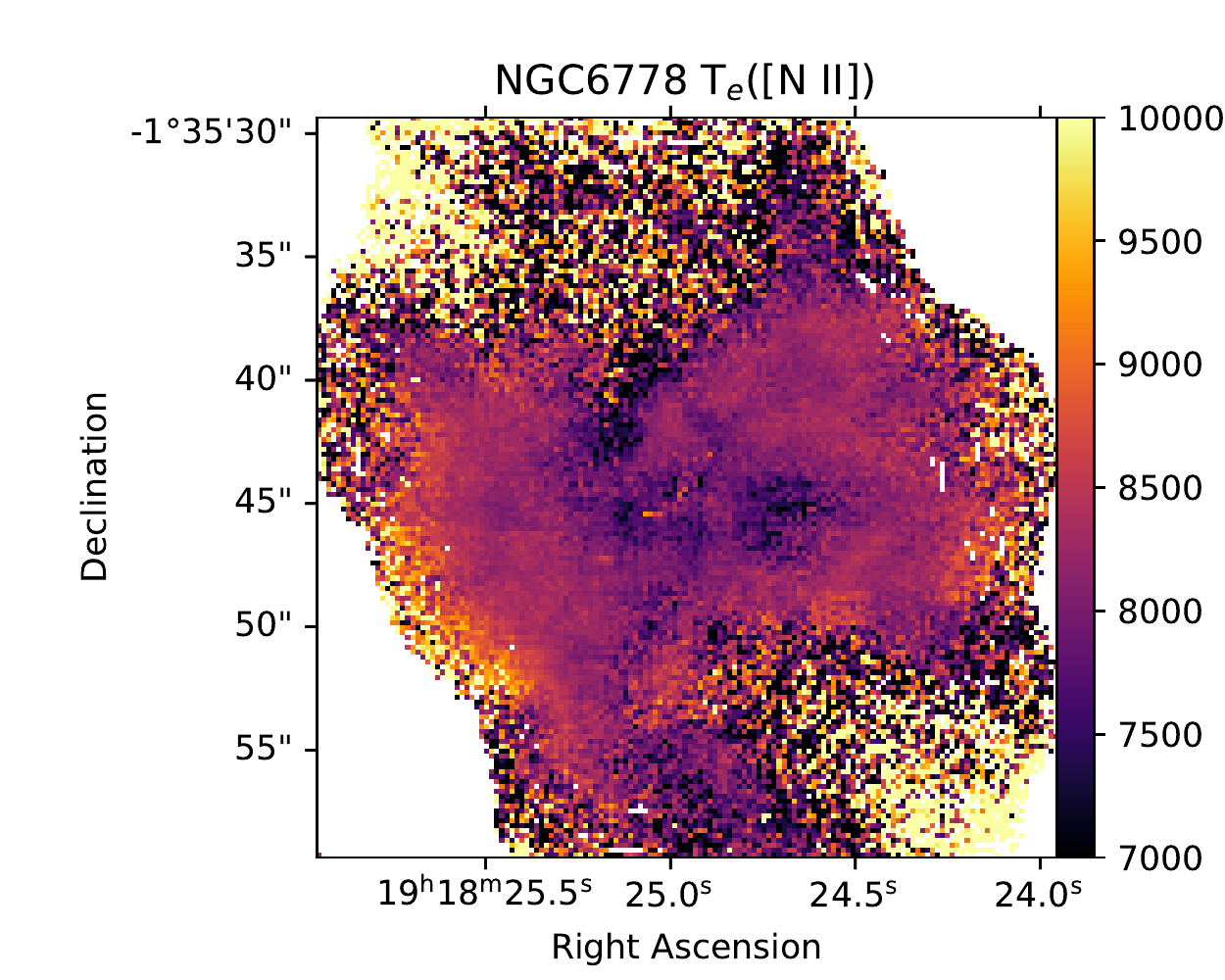}
\includegraphics[scale=0.33]{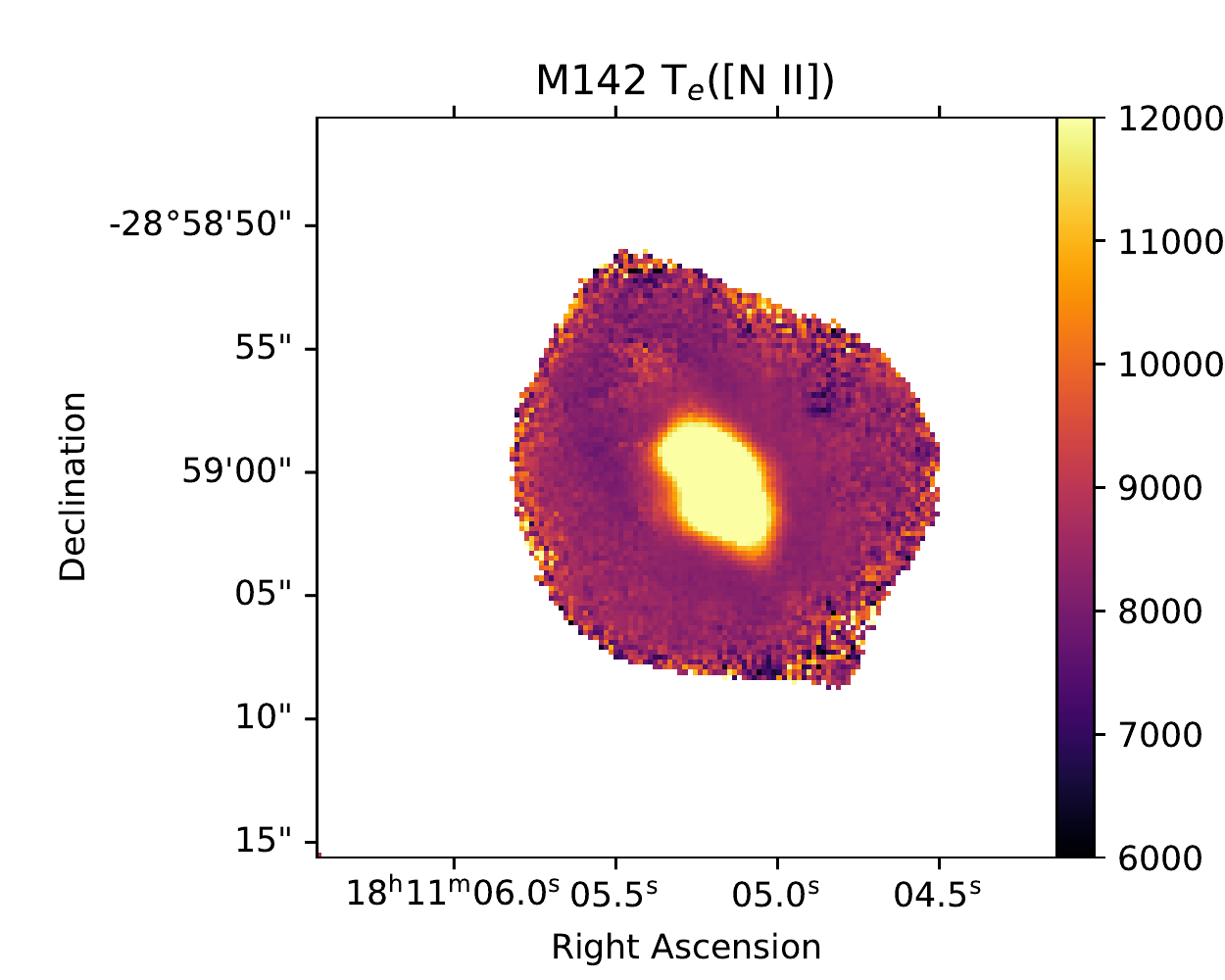}
\includegraphics[scale=0.33]{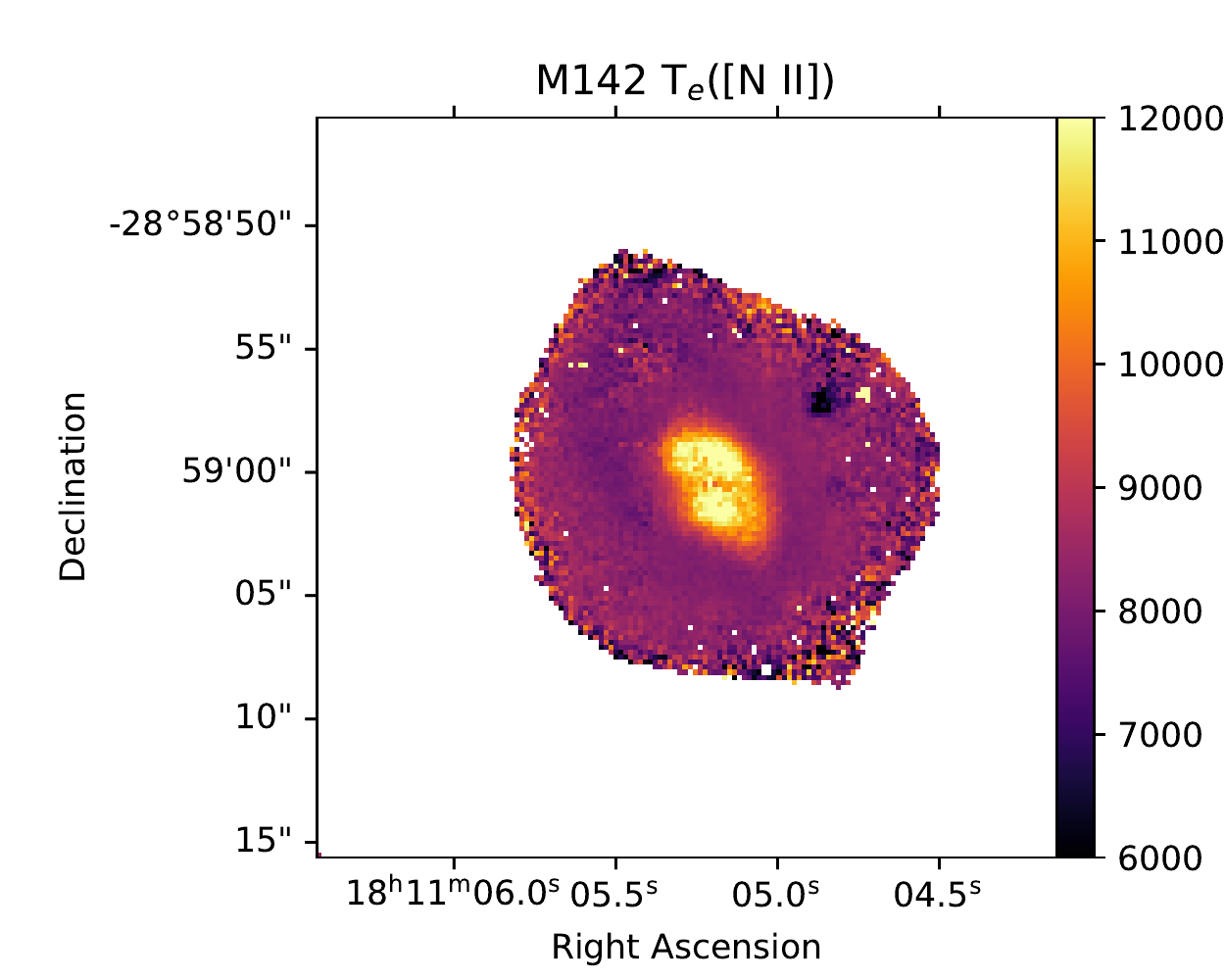}
\includegraphics[scale=0.33]{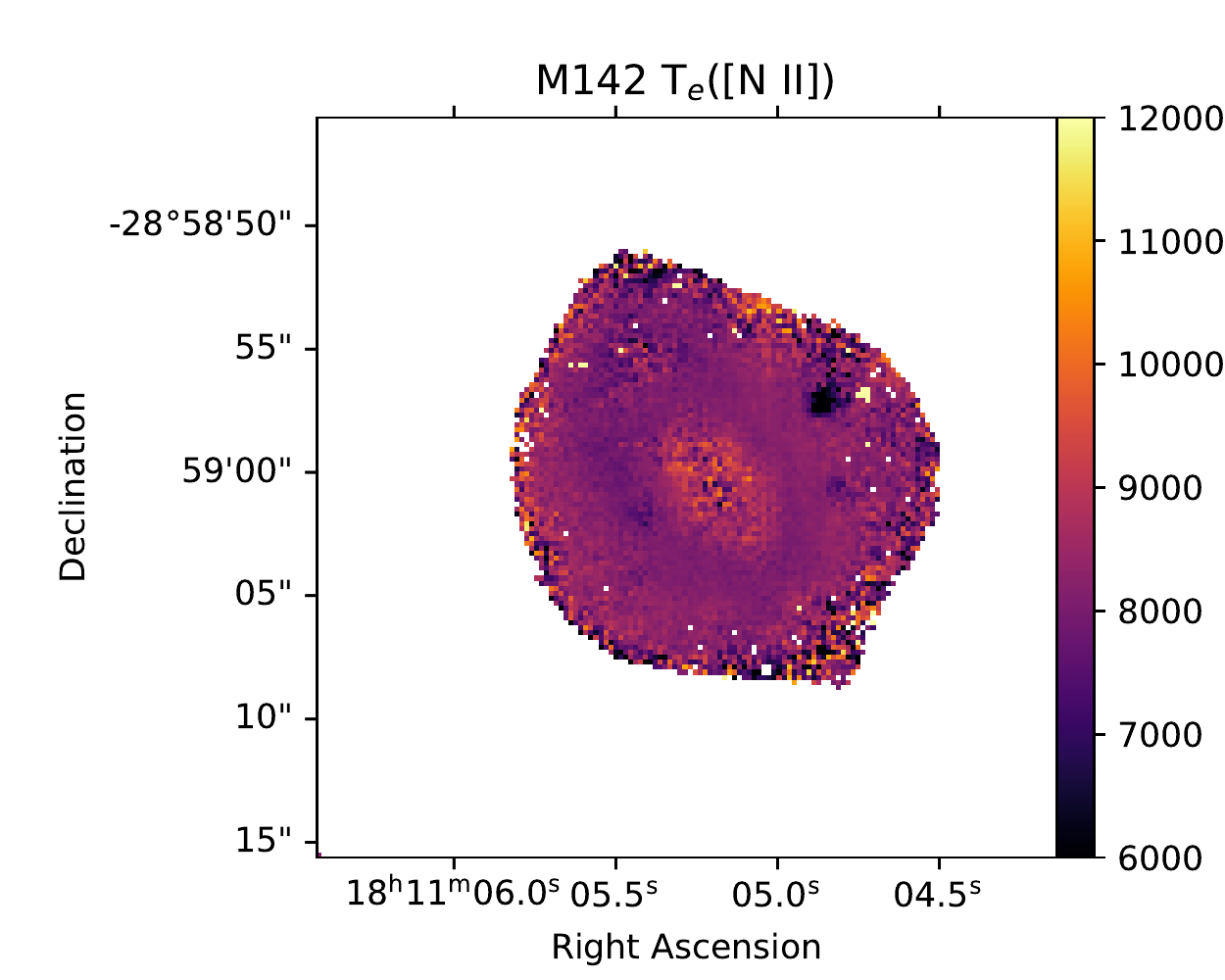}
\includegraphics[scale=0.33]{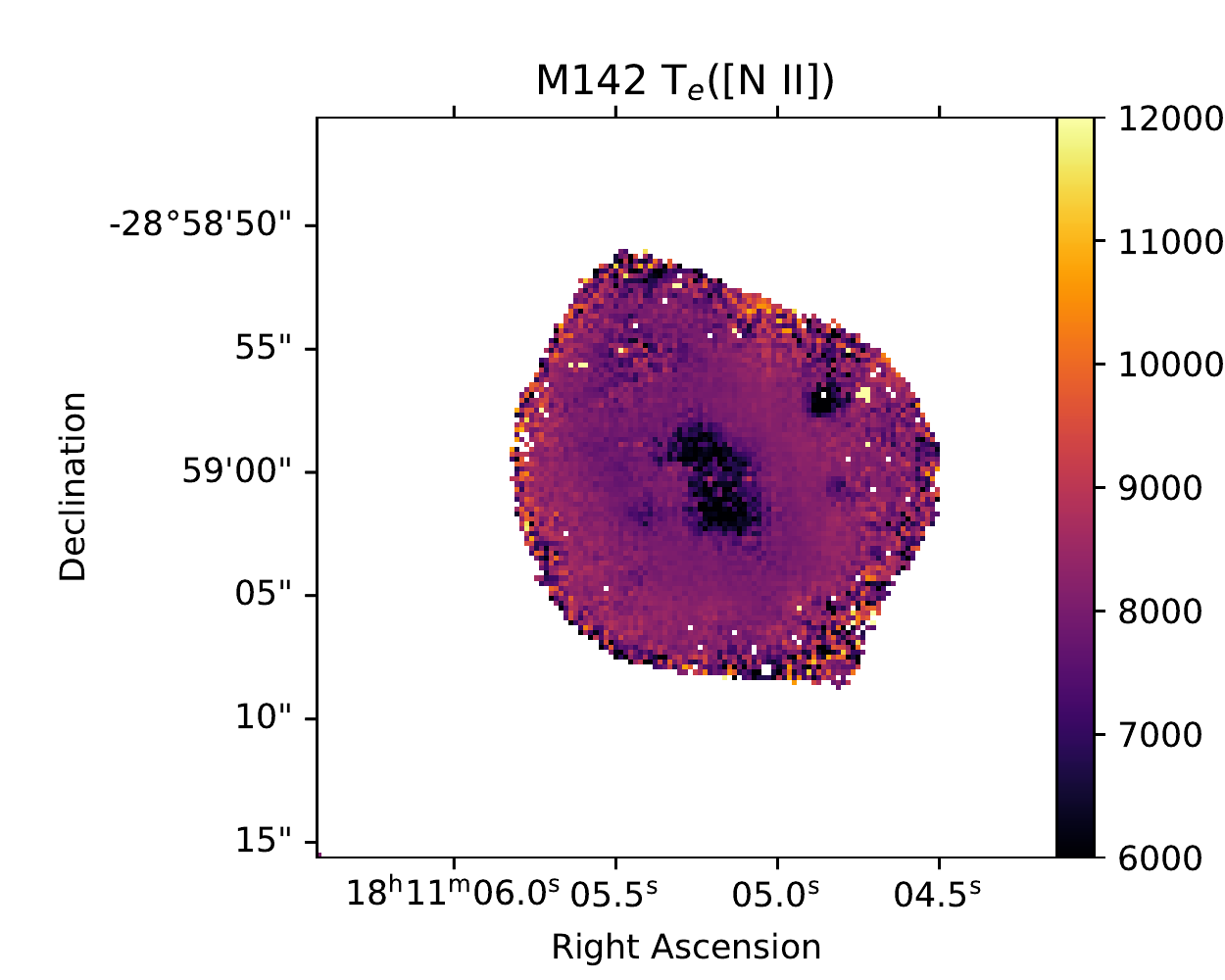}
\includegraphics[scale=0.33]{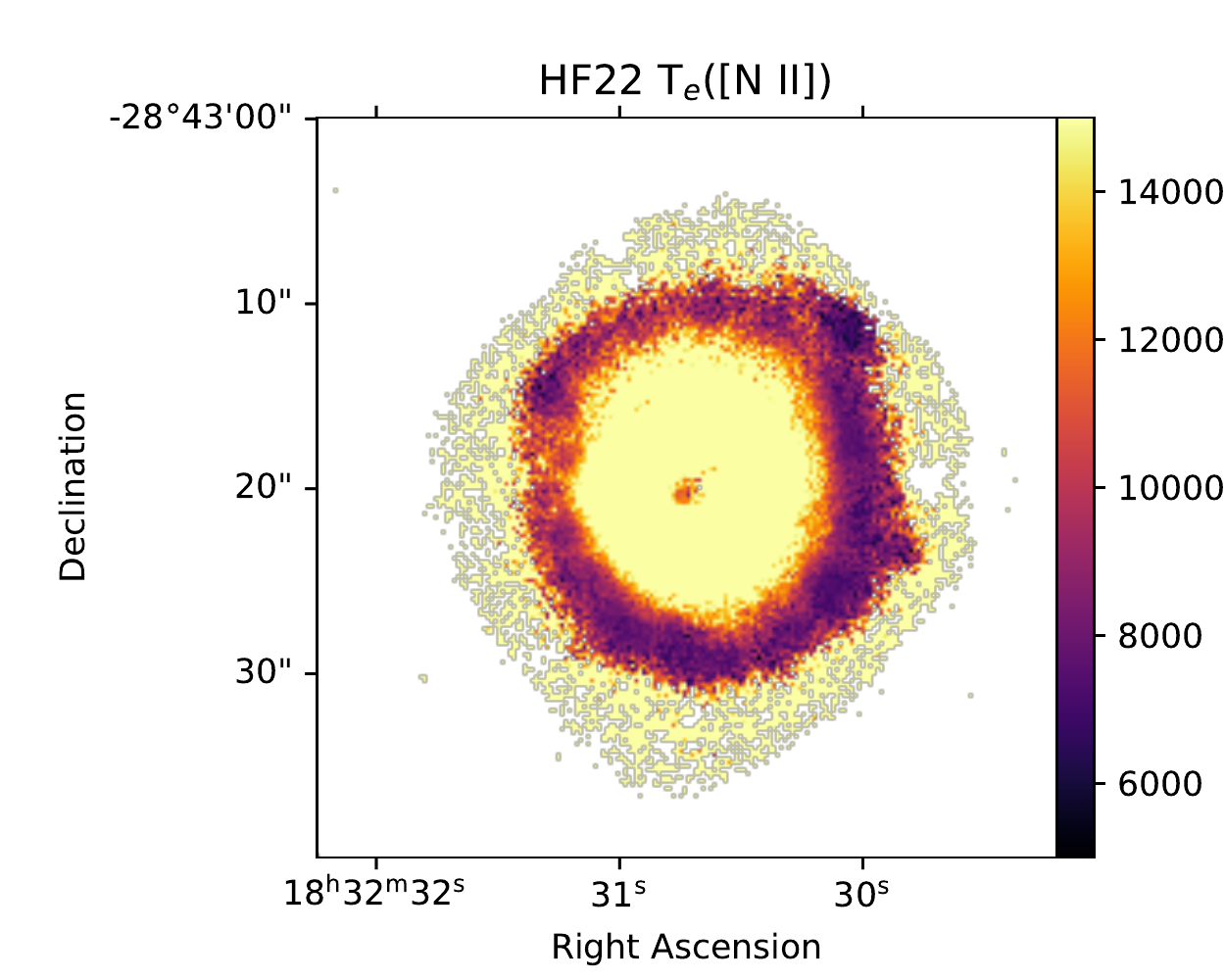}
\includegraphics[scale=0.33]{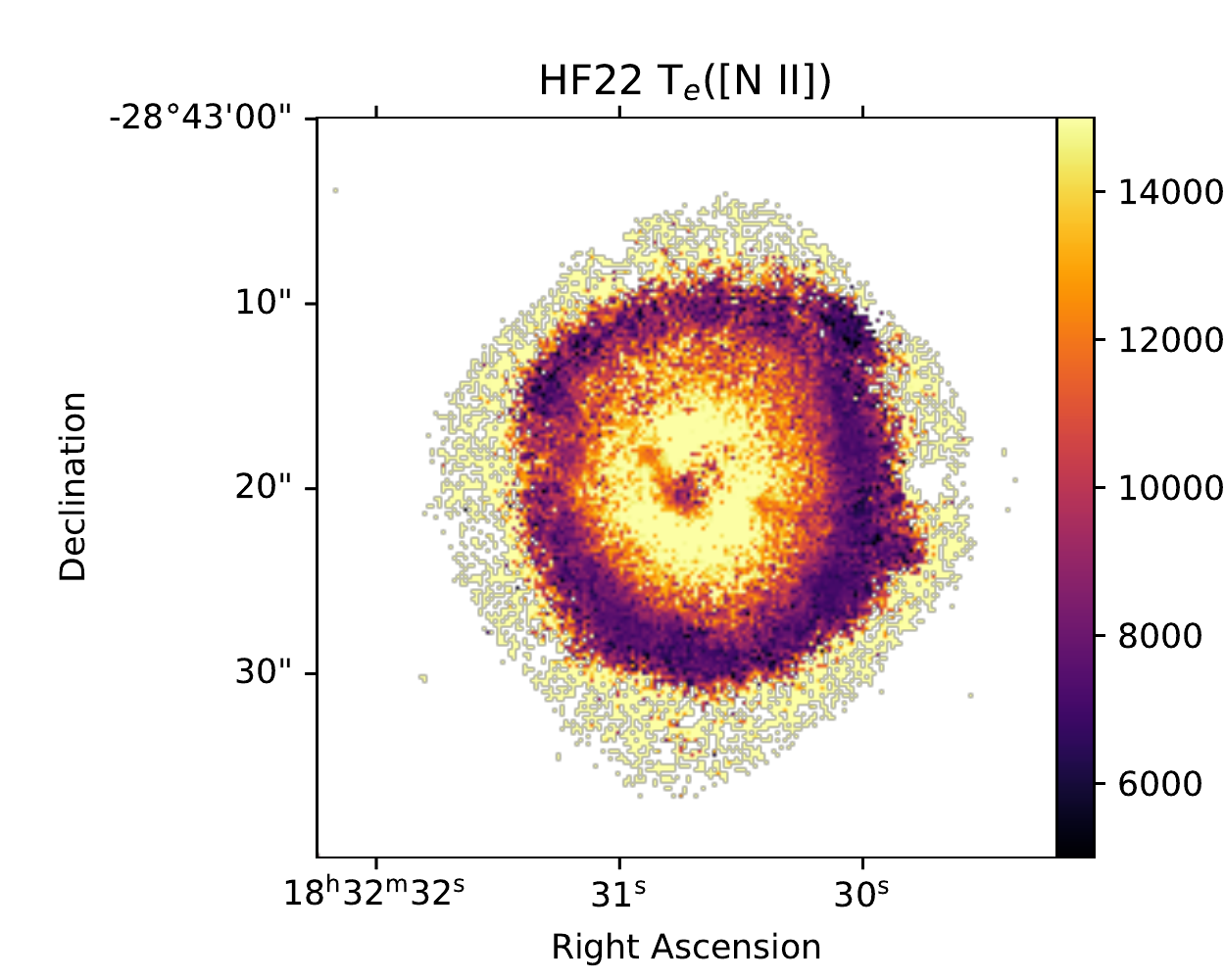}
\includegraphics[scale=0.33]{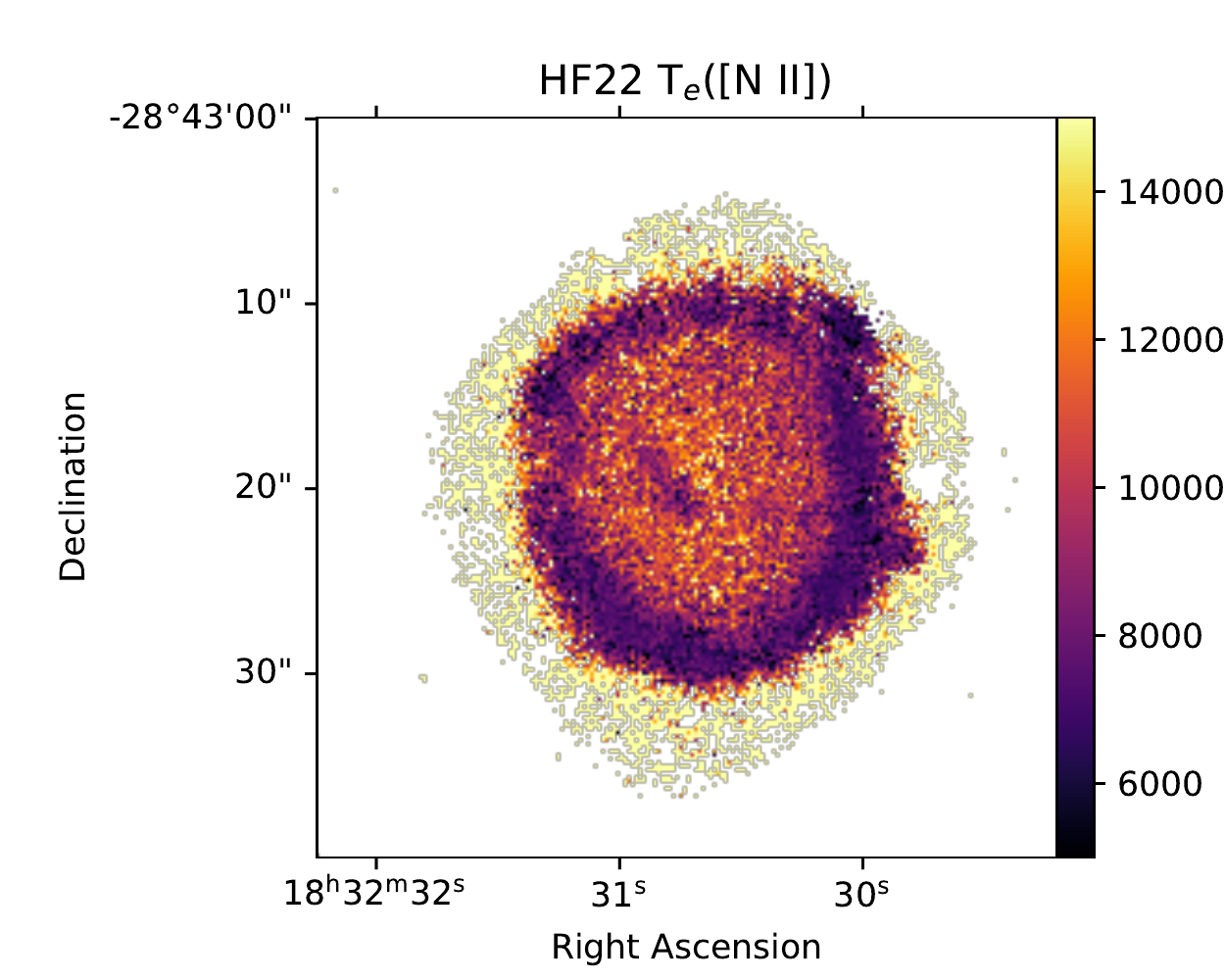}
\includegraphics[scale=0.33]{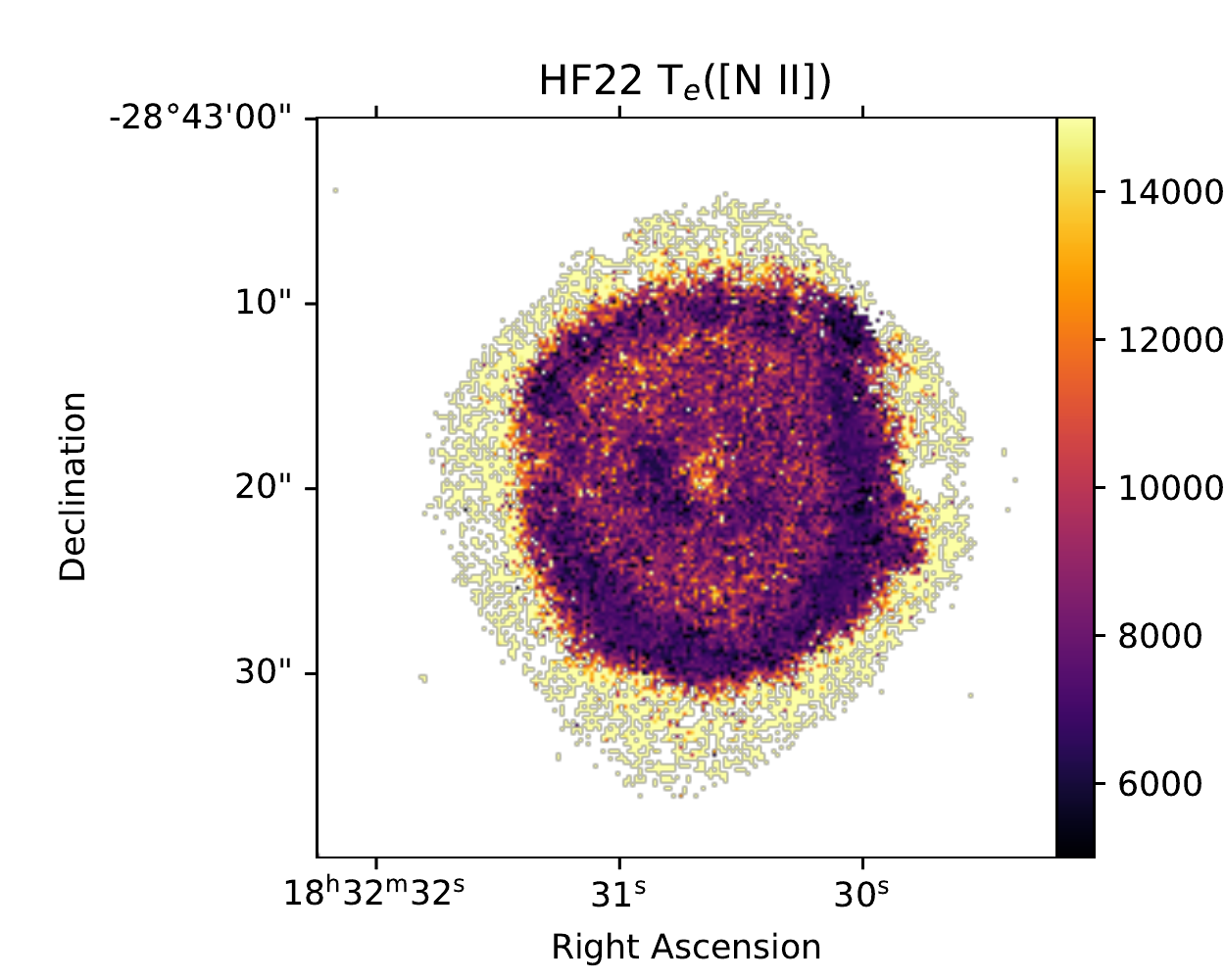}
\caption{{\te}({\fnii}) maps of the three PNe showing its variation from no recombination contribution correction (first column panels) to recombination contribution corrections assuming different temperatures for the recombination zone emission:  $T_{\rm e} = 1\,000$, 4\,000, 8\,000\,K (second, third and last column panels, respectively). The temperature scale is the same for the four cases.    
\label{fig:Te_NII}}
\end{figure*}

As we have described above, the recombination contribution to the auroral {\fnii} $\lambda$5755 line strongly affects the measured line flux in high-ADF PNe. In Fig.~\ref{fig:Te_NII}, we show the {\te}({\fnii}) maps with no recombination contribution correction (first column panels) and after applying recombination contribution corrections using different temperatures for the recombination emission (1\,000, 4\,000 and 8\,000\,K; second, third and last column panels, respectively) and adopting {\elecd}({\fsii}). For each object, we present all the maps with the same temperature scale to emphasize the {\te} differences between them. It is apparent from Fig.~\ref{fig:Te_NII} that neglecting the recombination correction for the {\fnii} $\lambda$5755 line translates in extremely high {\te} determinations, especially in the central zones of M\,1--42 and Hf\,2--2, where values over 20\,000\,K are reached. Such large values are not obtained when the \fsiii\ line ratio is used (see Fig.~\ref{fig:NGC6778_TeNe_4000} for NGC\,6778 and corresponding figures in the Appendix for the other two PNe). We do not see any physical justification for the presence of low ionization gas (emitting \fnii\ lines) at high temperature, when the intermediate ionization gas traced by the \fsiii\ emission is almost at a uniform temperature. We then attribute this high temperature to the lack of correction for the recombination contribution to the {\fnii} $\lambda$5755 line.  
Assuming a very low $T_{\rm e} =1\,000$\,K for the recombination zone (second column panels in Fig.~\ref{fig:Te_NII}) does not fix the problem: the {\te}({\fnii}) is still too high in the central zones of each nebula, implying an abnormal behaviour in the component of the gas where the bulk of CEL emission arises. On the other hand, adopting a similar {\te} for the gas where recombination emission is produced has the opposite effect, and {\te} significantly drops in the inner parts. Finally, adopting a value of $T_{\rm e} =4\,000$\,K for the recombination emission zone provides a {\te}({\fnii}) map whose structure is quite similar to that of the {\te}({\fsiii}) map for NGC\,6778 and M\,1--42 (see Figs.~\ref{fig:NGC6778_TeNe_4000} and~\ref{fig:M142_TeNe_4000}). For Hf\,2--2, the {\te}({\fnii}) and {\te}({\fsiii}) maps still seem quite dissimilar. However, the uncertainties in the central parts of this PN are extremely high as the {\fnii} $\lambda$5755 emission line is completely dominated by recombination in the whole nebula, with the exception of its outermost zones (see Fig.~\ref{fig:RecCont}), where, in fact, the obtained {\te}({\fnii}) and {\te}({\fsiii}) best agree. We therefore adopt $T_{\rm e} =4\,000$\,K as the characteristic {\te} to correct for the recombination contribution in the three PNe under study. 
It is important to note here that the value for the temperature adopted to compute the {\fnii} $\lambda$5755 recombination contribution may not be fully related to the electron temperature in the cold region: the correction is obtained by combining emissivities determined using atomic data from \citet{pequignotetal91} and \citet{fangetal11} (see previous Section), that may suffer from high uncertainties. The exact temperature for the cold region may actually be much lower than the adopted value. Therefore, it should be considered as a tunable parameter for the recombination correction. 

Adopting $T_{\rm e} =4\,000$\,K is a good compromise for our PNe- It is very similar to the {\te} values obtained in their central zones from {\te} recombination diagnostics (see next Section), which only provide qualitative information on {\te} in the low-temperature plasma, as they are based on {\hi} and {\hei} emission that is present in both components and strongly depends on the relative weight of the low-temperature region (see Section~\ref{sec:discuss}). The only way to break the introduced degeneracy is to have a {\te} diagnostic of the cold region, such as {\oii} $\lambda$4089/$\lambda$4649 \citep[see][]{storeyetal17}, which unfortunately is missing in the MUSE data owing to the lack of wavelength coverage below 4600\,\AA. 

Although the effect of the recombination contribution to the auroral lines has been proven to be dramatic when mapping physical conditions, it is mandatory to check this effect when integrating the whole or a significant part of the volume of the nebula, as the effect could be diluted owing to the radically different weight of recombination emission in the central and external zones of the nebula. In Section~\ref{sec:collapsed} we will discuss how the recombination contribution affects the physical conditions and chemical abundance determinations in the integrated spectra.

\subsection{Electron temperatures from {\hei} recombination lines and Paschen continuum}
\label{sec:te_rec_paschen}

The electron temperature can also be determined from the {\te} sensitive {\hei} $\lambda$7281/$\lambda$6678 recombination line ratio, following the procedure in \citet{zhangetal05}. We have used the fitting coefficients computed by \citet{mendezdelgadoetal21} to account for weak density dependencies and have then interpolated linearly (their table~7) the coefficients needed to solve their equation~4. For the densities considered in this work, the sensitivity of the {\hei} $\lambda$7281/$\lambda$6678 line ratio to \te\ is rather low for {\te} lower than 2\,000\,K 

For the determination of the electron temperature from the Paschen jump (PJ), we define the jump as the difference between the continuum measured at 8100 and 8400 \AA, normalized to the intensity of the {\hi} 9-3 line. We tested the following two methods to derive {\te}. 

The first one uses the {\sc pyneb} \verb!Continuum.BJ_HI! method to generate a table of continuum jump values as a function of the temperature. Interpolation over this 1D table is then performed to obtain the temperatures from the measured continuum jump values. This requires to fix the other parameters needed to compute the continuum jump: the electron density is fixed to $10^3$\,cm$^{-3}$ and the values of He$^+$/H and He$^{2+}$/H are set to 0.095 and 0.005, respectively.

In the second method,  the resulting jump values (obtained using the same \verb!Continuum.BJ_HI! method) are interpolated over a 3D grid of {\te}, {\elecd} and He$^+$/He values. The grid contains 5\,000 randomly generated values of {\te}, {\elecd} and He$^+$/He in the [$500 - 30\,000$\,K], [$100 - 10\,000$\,cm$^{-3}$] and [$0.0 - 1.0$] ranges, respectively (we set ${\rm He/H}=0.12$, anticipating the results described below). The interpolation in this 3D grid is performed using an ANN similar to the one described in Section~\ref{sec:te_ne_cels}. This method must be called once {\elecd} and the ionic fraction of He are known.

\begin{figure*}
\includegraphics[scale=0.45,trim={3cm 2cm 0 3cm}, clip]{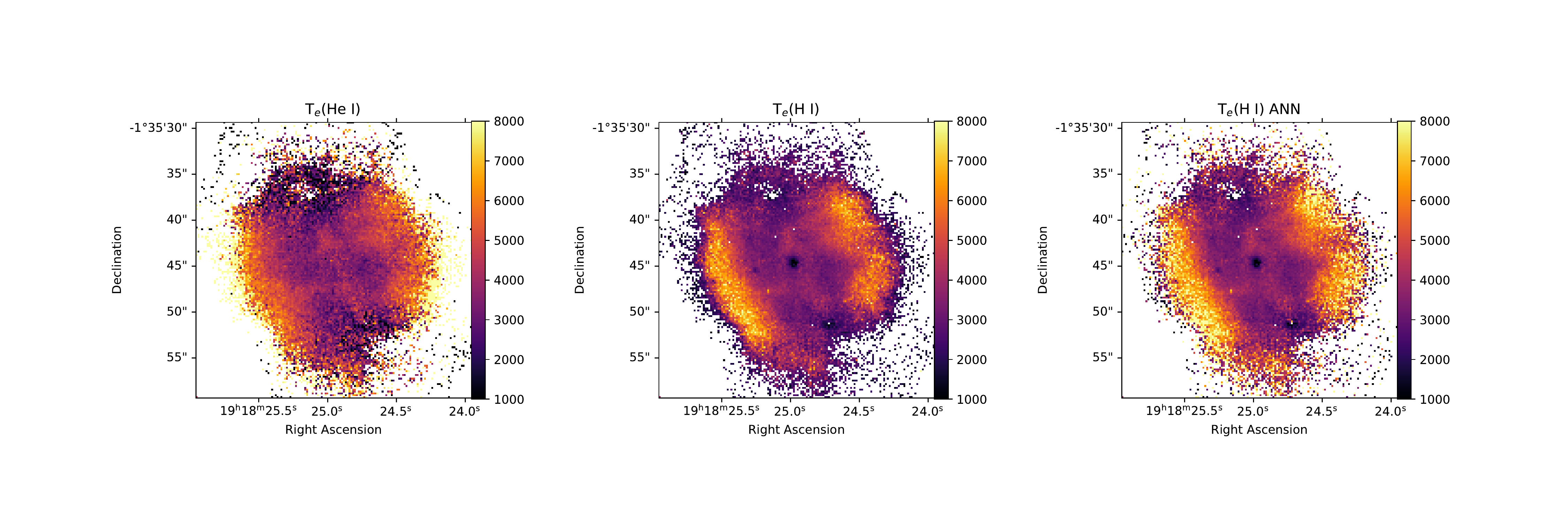}
\includegraphics[scale=0.45,trim={3cm 2cm 0 3cm}, clip]{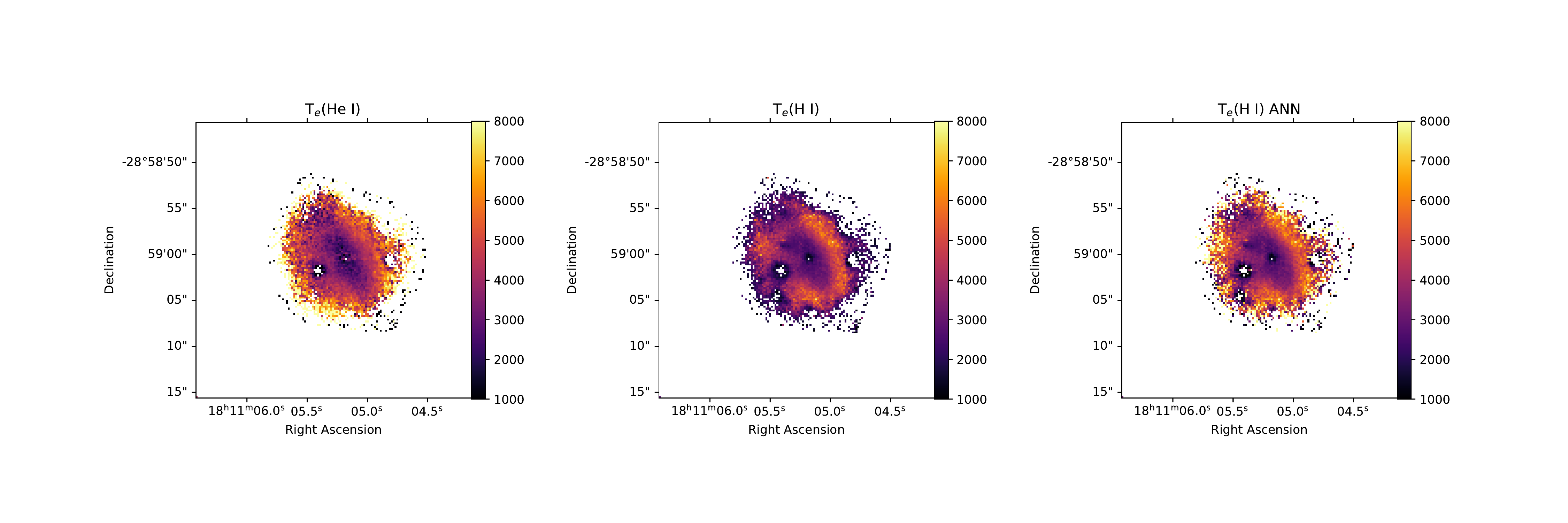}
\includegraphics[scale=0.45,trim={3cm 2cm 0 3cm}, clip]{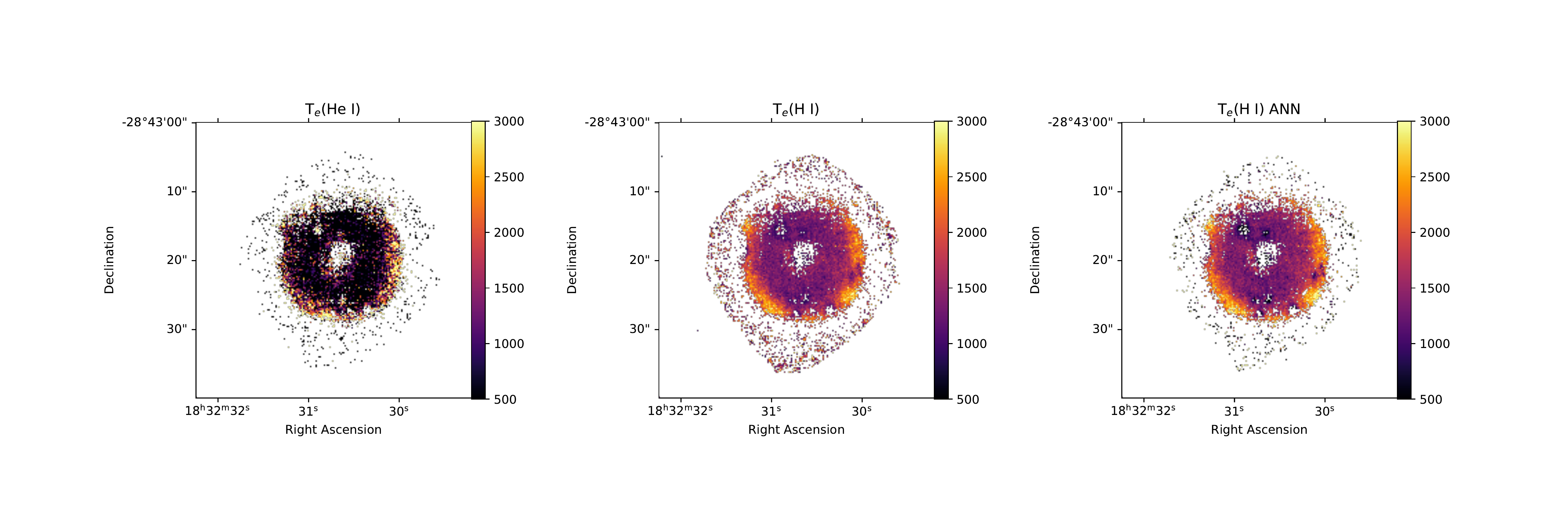}
\caption{{\te} maps obtained from the {\hei} $\lambda$6678/$\lambda$7281 line ratio (left panels), the Paschen jump relative to {\hi} P9 $\lambda$9229 line (middle panels) and the Paschen jump derived using an Artificial Neural Network (ANN) technique (right panels, see text for details) for the three objects in our sample (NGC\,6778, M\,1--42 and Hf\,2--2 from top to bottom). Note the extremely low values obtained for Hf\,2--2 (see text).
\label{fig:tpas_maps}}
\end{figure*}

In Fig.~\ref{fig:tpas_maps}, we show the {\te}({\hei}) (left panels) and {\te}(PJ) maps using method 1 (middle panels) and method 2 (right panels) for our three objects. The {\te}(PJ) maps derived with methods 1 and 2 are very similar, the only noticeable difference being an increasing {\te} in the external parts of the nebulae when using method 2, a behaviour also seen in the {\te}(\hei) maps. It is remarkable that all maps follow a very similar behaviour, with {\te} decreasing in the inner zones of the nebulae. This is an important result because, for the first time, low temperature zones can be directly related to high ADFs in PNe (see Section~\ref{sec:adf}).

\section{Chemical abundances}
\label{sec:ionic}

We constructed the ionic abundance maps using {\sc pyneb} and the atomic data  set contained in Table~\ref{tab:atomic_data}. For each ion, the lines with the highest signal-to-noise ratios were used. As the fluxes of the triplet {\hei} lines, such as $\lambda$5876 and $\lambda$7065, can be affected by the metastability of the $2^3$S level, only the {\hei} $\lambda$6678 line was used for abundance determination because it is the brightest singlet line in the observed wavelength range. The complete list of lines considered for building the ionic abundance maps is shown in Table~\ref{tab:ion_abundances}.

\subsection{Ionic abundances from CELs}
\label{sec:ionic_cels}

As described in Section~\ref{sec:phys_cond}, for each nebula a two-zone scheme was adopted, with a single {\elecd} value given by the {\fsii} diagnostic ratio in the two zones, and {\te}({\fnii}) for ions with ${\rm IP} \leq 17$\,eV (i.e. C$^0$, N$^0$, N$^+$, O$^0$, O$^+$ and S$^+$) and {\te}({\foiii}) for ions with ${\rm IP} > 17$\,eV (i.e. O$^{2+}$, S$^{2+}$, Cl$^{2+}$, Cl$^{3+}$, Ar$^{2+}$, Ar$^{3+}$ and Ar$^{4+}$). 

In Figs.~\ref{fig:abund_ngc6778_none} to~\ref{fig:abund_hf22_none}, we present the ionic abundance maps computed with no correction for the recombination contribution to the {\fnii} $\lambda$5755 and {\foii} $\lambda\lambda$7320+30 CELs. In Figs.~\ref{fig:abund_ngc6778_4000} to~\ref{fig:abund_hf22_4000}, the ionic abundance maps with the recombination correction for $T_{\rm e}=4\,000$\,K are shown. Accounting for the recombination contribution slightly affects the emission maps of low-ionization species in NGC\,6778 and M\,1--42, in particular in their central zones, where {\te}({\fnii}) is significantly lower after applying the correction. In both PNe, the {\foii} $\lambda\lambda$7320+30 abundance map is the most affected one, because the recombination contribution correction have an effect on both the {\te} determination and the computed emissivity of the {\foii} lines. For Hf\,2--2 the effect is dramatic for several lines (Fig.~\ref{fig:RecCont}): recombination emission dominates over collisional excitation in the measured flux of the {\fnii} $\lambda$5755 line, and the same happens for the {\foii} $\lambda\lambda$7320+30 lines. This translates into an extremely high {\te} determination in the central parts of the nebula. Given that recombination coefficients for H$^+$ have only been computed for $T_{\rm e} \leq 30\,000$\,K \citep{storeyhummer95}, when no recombination correction is applied, our pipeline skips the calculation of abundances in spaxels where {\te} exceeds this limit, hence the apparent blank gaps in the central parts of the ionic abundance maps of low ionization species illustrated in Fig.~\ref{fig:abund_hf22_none}. This results in an apparently  weird behaviour of the abundance maps. Contrarily, when the strong recombination contribution is accounted for, the abundance maps of the different low-ionization species uniformly show low abundances in the central regions of the PN and remarkably higher abundances in the necklace-like structure at the outer edge of Hf\,2--2.

\subsection{Ionic abundances from ORLs}
\label{sec:ionic_orls}

For consistency, we built the ORL abundance maps adopting the electron temperature used for the recombination correction. We have checked that using {\te} diagnostics maps, such as {\te}({\fsiii}) or {\te}(PJ), does not affect the spatial distribution of the abundance maps, and differences in the absolute abundance values are below 0.05 dex in O$^{2+}$/H$^+$. This is the expected behaviour, as ORL ionic abundances have a small dependence on {\te}. 

Regarding He$^+$, as the recombination coefficients by \citet{porteretal12, porteretal13} were only computed for $T_{\rm e} \geq 5\,000$~K, extrapolation  to lower {\te} was applied according to an inverse law in {\te} for the emissivity of these lines. This nicely reproduces the emissivity dependence of {\hei} $\lambda$6678 and $\lambda$7281 with {\te} computed by \citet{porteretal12, porteretal13} for $T_{\rm e} > 5\,000$~K.

The atomic data shown in Table~\ref{tab:atomic_data} were used to compute the C$^{2+}$, N$^{2+}$, O$^+$ and O$^{2+}$ abundances from ORLs, assuming case A for O$^+$ and C$^{2+}$ and case B for O$^{2+}$ and N$^{2+}$\footnote{For a detailed description of cases A and B, see \citet{osterbrockferland06}.}.

\begin{figure}
\includegraphics[scale=0.36,trim={0.5cm 3cm 0 3cm}, clip]{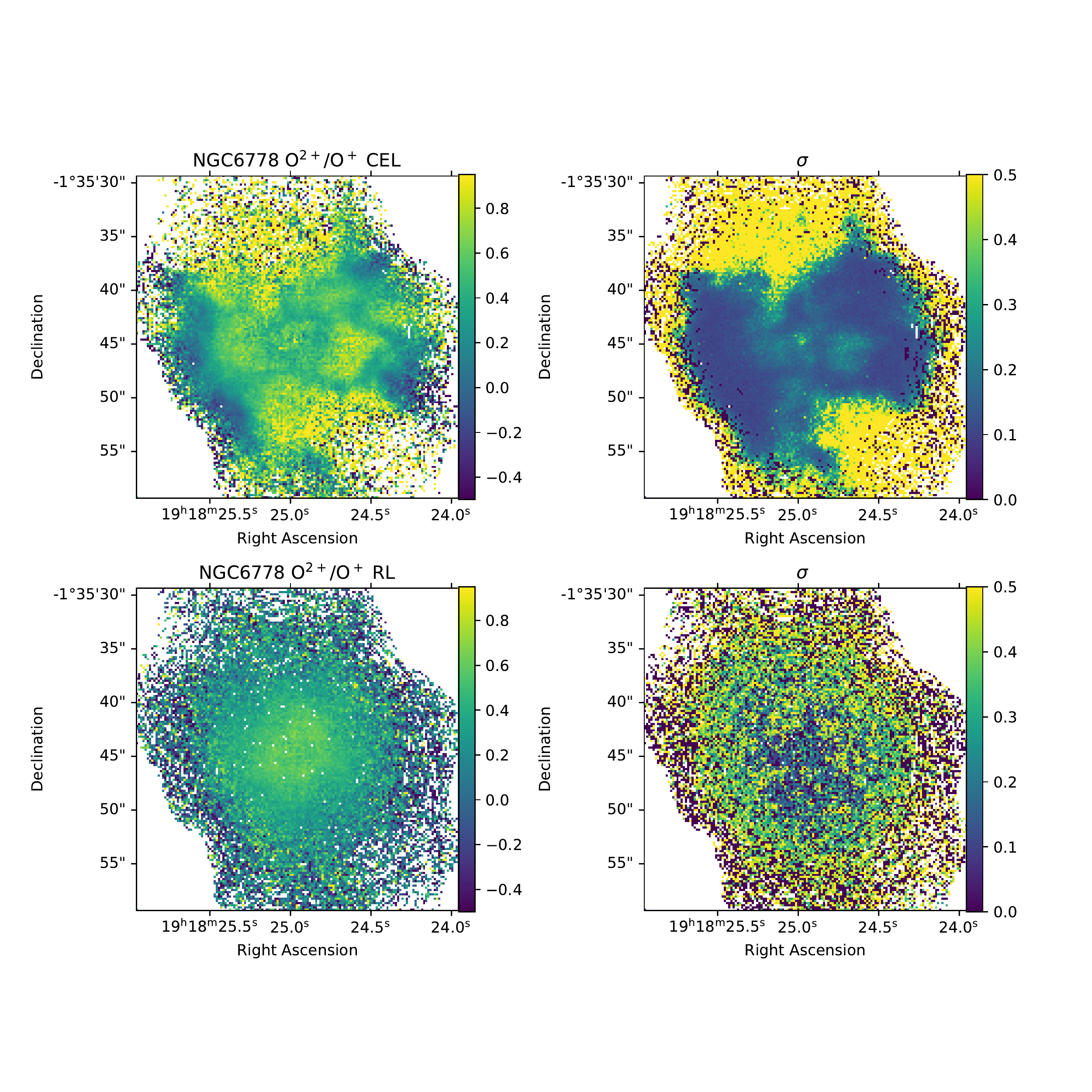}
\caption{CEL and ORL spatial distributions of  $\log{\rm (O^{2+}/O^{+})}$ in NGC\,6778 and their corresponding fractional error maps.
\label{fig:ion_degree_ngc6778}}
\end{figure}

Figs.~\ref{fig:maps_ngc6778} to~\ref{fig:maps_hf22} clearly show that the spatial distribution of the heavy element ORL emission is concentrated in the inner regions of each nebula, and that such spatial distribution is incompatible with the CEL emission of the corresponding ions (see e.g. the {\oii} ORL and {\foiii} CEL emission maps). This well-known behaviour was first addressed by \citet{garnettdinerstein01} in the PN NGC\,6720 and later confirmed by several authors \citep[e.g.][]{corradietal15,garciarojasetal16,jonesetal16,wessonetal18}. In this work, we show for the first time the spatial distribution of ionic chemical abundances from ORLs for different ions:  C$^{2+}$, N$^{2+}$, O$^+$ and O$^{2+}$. We find that all of them nearly spatially coincide.  This would be expected for ions with similar ionization potentials as C$^{2+}$, N$^{2+}$ and O$^{2+}$, but not for O$^+$ for which it should be seen some ionization stratification as compared to O$^{2+}$.
This result may indicate that the spatial distribution of heavy-element ORL emission would not provide information on the large scale ionization structure of the cold-gas component, but it would rather trace the distribution of unresolved nebular components  where the bulk of heavy-element ORLs is produced. Each of these small structures would be (almost) optically thick to the ionizing radiation, as in some of the models presented by \citet{yuanetal11} and \citet{gomezllanosmorisset20}, and would have its specific ionization structure.

Another parameter to explore using the O$^+$ and O$^{2+}$ ORL abundance maps is the ionization degree of the gas, which  seems to be remarkably different when computed with ORLs or with CELs.  In Fig.~\ref{fig:ion_degree_ngc6778}, we show the ionization degree maps for NGC\,6778 traced by the O$^{++}$/O$^+$ obtained using CELs (upper left panel) and ORLs (bottom left panel), as well as the corresponding fractional uncertainties (right panels). This figure illustrates how the ORL emitting gas seems to be less ionized than the CEL emitting gas. For M\,1--42 and Hf\,2--2 the differences are even larger (see Figs.~\ref{fig:ion_degree_m142} and~\ref{fig:ion_degree_hf22}  in the online material). 

The ionization of a given X$^i$ ion into X$^{i+1}$ is controlled by the ratio between the number of photons with energy above the X$^i$-ionizing energy and the number of available X$^i$ ions to ionize. As this ratio strongly decreases in the metal-rich region because X/H is increased, the ionization into X$^{i+1}$ decreases. In other words, the decrease in oxygen ionization in the ORL-bright region is due to an increase in the oxygen density, rather than a reduction of the hydrogen density, leading to an increase of the O/H abundance. This behaviour is actually predicted by the models \citep[see fig.~2 of][and the related discussion]{gomezllanosmorisset20}. 


\subsection{The abundance discrepancy}
\label{sec:adf}

\begin{figure}
\includegraphics[scale=0.36,trim={0.5cm 3cm 0 3cm}, clip]{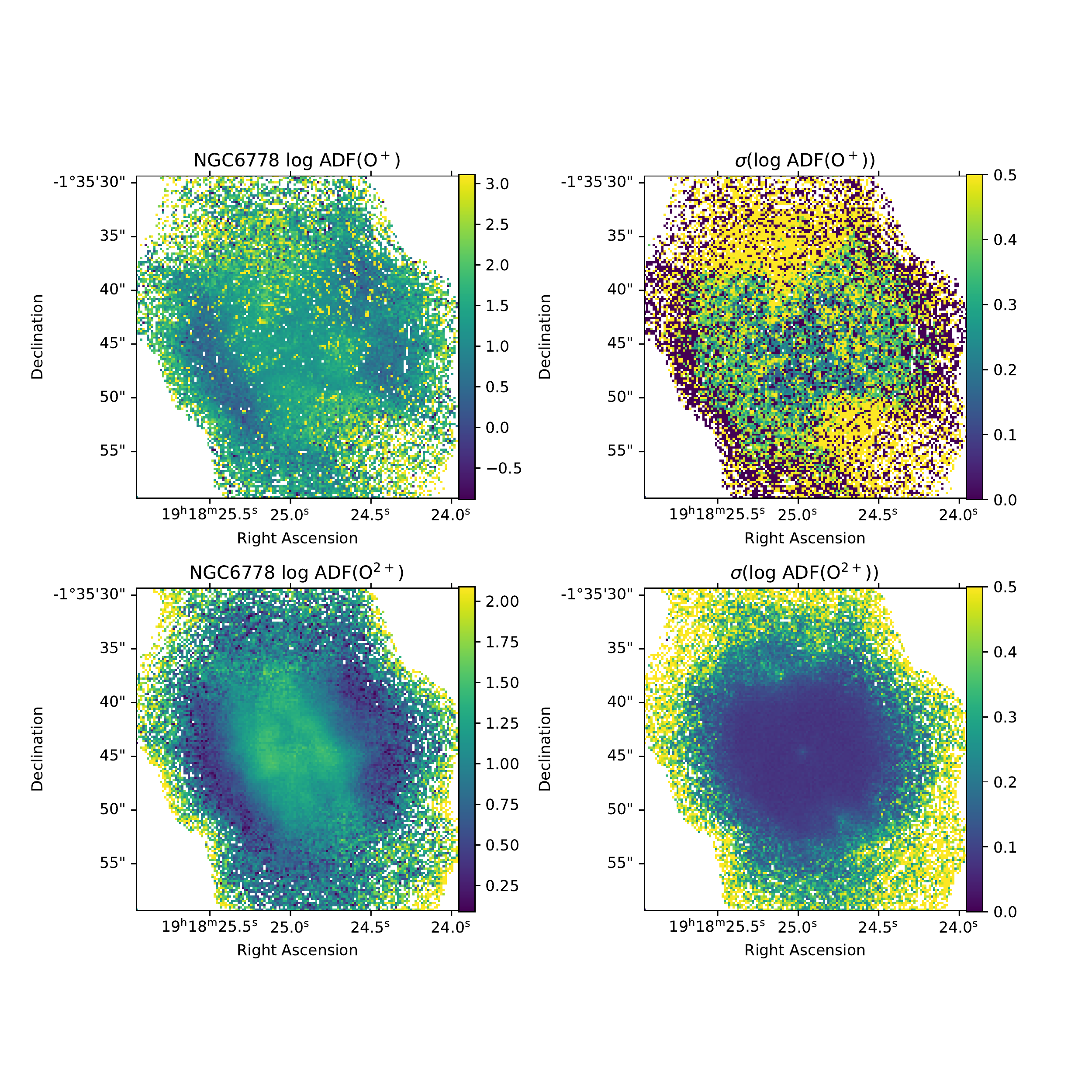}
\caption{Spatial distributions of  $\log {\rm [ADF(O^{+})]}$ and $\log {\rm [ADF(O^{2+})]}$ in NGC\,6778 and their corresponding fractional error (ADF error/ADF ratio) maps.
\label{fig:adfs_NGC6778}}
\end{figure}

\begin{figure}
\includegraphics[scale=0.36,trim={0.5cm 3cm 0 3cm}, clip]{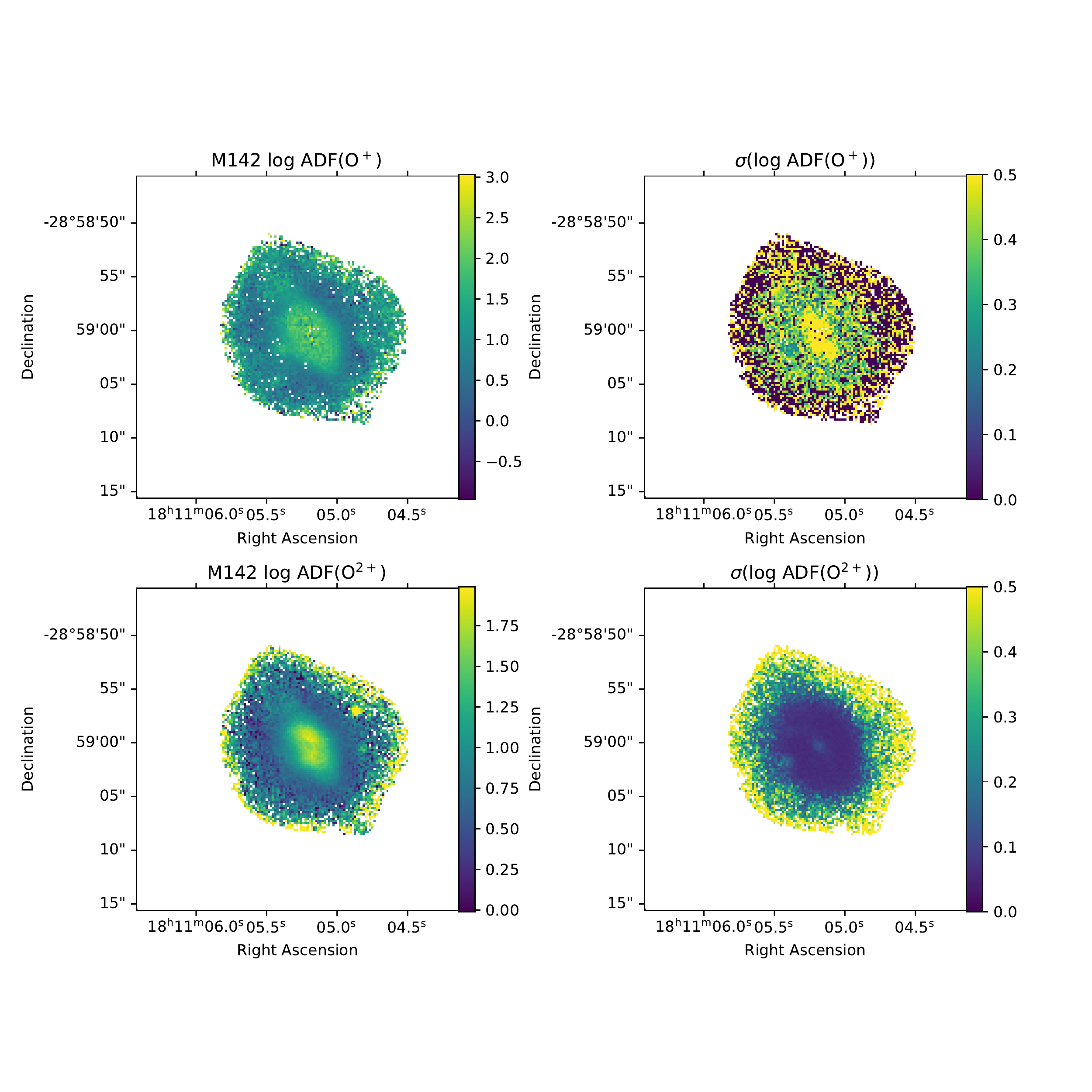}
\caption{Same as Fig.~\ref{fig:adfs_NGC6778}, but for M\,1--42.
\label{fig:adfs_M142}}
\end{figure}

\begin{figure}
\includegraphics[scale=0.36,trim={0.5cm 3cm 0 3cm},clip]{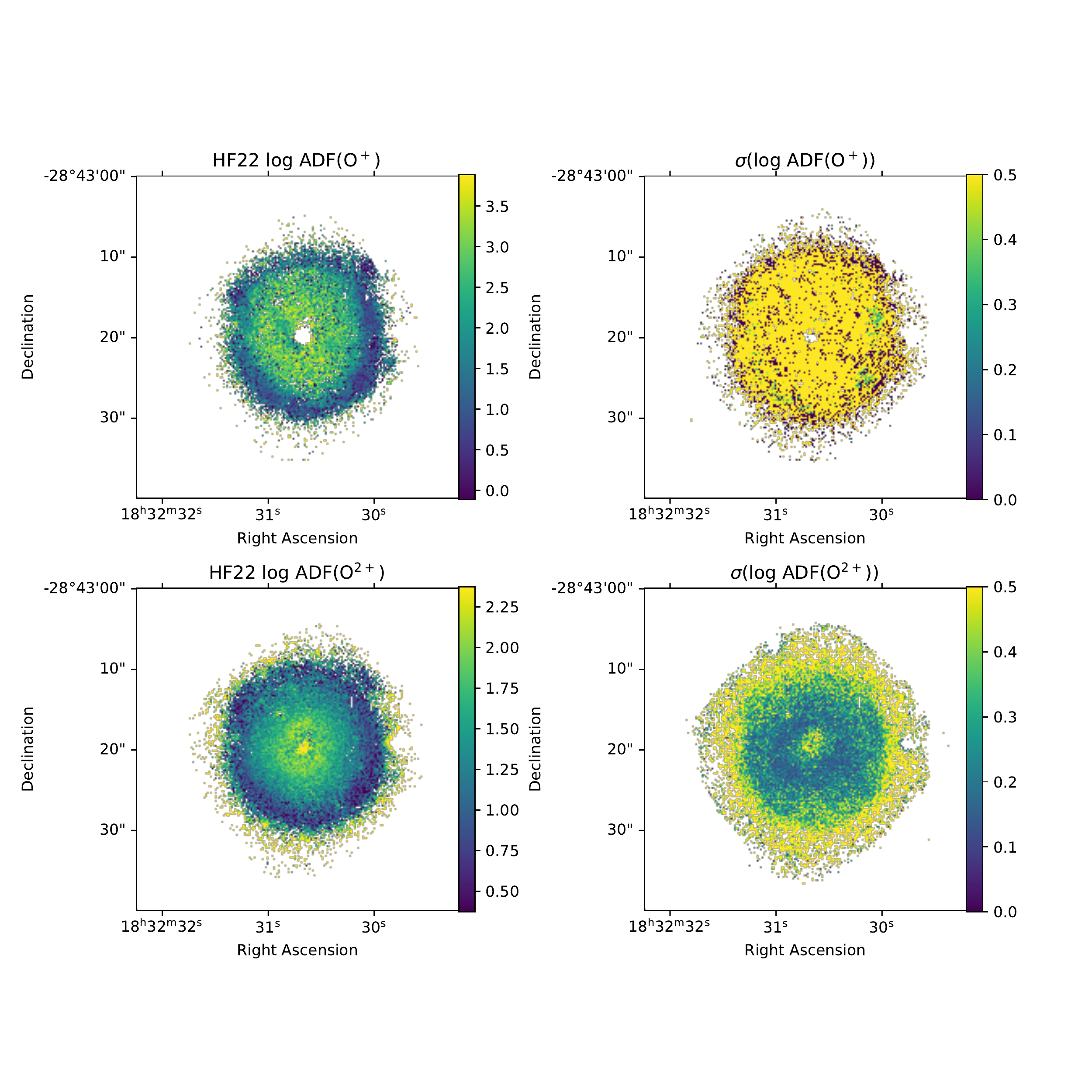}
\caption{Same as Fig.~\ref{fig:adfs_NGC6778}, but for Hf\,2--2.
\label{fig:adfs_HF22}}
\end{figure}

Once the O$^+$ and O$^{2+}$ ORL abundance maps are available, the next obvious step is to build the ADF maps. This is the first time that the O$^+$ and O$^{2+}$ ADF are mapped at the high spatial resolution provided by MUSE (0.2\,arcsec). \citet{tsamisetal08} built ADF maps for the high-ADF PNe NGC\,7009 \citep[${\rm ADF} \approx 5$,][]{liuetal95, fangliu11} and NGC\,6153 \citep[${\rm ADF} \approx 10$,][]{liuetal00, mcnabbetal16} using FLAMES-ARGUS, which provided much higher spectral resolution than MUSE, but a smaller field of view ($11.5 \times 7.2$\,arcsec) and a poorer spatial resolution.

Figs.~\ref{fig:adfs_NGC6778} to~\ref{fig:adfs_HF22}  present the ADF(O$^+$) and ADF(O$^{2+}$) along with their corresponding uncertainties (expressed in terms of the fractional uncertainty, i.e. ADF uncertainty/ADF) for the three PNe. The NGC\,6778 and M\,1--42 fractional uncertainty maps indicate that the uncertainties in the ADF(O$^{2+}$) maps are $\leq 10$\, per cent in the central parts of each PN. In Hf\,2--2, they are larger, of the order of ($20-30$\,per cent). The ADF(O$^+$) uncertainties are significantly larger: $\simeq 20-30$\,per cent in the central parts of NGC\,6778 and M\,1--42 and $>50$\,per cent in Hf\,2--2. 

Figs.~\ref{fig:adfs_NGC6778} to~\ref{fig:adfs_HF22} indicate that the ADF(O$^{2+}$)  centrally peaks for both ions in the three objects, which is in agreement with the results previously obtained in high-ADF PNe from long-slit spectra \citep{corradietal15, jonesetal16, garciarojasetal16, wessonetal18} and the aforementioned IFU study by \citet{tsamisetal08}. The ADF(O$^{2+}$) spatial distribution resembles that of the ORL O$^{2+}$/H$^+$ ratio. This is expected, given the relative homogeneity of the CEL O$^{2+}$/H$^+$ ratio in our three PNe (see Figs.~\ref{fig:abund_ngc6778_4000},~\ref{fig:abund_m142_4000} and~\ref{fig:abund_hf22_4000}).

This is also the first time that the ADF(O$^{+}$) spatial distribution is presented in a PN. It is slightly different to that of ADF(O$^{2+}$), given the different spatial distribution of the O$^+$ and O$^{2+}$ abundances obtained from CELs. Even so, it  clearly peaks in the inner nebular regions too, but the uncertainties are higher than for ADF(O$^{2+}$), particularly in the case of Hf\,2--2, where the observed {\oi} lines are very faint and the uncertainties in the ADF are beyond 50 per cent for the great majority of the spaxels (see Fig.~\ref{fig:adfs_HF22}). 

\subsection{Helium abundances}
\label{sec:Heabundance}
We first computed the total abundance maps for He. Note that no ICF is needed as its total abundance is computed by simply adding He$^+$ and He$^{2+}$. The  He/H abundance maps are displayed in Fig.~\ref{fig:ab_he}. It is  clear that the total abundance is not homogeneous and there exists a positive He/H abundance gradient toward the central parts of each nebula, which is relatively mild in NGC\,6778, but is much more evident in M\,1--42 and Hf\,2--2. This points to He abundance inhomogeneities in these PNe. If two plasmas at different temperatures coexist, as proposed, the observed behaviour is consistent with the cold component being not only metal-rich, but also He-enriched relative to H.

\subsection{Elemental abundances from the integrated spectra}
\label{sec:collapsed}

\begin{table}
\caption{Physical conditions derived for the integrated spectra. \label{tab:phys_cond}}
\resizebox{\columnwidth}{!}{%
\begin{tabular}{llccc}
\hline
 & & \multicolumn{1}{c}{NGC\,6778} & \multicolumn{1}{c}{M\,1--42} & \multicolumn{1}{c}{Hf\,2--2}  \\
\hline
{\te} (K) 	& {\fnii} $\lambda$5755/$\lambda$6548$^{\rm a}$ & 8345 $\pm$ 250 & 8210 $\pm$ 210  & 8200 $\pm$ 530  \\
            &  {\fsiii} $\lambda$6312/$\lambda$9069 & 8290 $\pm$ 310 &  7700 $\pm$ 200 & 6820 $\pm$ 210    \\
            & {\hei} $\lambda$6678/$\lambda$7281	&   4240 $\pm$ 710 &  4090 $\pm$ 690 & 670 $\pm$ 540  \\
            & PJ        &  3810 $\pm$ 850 & 3400 $^{+3900}_{-3400}$ & 1260 $\pm$ 450 \\
            & PJ (ML)   &  3850 $\pm$ 900 & 3450:$^{\rm b}$ & 1330 $\pm$ 460 \\
\noalign{\vskip 2mm}
{\elecd} (cm$^{-3}$)& {\fsii} $\lambda$6716/$\lambda$6731&   860 $\pm$ 220 &   1065 $\pm$ 280    &  350 $\pm$ 110   \\
            & {\fcliii} $\lambda$5518/$\lambda$5538 &   1035 $\pm$ 370 & 1710 $\pm$ 480 &  680 $\pm$ 340 \\
            & {\fariv} $\lambda$4711/$\lambda$4740 &   4970 $\pm$ 1170    &  8300:$^{\rm b}$ &  ---   \\
\hline
\end{tabular}
}
\begin{description}
\item $^{\rm a}$ {\fnii} $\lambda$5755 line corrected for recombination contribution assuming {\te} = 4\,000\,K 
\item $^{\rm b}$ Colons indicates large uncertainties  
\end{description}
\end{table}

\begin{table}
\caption{Ionic abundances in units of $12+\log({\rm X}^{\rm{n+}}/{\rm H}^+)$ in the integrated spectra. \label{tab:ion_abundances}}
\resizebox{\columnwidth}{!}{%
\begin{tabular}{llcccc}
\hline
Ion & Line & NGC\,6778 & M\,1--42 & Hf\,2--2  \\
\hline
He$^+$ 	 & {\hei} $\lambda$6678      & 11.13 $\pm$ 0.03 & 11.14 $\pm$ 0.02 & 11.14 $\pm$ 0.03  \\
He$^{2+}$ & {\heii} $\lambda$4686    &  9.56 $\pm$ 0.03 &  9.80 $\pm$ 0.04 &  9.17 $\pm$ 0.03  \\
C$^0$ 	& {\fci} $\lambda$8728      &  5.69 $\pm$ 0.08 &  6.10 $\pm$ 0.06 &  5.68 $\pm$ 0.15  \\
C$^{2+}$ 	& {\cii} $\lambda$5342   &  8.93 $\pm$ 0.03 &  9.03 $\pm$ 0.03 &  9.45 $\pm$ 0.03  \\
 		& {\cii} $\lambda$6462       &  9.07 $\pm$ 0.03 &  9.15 $\pm$ 0.03 &  9.53 $\pm$ 0.03  \\
N$^0$ 	& {\fnitroi} $\lambda$5199+     &  6.74 $\pm$ 0.07 &  6.91 $\pm$ 0.06 &  6.25 $\pm$ 0.14  \\
N$^{+}$ 	& {\fnii} $\lambda$5755     &  7.95 $\pm$ 0.06 &  8.05 $\pm$ 0.05 &  7.09 $\pm$ 0.10  \\
         	& {\fnii} $\lambda$6548     &  7.96 $\pm$ 0.06 &  8.06 $\pm$ 0.05 &  7.10 $\pm$ 0.10  \\
N$^{2+}$ 	& {\nii} $\lambda$5676   &  9.08 $\pm$ 0.03 &  ---             &  9.19 $\pm$ 0.04  \\
 		& {\nii} $\lambda$5679       &  9.29 $\pm$ 0.03 &  9.38 $\pm$ 0.03 &  9.37 $\pm$ 0.03  \\
O$^{0}$ 	& {\foi} $\lambda$6300  &  7.30 $\pm$ 0.06 &  7.45 $\pm$ 0.05 &  6.26 $\pm$ 0.13  \\
 		& {\foi} $\lambda$6364      &  7.31 $\pm$ 0.06 &  7.45 $\pm$ 0.05 &  6.23 $\pm$ 0.13  \\
O$^{+}$ 	& {\oi} $\lambda$7773+   &  9.11 $\pm$ 0.04 &  9.11 $\pm$ 0.06 &  9.65 $\pm$ 0.41  \\
 		& {\foii} $\lambda$7319+    &  8.02 $\pm$ 0.13 &  8.09 $\pm$ 0.10 &  7.78 $\pm$ 0.39  \\
 		& {\foii} $\lambda$7330+    &  8.02 $\pm$ 0.14 &  8.11 $\pm$ 0.10 &  7.79 $\pm$ 0.39  \\
O$^{2+}$ 	& {\oii} $\lambda$4649+   &  9.60 $\pm$ 0.03 &  9.61 $\pm$ 0.04 &  9.76 $\pm$ 0.03  \\
 		& {\oii} $\lambda$4661       &  9.41 $\pm$ 0.04 &  9.48 $\pm$ 0.04 &  9.57 $\pm$ 0.04  \\
 		& {\foiii} $\lambda$4959    &  8.54 $\pm$ 0.06 &  8.65 $\pm$ 0.05 &  8.42 $\pm$ 0.06  \\
S$^{+}$ 	& {\fsii} $\lambda$6716 &  6.29 $\pm$ 0.05 &  6.38 $\pm$ 0.04 &  5.60 $\pm$ 0.09  \\
 		& {\fsii} $\lambda$6731     &  6.30 $\pm$ 0.05 &  6.39 $\pm$ 0.04 &  5.60 $\pm$ 0.09  \\
S$^{2+}$ 	& {\fsiii} $\lambda$6312&  6.79 $\pm$ 0.08 &  6.99 $\pm$ 0.05 &  6.57 $\pm$ 0.09  \\
 		& {\fsiii} $\lambda$9069    &  6.80 $\pm$ 0.08 &  7.00 $\pm$ 0.05 &  6.57 $\pm$ 0.08  \\
Cl$^{2+}$ & {\fcliii} $\lambda$5518 &  5.13 $\pm$ 0.06 &  5.24 $\pm$ 0.05 &  4.98 $\pm$ 0.06  \\
 		& {\fcliii} $\lambda$5538   &  5.14 $\pm$ 0.06 &  5.29 $\pm$ 0.05 &  5.02 $\pm$ 0.07  \\
Cl$^{3+}$ & {\fcliv} $\lambda$7531  &  4.66 $\pm$ 0.07 &  4.81 $\pm$ 0.04 &  4.25 $\pm$ 0.08  \\
 		& {\fcliv} $\lambda$8046    &  4.43 $\pm$ 0.07 &  4.60 $\pm$ 0.04 &  3.91 $\pm$ 0.08  \\
Ar$^{2+}$ & {\fariii} $\lambda$7136 &  6.34 $\pm$ 0.07 &  6.54 $\pm$ 0.04 &  6.09 $\pm$ 0.07  \\
 		& {\fariii} $\lambda$7751   &  6.32 $\pm$ 0.07 &  6.52 $\pm$ 0.04 &  6.07 $\pm$ 0.07  \\
Ar$^{3+}$ & {\fariv} $\lambda$4740    &  5.64 $\pm$ 0.07 &  5.98 $\pm$ 0.05 &  5.51 $\pm$ 0.06  \\
Ar$^{4+}$ 	& {\farv} $\lambda$7005 &  3.53 $\pm$ 0.07 &     ---          &      ---          \\
\hline
\end{tabular}
}
\end{table}

\begin{figure}
\includegraphics[scale = 0.6, clip]{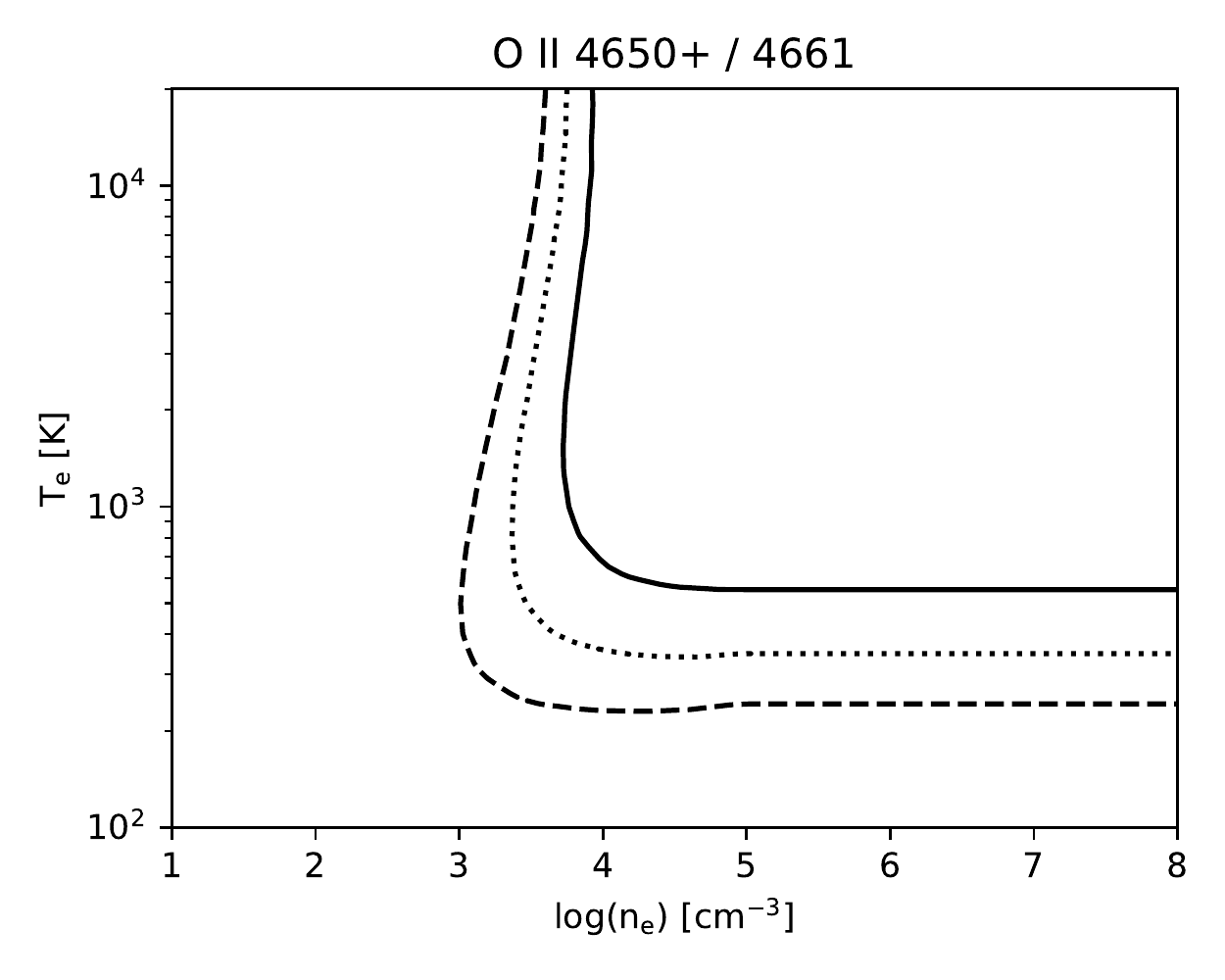}
\caption{Diagnostic diagram obtained from \permr{O}{ii}{4650+}{4661} for the observed ratios of NGC\,6778 (solid line),  M\,1--42 (dotted line) and Hf\,2--2 (dashed line).
\label{fig:OII_diag}}
\end{figure}

\begin{figure*}
\includegraphics[scale = 0.45]{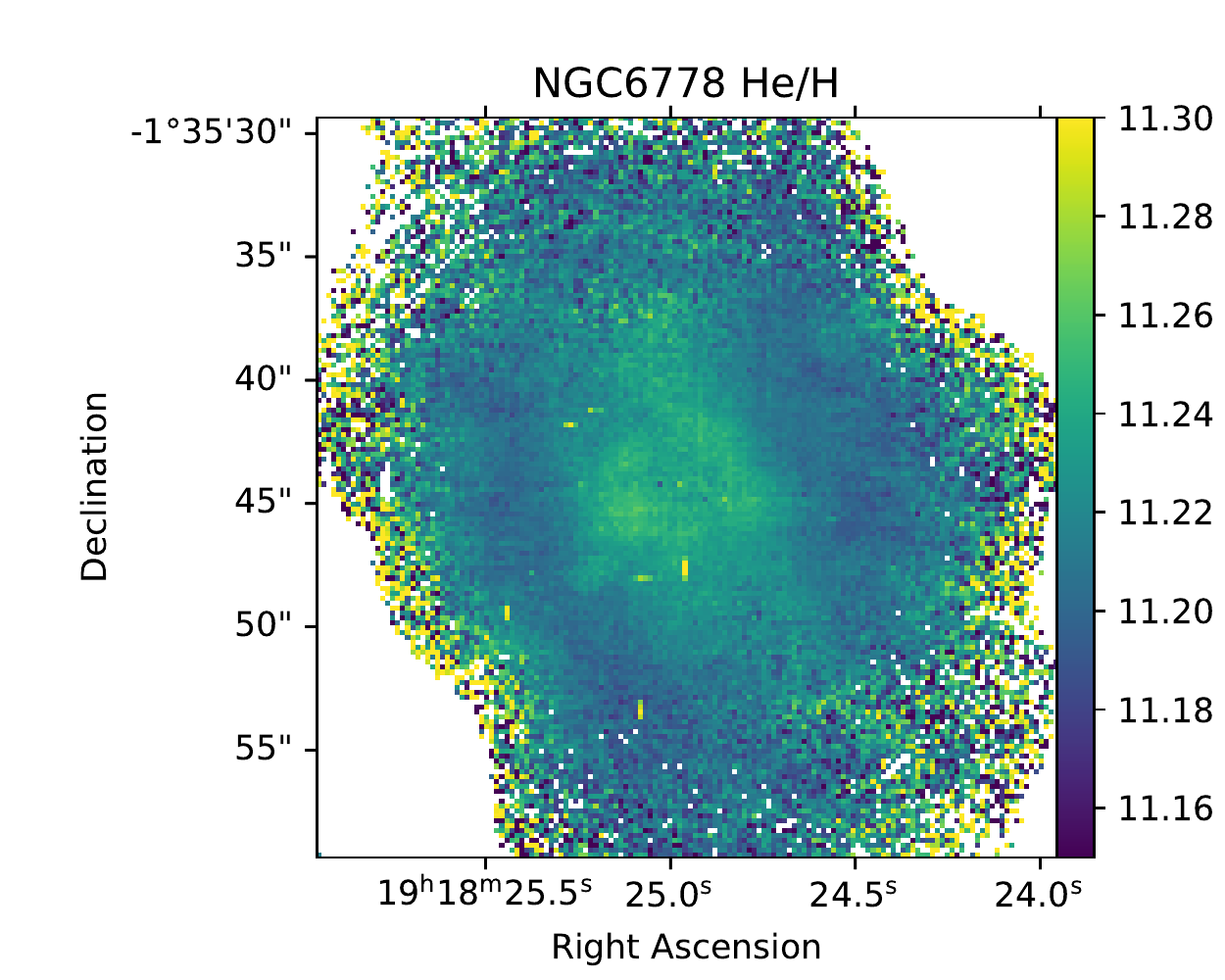}
\includegraphics[scale = 0.45]{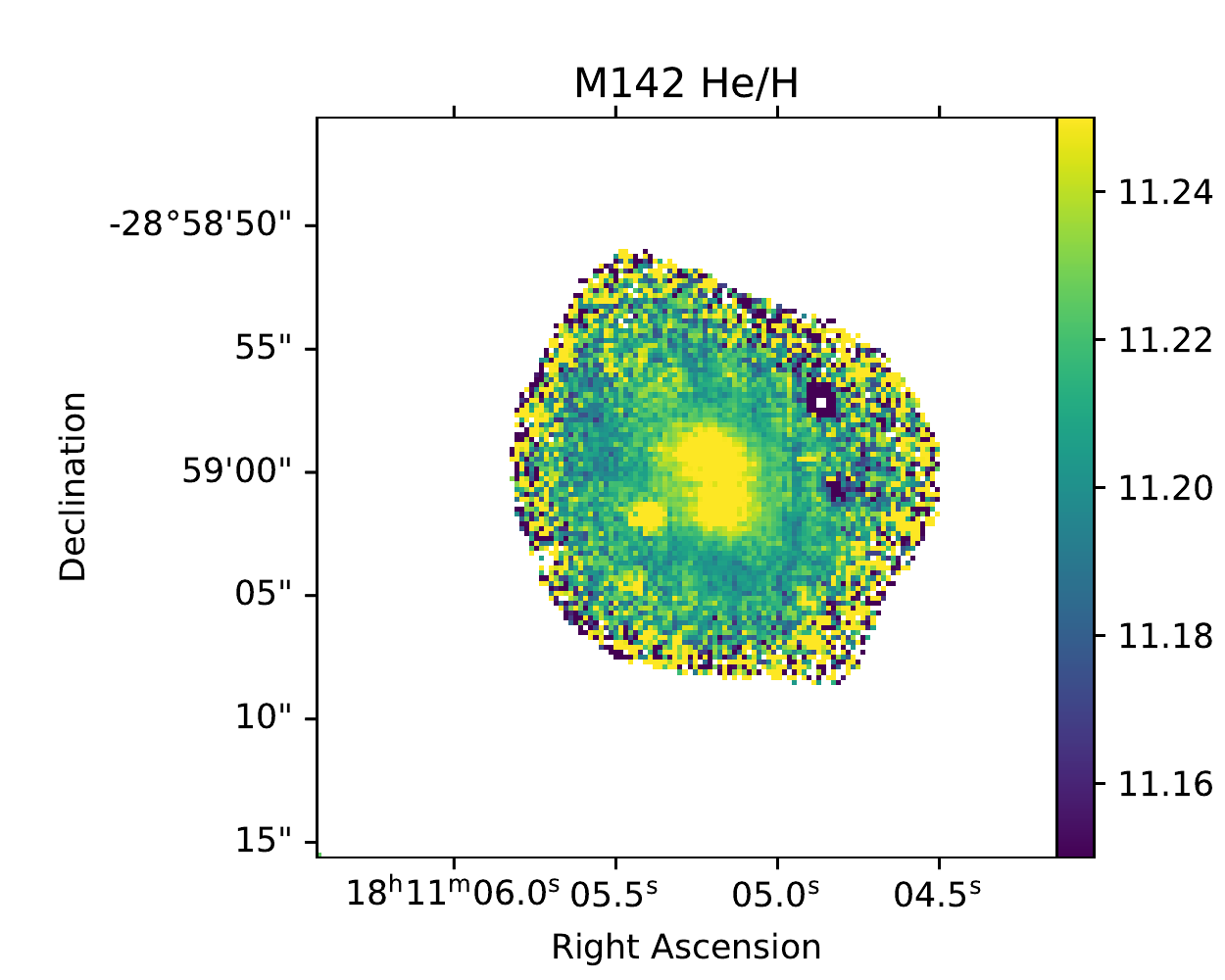}
\includegraphics[scale = 0.45]{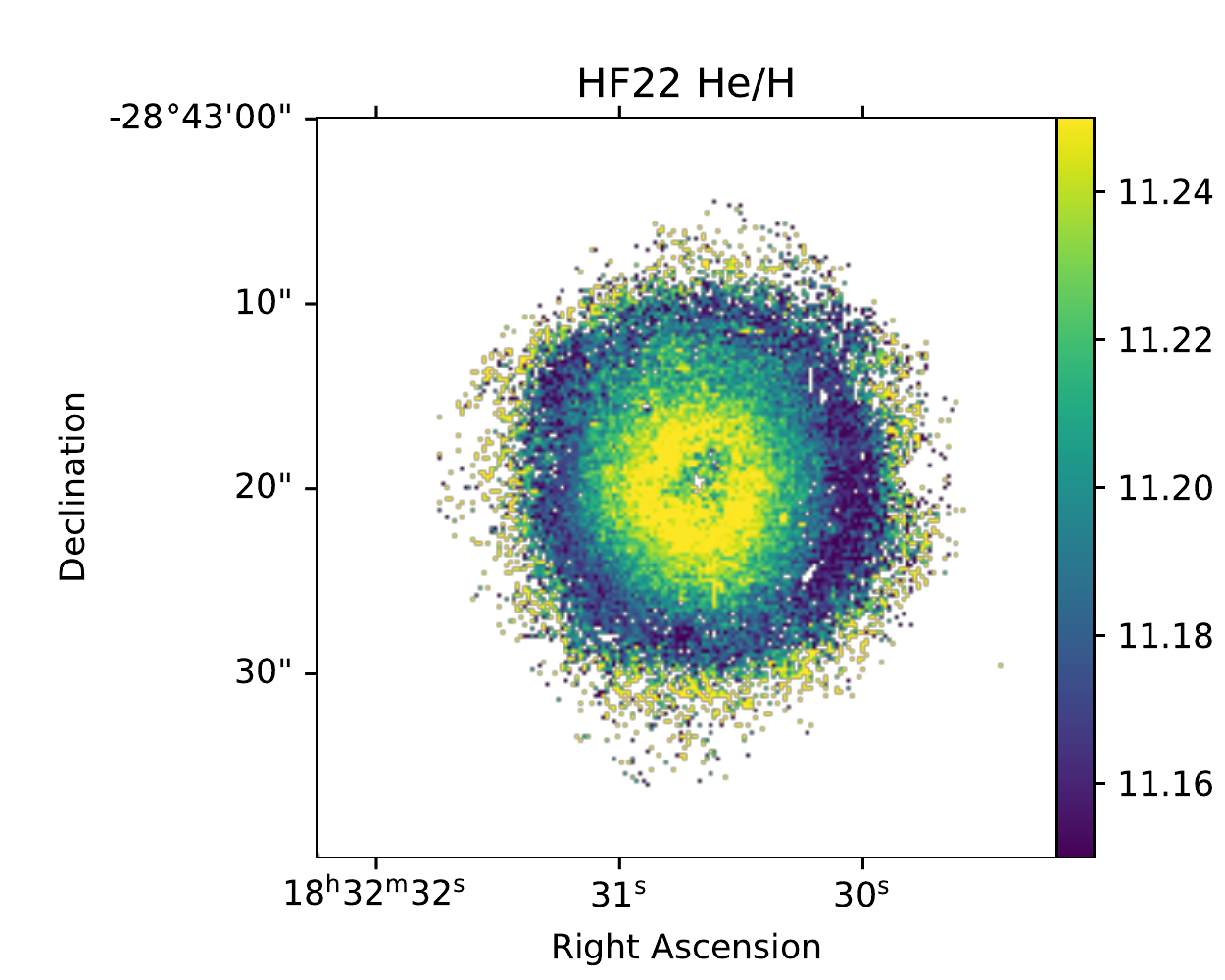}
\caption{Total 12+$\log$(He/H) abundance maps, computed from the sum of  He$^+$/H$^+$ and He$^{2+}$/H$^+$ ratios for the three planetary nebulae of our sample
\label{fig:ab_he}}
\end{figure*}

To compute the elemental abundances of elements heavier than helium, ICFs must be accounted for. The available ICFs in the literature were built using sets of PN photoionization models that are theoretical representations of entire objects, in which relations between the observed and unseen ionic species are obtained as a function of the ionization degree. Therefore, it does not make sense to apply these relations on a spaxel-by-spaxel basis as the observed ionization degree in a single spaxel is not representative of the whole nebula.

For this reason, the abundances of elements other than He were determined from the integrated spectrum within the mask described in Section~\ref{sec:maps}. In Table~\ref{tab:line_fluxes}, we present the unreddened and extinction corrected line fluxes along with their uncertainties. The extinction corrected fluxes of {\fnii} $\lambda$5755 and {\foii} $\lambda\lambda$7320+30 reported in the table correspond to a recombination correction assuming {\te} = 4\,000\,K. In Table~\ref{tab:phys_cond}, we show the physical conditions computed for the integrated spectra after assuming {\te} = 4\,000 K for the recombination correction, and in Table~\ref{tab:ion_abundances} we show the ionic abundances obtained for the different ions and emission lines.

It can be seen from Table~\ref{tab:ion_abundances} that there is an overall good agreement between abundances derived from different lines of the same ion. The exception are the abundances from the recombination lines of {\cii}, {\nii} and {\oii}, where the abundances determined from the brightest lines ({\cii} $\lambda$6462, {\cii} $\lambda$5679 and {\oii} $\lambda$4649+50) seem to be systematically higher. In the case of the {\oii} ORLs, these differences could be attributed to {\ciii} $\lambda$4649 contamination of the $\lambda$4649+50 feature. \citet{mcnabbetal16} obtained deep high resolution spectra for M\,1--42 and Hf\,2--2 and did not report any detection of {\ciii} lines. Similarly, for NGC\,6778, \citet{jonesetal16}, although with lower spectral resolution, did not report the detection of any of the {\ciii} multiplet V1 lines. Hence, line contamination is unlikely to cause  these systematic differences. The observed differences could also be attributed to departure from local thermodynamic equilibrium (LTE) from the upper levels of the transitions of this multiplet at the relatively low {\elecd} of these PNe. These effects were taken into account in the effective recombination coefficient computations by \citet{storeyetal17}, but the lack of a reliable {\te}({\oii}) diagnostic precludes the computation of a reliable {\elecd} for the {\oii} emitting plasma. Fig.~\ref{fig:OII_diag} illustrates the dependence of the  observed {\oii} $\lambda$4649+50/$\lambda$4661 ratios with {\te} and {\elecd}. To compute the O$^{2+}$/H$^+$ ratio for each PN,  we assumed $T_{\rm e} = 4\,000$\,K and {\elecd}({\fsii}). It is quite clear that when fixing $T_{\rm e} = 4\,000$\,K, the observed ratios are compatible for {\elecd} in the range $10^{3.3}$\,$\lesssim$ {\elecd}~[cm$^{-3}$] $\lesssim$ 10$^4$, larger than those adopted ({\elecd} $\lesssim$ 10$^3$\,cm$^{-3}$). Regardless of the origin of these differences, we have adopted as representative values those given by the brightest line in each case, i.e. {\cii} $\lambda$6462, {\nii} $\lambda$5679 and {\oii} $\lambda$4649+50.

\subsubsection{Ad-hoc ICFs}

We considered the ICFs obtained from the literature and computed for each object from a grid of photoionization models described in \citet{delgadoingladaetal14}. We use an extended version of this grid with a larger effective temperature range and run with version c17.01 of {\sc cloudy} \citep{ferlandetal17}, hold in the \verb!3MdB_17! database\footnote{See full description at \url{https://sites.google.com/site/mexicanmillionmodels/the-different-projects/pne_2014}} \citep{morissetetal15} under the reference \verb!PNe_2020!. 

From the more than 700\,000 models in the database, we select those which fit the observed values of the following ionic abundance ratios within 1 dex, for each object: He$^{2+}$/He$^{+}$, O$^{2+}$/O$^{+}$, S$^{2+}$/S$^{+}$, Cl$^{3+}$/Cl$^{2+}$ and Ar$^{3+}$/Ar$^{2+}$. This set of models were used to train seven XGBoost \citep{chenguestrin16} regressors, using the logarithmic values of the above ionic abundance ratios (scaled using the \verb!StandardScaler! routine from {\sc sklearn}) as input and the following seven ICFs: C/C$^+$, N/N$^+$, (O$^+$/O).(N/N$^+$), O/(O$^+$ + O$^{2+}$), S/(S$^+$ + S$^{2+}$), Cl/(Cl$^{2+}$ + Cl$^{3+}$) and Ar/(Ar$^{2+}$ + Ar$^{3+}$) as output. 80~per cent of the model set was dedicated to the training, while the remaining models were used as a test set. The hyper-parameters were set to the following values: \verb!learning_rate!~$=0.1$, \verb!n_estimators!~$=500$ and \verb!max_depth!~$=10$, which give a standard deviation of less than 1\,per cent for the difference between the predicted and test values in the test set. This process was applied for each object separately. 

These ad-hoc ICFs were then used to determine the elemental abundances in each object and to compare the results with the values obtained using the ICFs from \citet{peimbertcostero69}, \citet{kingsburghbarlow94} and \citet{delgadoingladaetal14}. The abundances are listed in Table~\ref{tab:comp_lit}. 


Comparison between the ICFs obtained using ML techniques and those from the literature\footnote{We have used the ICFs from \citet{delgadoingladaetal14} for He (eq. 10), C (eq. 39), O (eq. 12), S (eq. 26), Cl (eq. 29) and Ar (eq. 35), and from \citet{peimbertcostero69, kingsburghbarlow94} for N.} results in elemental abundances that are in excellent agreement for all cases (within 0.04~dex at most), except for N in Hf\,2--2. For this PN, the ICF computed using ML provides $12+\log({\rm N/H})= 8.14 \pm 0.18$, while the classical ICF (N/O=N$^+$/O$^+$, e.g. $\log[{\rm ICF(N}^+/{\rm O}^+)]=0.0$) yields $7.83 \pm 0.28$. However, even in this case, the difference is still within 1-$\sigma$. For Hf\,2--2, the He ionization is rather low (He$^{2+}$/(He$^{+}$ + He$^{2+}$) $\sim$ 0.01) and this classical ICF is not correct \citep[see panel (a) of fig. 6 in][]{delgadoingladaetal14}. Our ad-hoc ML method leads to $\log[{\rm ICF(N^+/O^+)]} = 0.31 \pm 0.05$\,dex. The values from \citet{delgadoingladaetal14} are 0.12$^{+0.40}_{-0.26}$ when \heii\ is seen, and 0.52$^{+0.40}_{-0.26}$ instead, enclosing our ML value.

The ICFs needed to obtain the final element abundances are mainly based on ionic fractions of He and O when derived from \citet{delgadoingladaetal14}, and ionic fractions of He$^{2+}$/He$^{+}$, O$^{2+}$/O$^{+}$, S$^{2+}$/S$^{+}$, Cl$^{3+}$/Cl$^{2+}$, and Ar$^{3+}$/Ar$^{2+}$ in the case of our ad-hoc ICFs from ML. The metal ionization states are obtained from CELs, but for helium we rely on RLs, without a clear knowledge of where theses lines are coming from. If a significant proportion of helium lines originate from the ``cold'' region with a different ionization state than the ``warm'' region, the ICFs depending on He$^{2+}$/He$^{+}$, and the corresponding abundances, may be wrong.

\subsubsection{Comparison with previous works}

\begin{table*}
\caption{Comparison between the elemental abundances$^{\rm a}$ determined in this work, using the ICFs obtained with machine learning techniques, and those obtained in the literature$^{\rm b}$. \label{tab:comp_lit}}
\begin{tabular}{lcccccccc}
\hline
 & \multicolumn{2}{c}{NGC\,6778} & \multicolumn{3}{c}{M\,1--42} & \multicolumn{3}{c}{Hf\,2--2}  \\
\hline
 & This work & JWG16  &  This work  &  LLB01 &  MFL16 & This work & LBZ06 & MFL16  \\
\hline
He 	 & 11.15 $\pm$ 0.02 &  11.20  &  11.16 $\pm$ 0.02   & 11.17 &  11.09   & 11.08 $\pm$ 0.03  &  11.02  & 11.07   \\
C (RLs)$^{\rm c}$ & 9.14 $\pm$ 0.03 &  9.52 &   9.22 $\pm$ 0.03  & 9.35  & 9.40   & 9.64 $\pm$ 0.27  & 9.63 & 9.62  \\
N 	&   8.59 $\pm$ 0.06 &  8.60    &  8.76 $\pm$ 0.06 &  8.68  & 8.77  & 8.14 $\pm$ 0.18 & 7.77 & 8.00  \\
O 	&  8.69 $\pm$ 0.06    &  8.53   & 8.79 $\pm$ 0.04 &  8.63  & 8.75  & 8.52 $\pm$ 0.09  &  8.11 &  8.35  \\
O (RLs)$^{\rm c}$  &  9.73 $\pm$ 0.03  & 9.78  & 9.74 $\pm$ 0.03   & 9.79 & 9.56 &  10.01 $\pm$ 0.26  & 9.94 & 9.72  \\
S 	&   7.00 $\pm$ 0.07 &  6.53   &    7.20 $\pm$ 0.04   &   7.08  & 6.90 &  6.74 $\pm$ 0.08  & 6.37 & 6.69 \\
Cl &   5.33 $\pm$ 0.06 & --- &   5.47 $\pm$ 0.04  & 5.26  & 5.52 &  5.10 $\pm$ 0.06 & --- & 5.64 \\
Ar  &   6.44 $\pm$ 0.06   &  5.67   &  6.67 $\pm$ 0.04 & 6.56 & 6.10 &  6.21 $\pm$ 0.07  & 6.13 & 5.78 \\
\hline
\end{tabular}
\begin{description}
\item $^{\rm a}$ Abundances in units of $12+\log{\rm (X/H)}$. \\
\item $^{\rm b}$ References: JWG16: \citet{jonesetal16}; LLB01: \citet{liuetal01}; MFL16: \citet{mcnabbetal16}; LBZ06: \citet{liuetal06}.  \\
\item $^{\rm c}$ In these cases, the ICFs were not computed using machine learning techniques given the lack of ionization state diagnostics for the ``cold'' gas component. Instead, we used the ICFs from \citet{delgadoingladaetal14}: C (eq. 39) and O (eq. 12). However, these recipes should be taken with caution (see text). The influence of a cold region on the H\,{\sc i} emission lines is not taken into account here (see Section~\ref{sec:cold_region}).\\

\end{description}
\end{table*}

The most relevant previous works are those by \citet{jonesetal16} for NGC\,6778, \citet{liuetal01} and \citet{mcnabbetal16} for M\,1--42 and \citet{liuetal06} and \citet{mcnabbetal16} for Hf\,2--2. 
Table~\ref{tab:comp_lit} presents the elemental abundances obtained from the integrated spectra of the three PNe in this work as well as those reported in the literature. 

Ideally, comparison of chemical abundances obtained from different data sets should be done by recomputing abundances using the same methodology. However, this is not always an easy task, especially given the limited wavelength coverage of the MUSE data, that miss some important lines at wavelengths bluer than 460 nm (e.g. {\foii} $\lambda$3736+29 and {\foiii} $\lambda$4363). Therefore, we have decided to carry out a critical comparison of the most relevant elemental abundances. 

For NGC\,6778, there are significant differences between our abundances and those derived by \citet{jonesetal16}, with the exception of O from RLs and N from CELs, which are largely consistent. We find an excellent agreement in the total He abundance when computing the He$^{+}$ abundance adopting {\te}({\hei}). Regarding heavy-element abundances from ORLs, C abundances disagree by almost 0.4 dex. This difference is mainly driven by the different C$^{2+}$ abundance obtained by \citet{jonesetal16}, which is a factor of 1.5 higher than ours, and is caused by the combination of two factors: the {\cii} lines are slightly brighter in \citet{jonesetal16} and they adopted {\te}({\foiii}), which increases the abundances by $\approx 20$ per cent relative to using the {\te} obtained with ORLs diagnostics. On the other hand, the O abundance obtained by these authors from CELs is 0.16 dex lower than ours, but this difference is easily explained by the higher {\te}({\foiii}) adopted by \citet{jonesetal16}, that could be affected by the recombination contribution to the {\foiii} $\lambda$4363 auroral line \citep{gomezllanosetal20}, while our adopted {\te}({\fsiii}) is free from such effects (see Section~\ref{sec:rec_cont}). Similar arguments apply for S, although the differences found are larger and cannot be fully explained by the adopted {\te}. In this case, two factors can be at play: i) the S$^{2+}$ abundances by \citet{jonesetal16} were derived from the {\fsiii} $\lambda$6312 CEL, which is extremely dependent on the assumed {\te}; and ii) there was a misalignment between the blue and red arm observations of \citet{jonesetal16}, that, although treated with extreme caution, could have introduced systematic effects in the derived line ratios. Finally, the $\simeq 0.8$ dex lower Ar abundance in \citet{jonesetal16} is the result of accounting for Ar$^{3+}$ abundances only. Thus, the ICF is quite uncertain, especially given that Ar$^{2+}$ is the dominant ionic species in the nebula  \citep{delgadoingladaetal14}. 
Our newly derived O (and also Cl and Ar) abundances indicate that NGC\,6778 is an object with almost solar metallicity \citep{lodders19}. 

For M\,1--42, the results presented in the two other works mentioned are in relatively good agreement with ours, although some slightly larger differences ($>0.2$ dex) can be found for S, Cl and Ar from CELs, and for O from ORLs. However, the agreement is, in general, quite good for O and N, providing a value of ${\rm N/O} \simeq 0.93$, which reinforces the classification of this object as a type-I PN.

Finally, for Hf\,2--2, comparison with the works by \citet{liuetal06} and \citet{mcnabbetal16} confirms the high ADF(O$^{2+}$) found in this PN, although it seems to be not as high as previously reported. This is an obvious consequence of us adopting the relatively low {\te}({\fsiii})$=6\,820$\,K, which is $\simeq 1\,900$ and $2\,800$\,K lower than the {\te}({\foiii}) in \citet{liuetal06} and \citet{mcnabbetal16}, respectively. We cannot assign these differences to any specific reason other than the fact that the auroral {\fsiii} $\lambda$6312 line is relatively weak in this PN. However, after examination of the observed {\fsiii} line fluxes relative to H$\beta$ in the two previous works, we find that \citet{liuetal06} reported a {\fsiii} $\lambda$6312/H$\beta$ ratio that is 60 per cent higher than the reddened flux measured in this work from the integrated spectra. Unfortunately, these authors did not cover wavelengths redder than 720 nm, and hence they were unable to detect the nebular {\fsiii} lines.

The fluxes reported by \citet{mcnabbetal16} are deemed unreliable owing to two main reasons: i) they reported fluxes one order of magnitude higher for multiple faint lines, which points to some kind of mistake in transcribing the line flux tables, ii) the dereddened fluxes of the {\hi} Paschen lines are much brighter (more than a factor 2) than expected for the assumed physical conditions, while the {\fsiii} $\lambda$9069/H$\beta$ line ratio is $\approx 30$ per cent smaller than observed here, which points to problems with the flux calibration, extinction correction and/or telluric absorptions.  






\subsubsection{Effect of the recombination contribution correction on the integrated spectra }

We have investigated the effect of neglecting the recombination contribution correction on the integrated spectra. Although its effect on the physical conditions maps is dramatic, as we have shown in Section~\ref{sec:rec_cont}, it is somewhat mitigated in the integrated spectra of NGC\,6778 and M\,1--42, owing to the different weights of the ORL and CEL emitting zones in the integrated spectrum. However, it is still extremely significant in Hf\,2--2 and should be accounted for carefully. We find a recombination contribution to the {\fnii} $\lambda$5755 auroral line of $\approx 17$, $\approx 18$ and $\approx 64$ per cent (assuming $T_{\rm e}=4\,000$\,K) for NGC\,6778, M\,1--42 and Hf\,2--2, respectively. This translates into respective {\te}({\fnii}) decreases of $\simeq 550$, $\simeq 600$ and $\simeq 3950$\,K in NGC\,6778, M\,1--42 and Hf\,2--2, which are subsequently propagated to the abundances of low-ionization species.    
Similarly, these corrections are of $\approx 44$, $\approx 42$ and $\approx 75$ per cent for the {\foii} $\lambda\lambda$7320+30 trans-auroral lines in the three PNe under the same {\te} conditions. This illustrates the importance of accounting for the recombination contribution when these lines are the only ones available to compute the O$^+$ abundances. 

\begin{table*}
\caption{Differences between physical and chemical properties derived assuming different recombination contribution corrections with respect to the adopted ones. \label{tab:comparison}}
\begin{tabular}{lccc|ccc|ccc}
\hline
 & \multicolumn{3}{c}{NGC\,6778} & \multicolumn{3}{c}{M\,1--42} & \multicolumn{3}{c}{Hf\,2--2}  \\
\hline
 {\te} for Rec. Cont. & None & 1\,000  & 8\,000 &None & 1\,000  & 8\,000 & None & 1\,000  & 8\,000  \\
\hline
$\Delta$({\te}({\fnii})) (K) 	 &  $+$560  & $+$190  & $-$130  &  $+$590  & $+$200   & $-$150   &  $+$3960  &  $+$1465   & $-$1225 \\
$\Delta$({\elecd}({\fsii})) (cm$^{-3}$) &  $+$25  & $+$10  & $-$5  &  $+$30  & $+$10  & $-$10   &  $+$60 & $+$25  &  $-$20 \\
$\Delta$(log(N$^+$/H$^+$))	        & $-$0.09   & $-$0.03  & $+$0.02  & $-$0.09   & $-$0.03  &   $+$0.02  &  $-$0.45  & $-$0.20  &  $+$0.24 \\
$\Delta$(log(O$^+$/H$^+$)) 	   & $+$0.04   & $+$0.02  & $-$0.03  & $+$0.01   & $+$0.01  & $-$0.01  & $-$0.49   & $-$0.21  &  $+$0.17 \\
$\Delta$(log(N/H)) 	             & $-$0.08   & $-$0.05  & $+$0.03  & $-$0.07  & $-$0.03  &  $+$0.403  &  $-$0.01 & $-$0.03  & $-$0.04 \\
$\Delta$(log(O/H)) 	             & $+$0.01   & =  &  $-$0.01  & $-$0.01  & =  & $-$0.01  & $-$0.07  & $-$0.04 & $+$0.04 \\
\hline
\end{tabular}
\begin{description}
\item ...
\end{description}
\end{table*}

\section{Discussion}
\label{sec:discuss} 

\subsection{On the influence of a cold region on {\hi} temperature and abundance determinations}
\label{sec:cold_region}

\begin{figure}
\includegraphics[scale=0.65]{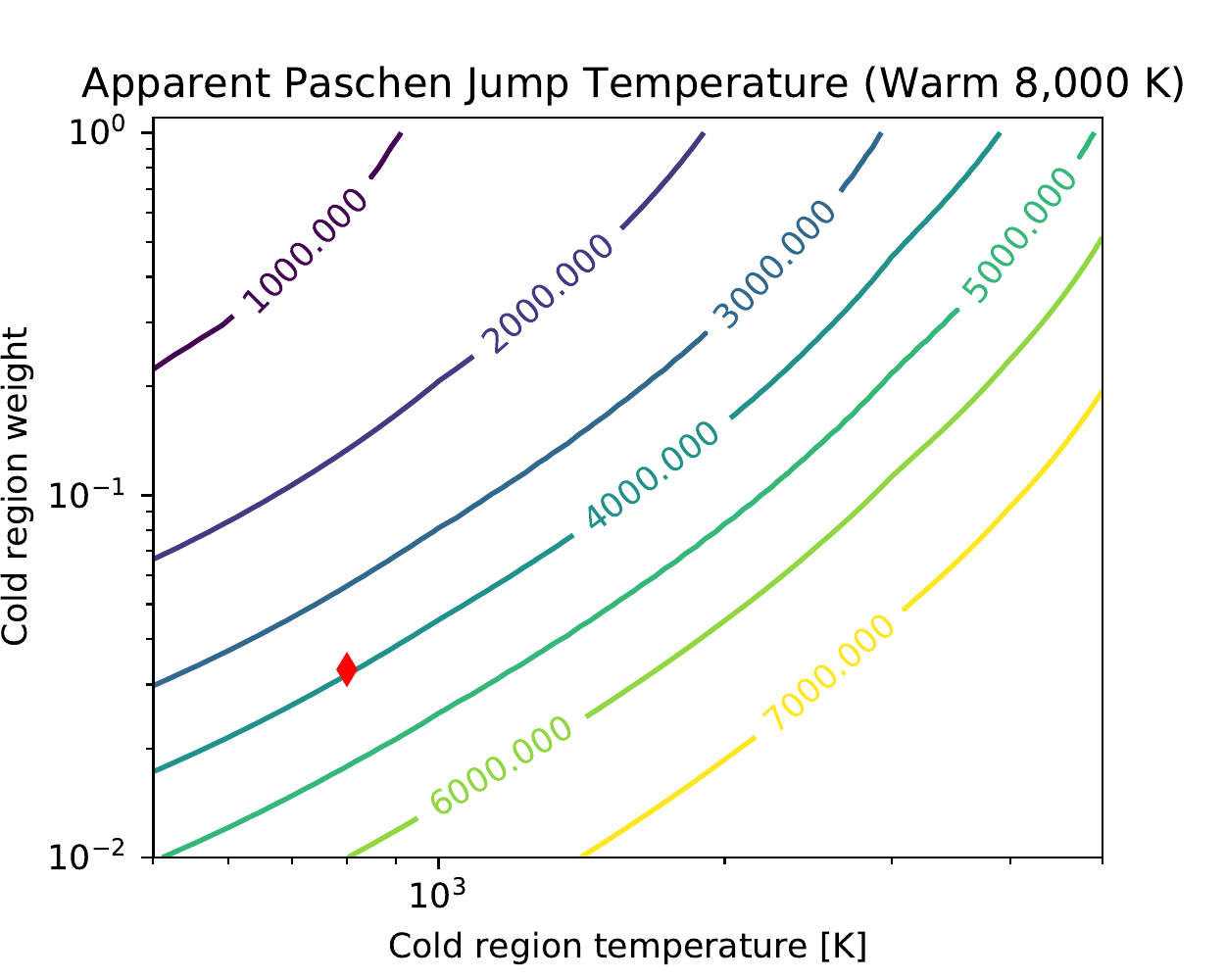}
\caption{Contour plot of the observed Paschen jump electron temperature as a function of the temperature of the cold region and its relative weight in the total \hi\ emission. For the warm gas, we assumed $T_{\rm e}=8\,000$\,K. The red diamond illustrates the case of a cold region at 800\,K contributing 3.3 per cent of the emission and leading to an apparent Paschen jump temperature of 4\,000\,K. 
\label{fig:T_PJ_cold_warm}}
\end{figure}

The main result of our analysis of the emission lines, that led to the determination of the physical and chemical parameters of the three PNe, points to the presence of two phases of gas, one of which is quite cold ({\te} of the order of $\sim 10^3$~K). This cold region shows metal abundances significantly larger than those obtained for the ``classical" region. Despite being poor in hydrogen, the cold region may substantially contribute to the observed \hi\ recombination line and continuum emission (this is clearly the case considering the low temperatures from the Paschen jump obtained in Section~\ref{sec:te_rec_paschen}). This opens an interesting discussion as to the way ionic (and elemental) abundances are derived. Let us consider a given emission line at wavelength $\lambda$ emitted by the two different phases of the gas (the warm ``classical" region and the cold, metal-rich region), identified by the $w$ and $c$ superscripts, respectively. The resulting line intensity, normalized to \hb, is the following:

\begin{equation}
    \frac{I_\lambda}{I_\beta} = 
    \frac{I_\lambda^{\rm w} + I_\lambda^{\rm c}}{I_\beta^{\rm w} +  I_\beta^{\rm c}} = 
    \frac{I_\lambda^{\rm w} + I_\lambda^{\rm c}}{(1-\omega)   I_\beta + \omega I_\beta}~,
\end{equation}

\noindent
where $\omega$ is the weight of the cold region in the total \hb\ (and recombination continuum) emission (the ``classical" region contribution being $1-\omega$). The abundance of the ion responsible for the line emission is commonly determined assuming it is directly proportional to ${I_\lambda}/{I_\beta}$. If the line is only emitted by one of the two regions, this method will underestimate the ion abundance. However, if it is emitted by the warm region alone (as is the case for CELs), $I_\lambda^{c} = 0$ and the abundance should be obtained from ${I_\lambda}^{w}/{I_\beta}^{w} = {I_\lambda}/(1-\omega) {I_\beta}$. The abundance is underpredicted by a factor $1/(1-\omega)$. In the case of a line only emitted by the cold region (e.g. ORLs), the underprediction factor is $1/\omega$. 

Under this hypothesis of two distinct gas phases, the ionic abundances in each region can be computed using the corresponding electron temperature. For the ``classical" region, it is obtained using ``classical" diagnostic line ratios like \forbr{O}{iii}{4363}{5007}. For the cold region, the exact value of the temperature is commonly considered as not necessary, as the emissivities of the metal and H recombination lines have almost the same dependence on the temperature and cancel each other in the abundance determination. However, {\hi} emission is produced in both regions, and therefore this argument no longer applies.

The weight $\omega$ of the cold region in the total \hb\ emission can be estimated using the determination of the temperature from the Paschen jump (see Section~\ref{sec:rec_cont_te}), in a way inspired by \citet{gomezllanosmorisset20}. In Fig.~\ref{fig:T_PJ_cold_warm} we show the apparent Paschen jump temperature determined from an observation where the emission of the continuum and of the \hi\ lines results from emission in both a warm region (here at T$_e^w\sim$ 8\,000~K) and a cold region of temperature $T_{\rm e}^{\rm c}$ ($x$-axis). The $y$-axis is the cold region weight $\omega$. We see that if the cold region is, for example, at 800~K, an apparent Paschen jump temperature of 4\,000~K is obtained if the cold region contribution to \hi\ is as small as 3.3 per cent (red diamond in Fig.~\ref{fig:T_PJ_cold_warm}). This implies that any determination of abundances obtained from emission lines mainly emitted by the cold region needs to be corrected by a factor of 1/0.033 = 30! The determination of the abundances from CELs will only be underestimated by a factor $1/(1-0.033)$, that is, $\sim 3.4$\,per cent. This obviously depends on the adopted values for the warm and cold region temperatures. In addition, the \hei\ temperature may not be relevant to obtain the cold region temperature if strictly metal-rich (and then H- and He-poor).

We can then derive a correction to the ionic abundances and obtain a more realistic value of e.g. O$^{2+}$/H$^+$ in the cold and warm regions. The ratio between these corrected abundances obtained from ORLs and CELs is close to what \citet{gomezllanosmorisset20} refer to as the Abundance Contrast Factor (ACF). The relation between the ACF and the observed ADF is:
\begin{align}
\label{eq:ACF}
    {\rm ACF}(X^{i+}/H^+) &= \frac{(X^{i+}/H^+)_{rec}}{\omega} \cdot \frac{1-\omega}{(X^{i+}/H^+)_{col}} \nonumber \\ 
                    &= {\rm ADF}(X^{i+}/H^+) \cdot \frac{1-\omega}{\omega}~.
\end{align}

We explore the effect of the ``dual phase hypothesis'' on the determination of the ACF(O$^{2+}$/H$^+$) as defined by Eq.~\ref{eq:ACF} for our three PNe. We consider three different values for the cold region temperature: 500, 800 and 1\,000~K. The obtained ADF does not significantly change when T$_e^w$ varies from 8\,000 to 12\,000~K, and hence we fix the electron temperature of the warm phase to $T_{\rm e}^{\rm w}=8\,000$~K. Table~\ref{tab:ACF} gives the logarithmic ADF(O$^{2+}$/H$^+$) and ACF(O$^{2+}$/H$^+$) values that we obtain for each PN. In the case of PN Hf\,2--2, where the Paschen jump temperature is close to 1\,000~K, $T_{\rm e}^{\rm c}\approx 500$~K is preferred. We can see that the ACF increases for decreasing $T_{\rm e}^{\rm c}$. For NGC\, 6153, \citet{gomezllanosmorisset20} determined $\log[\rm ACF(O/H)]$ in the range $2.1-2.7$\,dex, with $T_{\rm e}^{\rm c}$ close to 500~K. These values are very similar to the ones obtained here. It is interesting to note that the highest ACFs are not obtained for the highest ADFs, which points to both parameters not being connected by a simple relation.

\begin{table}
\caption{Logarithmic ADF(O$^{2+}$/H$^+$) and  ACF(O$^{2+}$/H$^+$) values for the collapsed spectra. \label{tab:ACF}}
\resizebox{\columnwidth}{!}{%
\begin{tabular}{ccccc}
\hline
& NGC\,6778 & M\,1--42 & Hf\,2--2  \\
\hline
& \multicolumn{3}{c}{ADF}\\
& \multicolumn{3}{c}{(dex)}\\
\noalign{\smallskip}
& 0.97 $\pm$ 0.07 & 0.90 $\pm$ 0.06 & 1.26 $\pm$ 0.08\\
\hline
$T_{\rm e}^{\rm c}$ & \multicolumn{3}{c}{ACF}\\
(K) & \multicolumn{3}{c}{(dex)}\\
\noalign{\smallskip}
500 & 2.72 $\pm$ 0.19 & 2.54 $\pm$ 0.63  & 1.93 $\pm$ 0.41 \\
800 & 2.43 $\pm$ 0.20 & 2.24 $\pm$ 0.70  & 1.44 $\pm$ 0.50 \\
1\,000 & 2.27 $\pm$ 0.22 & 2.06 $\pm$ 0.78  & 0.81 $\pm$ 0.53 \\

\hline
\end{tabular}
}
\end{table}

\citet{2021Ueta_arXi} point out the effect of considering the ``true'' {\te} in the definition of the expected \hi\ line ratio when determining the extinction coefficient from 2D data. We use 10\,000~K to compute the emissivities of the \hi\ lines and obtained the classical value of 2.86 for \ha/\hb. However, we have already described that part of the \hi\ emission originates from a much colder region, which can increase this ratio to values up to 3.05 (equivalent to the theoretical \ha/\hb\ ratio obtained for \te=4\,500~K). We thus reran the whole pipeline to derive all the results presented in previous sections using this larger value of \ha/\hb\ (then reducing the reddening correction). We found that the \fnii\ and \fsiii\ electron temperatures decrease by 100 and 200~K, respectively, and an increase of the metal abundances by 0.02 to 0.07 dex, depending on the wavelength of the emission line. We also carried out a quick, more detailed exploration on the effects introduced in our computations of physical conditions and chemical abundances by using the spatial variations of {\te}(PJ) to compute the resolved c(\hb), and found no strong effect. Therefore, the main conclusions presented in previous sections remain unaltered.

\subsection{Oxygen content in the cold region}
\label{sec:oxy_fraction}

Following \citet{gomezllanosmorisset20}, we can estimate the oxygen mass ratio between the cold and the warm region, $M^{\rm c}/M^{\rm w}$. Considering that the oxygen ORLs are only emitted by the cold region and that the CELs come from the warm region, we have:
\begin{equation}
    \label{eq:oxypmass}
    \frac{M^{\rm c}}{M^{\rm w}}({\rm O}^{+}) = \frac{\epsilon_{[OII]}(T_{\rm e}^{\rm w}, n_{\rm e}^{\rm w})}{\epsilon_{OI}(T_{\rm e}^{\rm c}, n_{\rm e}^{\rm c})} \cdot \frac{I_{OI}}{I_{[OII]}} \cdot \frac{n_{\rm e}^{\rm w}}{n_{\rm e}^{\rm c}}
\end{equation}
and
\begin{equation}
    \label{eq:oxyppmass}
    \frac{M^{\rm c}}{M^{\rm w}}({\rm O}^{2+}) = \frac{\epsilon_{[OIII]}(T_{\rm e}^{\rm w}, n_{\rm e}^{\rm w})}{\epsilon_{OII}(T_{\rm e}^{\rm c}, n_{\rm e}^{\rm c})} \cdot \frac{I_{OII}}{I_{[OIII]}} \cdot \frac{n_{\rm e}^{\rm w}}{n_{\rm e}^{\rm c}}~,
\end{equation}
\noindent
where $\epsilon$ is the emissivity corresponding to the observed intensity $I$.
The results are presented in Table~\ref{tab:oxymass}. The values of ${M^{\rm c}}/{M^{\rm w}}$ are inversely proportional to the adopted $N_e^c$ and almost linearly dependent on the adopted $T_{\rm e}^{\rm c}$, with both parameters being highly uncertain. We used $T_{\rm e}^{\rm c} = 800\,\rm K$; higher temperatures (e.g. 4\,000\,K) would result in larger oxygen mass ratios by a factor of about 4. The oxygen mass fraction does not vary significantly from one object to another and is similar to the value obtained by \citet{gomezllanosmorisset20} for NGC\,6153 considering $T_{\rm e}^{\rm c} = 800$~K. Whatever the exact value of $T_{\rm e}^{\rm c}$, we found that the amount of oxygen contained in the cold and ``classical" warm regions  are of the same order of magnitude. 

We show in this Section and  Section~\ref{sec:cold_region} that a better knowledge of the properties of the three PNe presented here requires a good determination of either the chemical composition or the electron temperature of the metal-rich, cold region. We remind that the ADF does not provide any insight into the actual enrichment of the cold region. A detailed photoionization model, as the one obtained by \citet{gomezllanosmorisset20}, could be obtained by exploring large grids of photoionization models, varying O/H and the weight of the rich region, as well as its density and helium abundance. This is far beyond the scope of the present work, but is certainly necessary to place constraints on the physical and chemical properties of high-ADF PNe. More observational constraints, such as high spectral and spatial resolution observations that provide {\oii} ORL electron temperature and density diagnostic ratios, will also be helpful. 

\begin{table}
    \centering
    \begin{tabular}{cccc}
         \hline
          & NGC\,6778 & M\,1--42 & Hf\,2--2  \\
         \hline
         $T_{\rm e}^{\rm w}$~[K]  &  8\,300 & 8\,000 & 7\,000  \\
         $T_{\rm e}^{\rm c}$~[K]  &  800 & 800 & 800 \\
         $n_{\rm e}^{\rm w}$~[cm$^{-3}$]  &  800 & 1\,000 & 500  \\
         $n_{\rm e}^{\rm c}$~[cm$^{-3}$]  &  2\,000 & 2\,000 & 2\,000  \\
         ${M^{\rm c}}/{M^{\rm w}}({\rm O}^{+})$  & 0.7 & 0.7 &  1.0\\
         ${M^{\rm c}}/{M^{\rm w}}({\rm O}^{2+})$  & 0.6 & 0.7 &  0.9\\
         \hline
    \end{tabular}
    \caption{Parameters and determinations mass ratios between the cold and warm regions, derived for ${\rm O}^{+}$ and ${\rm O}^{2+}$, for each PN of our study.}
    \label{tab:oxymass}
\end{table}

\subsection{The link between high ADF and binarity}
\label{sec:interpret_maps}



It now seems relatively clear that the extreme ADFs observed in some PNe are due to the presence of a second, metal-rich gas phase \citep[e.g.;][]{liuetal06,wessonetal18}. The origin of this second gas phase is highly uncertain and, without knowledge of the properties of the two phases \citep[which cannot be inferred directly from observations;][]{gomezllanosetal20,gomezllanosmorisset20}, various (highly speculative) hypotheses that have been put forward in the literature.  

It has been highlighted the similarities between the observed ORL abundances and those in neon novae \citep[e.g.][]{wessonetal03,wessonetal08}, leading to speculation that the metal-enriched phase could be the product of some form of late reprocessing either as a result of mass transfer from a companion star to the white dwarf  \citep{wessonetal08novavul,jones19} or fallback of binary common-envelope material \citep{kuruwita16,reichardt19}. \citet{wessonetal18} argued in favour of a nova-like origin of the metal-enriched material, that may be strongly linked to the correlation between short binary orbital periods and high ADFs. In either case, the material ejected due to this reprocessing would be expected to have an appreciably larger expansion velocity than the ``normal'' nebular phase due to earlier mass loss. While some evidence exists for kinematic differences between the ORL and CEL emitting gas phases in PNe \citep{richeretal17}, nova-like velocities have not been detected so far in the H-deficient material in any high-ADF PNe. Unfortunately, the spectral resolution of our MUSE data is insufficient to discern any kinematic differences between the {\oii} and {\foiii} emitting material. Thus, detailed 2D kinematic studies of both gas components are still needed to address any solid conclusions.

The possibility that the link between extreme ADFs and common-envelope evolution is simply a consequence of the conservation of chemical stratification in the giant's envelope upon ejection should also be considered. While the ORL emitting gas is more centrally located in all the PNe analysed in this work, its spatial distribution (and that of the ADF, see Figs.~\ref{fig:adfs_NGC6778}, \ref{fig:adfs_M142} and \ref{fig:adfs_HF22}) varies relatively smoothly rather than appearing as a clearly distinct shell.  The extremely short time-scale of the common envelope \citep[$\sim 1$\,yr, e.g.][]{igoshev20}, as well as the limited mixing of the envelope over this time-scale \citep[due to its strong entropy and density gradient;][]{webbink08}, could mean that the chemical stratification in the envelope is preserved upon ejection.  In that case, the centrally peaked ADF and the distribution of the CEL-emitting material may simply be a consequence of this chemical stratification, whereby the internal parts of the giant's envelope are richer in the products of nucleosynthesis, and hence have a naturally higher metallicity. This allows the inner parts to cool more efficiently and leads to enhanced ORL emission.  

\section{Conclusions}
\label{sec:conclu} 

In this work we present deep integral-field unit (IFU) spectroscopy of three PNe (NGC\,6778, M\,1--42 and Hf\,2--2) with previously reported high abundance discrepancy factors (ADF > 20). The spectra were obtained with the MUSE spectrograph covering the wavelength range $4600-9300$\,\AA\ with effective spectral resolution from $R=1609$ to $R=3506$ for the bluest to the reddest wavelengths, respectively. The analysis is restricted to the central 900\,arcsec$^{2}$ in NGC\,6778 and M\,1--42 and the central 1600\,arcsec$^{2}$ in Hf\,2--2 with a spatial sampling of 0.2\,arcsec. These are the deepest IFU data for the three PNe currently available.

We produced emission line maps of more than 40 lines for each object, including several recombination lines of H$^{+}$, He$^{+}$, He$^{2+}$, C$^{2+}$, N$^{2+}$, O$^{+}$ and O$^{2+}$ and collisionally excited lines of C$^{0}$, N$^{0}$, N$^{+}$, O$^{0}$, O$^{+}$, O$^{2+}$, S$^{+}$, S$^{2+}$, Cl$^{2+}$, Cl$^{3+}$, Ar$^{2+}$, Ar$^{3+}$ and Ar$^{4+}$. From these we compute the extinction, physical conditions ({\te} and {\elecd}) from CEL and ORL diagnostics and ionic abundance maps using machine learning techniques, which significantly reduce the computational time needed to analyse such large data sets.

We find that the recombination correction plays a very important role in both the {\te}({\fnii}) and O$^+$ abundance maps. In addition, the obtained maps strongly depend on the assumed {\te} for the recombination emission. We applied the recombination correction to the auroral {\fnii} $\lambda$5755 and {\foii} $\lambda\lambda$7320+30 CELs assuming {\te}=1\,000, 4\,000, 8\,000\,K, as well as no correction at all. We find that adopting a recombination correction assuming {\te} = 4\,000 K is a good compromise for the PNe presented in this work, given the similarity with what is obtained for their central parts using recombination line {\te} diagnostics. We built CEL ionic abundance maps adopting different recombination contributions to the auroral {\fnii} and trans-auroral {\foii} lines, showing that the ionic abundance maps of low-ionization species are strongly affected if this contribution is not properly taken into account.  
The {\hi} continuum to line ratio and the {\hei} line ratio maps show {\te} values systematically lower than those obtained from CEL diagnostics. Furthermore, both types of map display very similar structures, the most remarkable being a decrease of {\te} toward the inner zones of the PNe, which is found to correlate with an enhancement of the ORL emission lines of heavy elements.

We computed elemental abundances using the integrated spectra of each PN and a large number of ionization constraints, such as He$^{2+}$/He$^{+}$, O$^{2+}$/O$^{+}$, S$^{2+}$/S$^{+}$, Cl$^{3+}$/Cl$^{2+}$, and Ar$^{3+}$/Ar$^{2+}$, from which ad hoc ICFs are calculated using a large database of photoioinization models and machine learning methods. For all the elements considered and objects, the agreement with the ICFs from the literature is excellent, with the exception of N in Hf\,2--2, where 0.3 dex differences are found.

From the ionic abundance maps of the {\cii}, {\nii}, {\oi} and {\oii} ORLs we find that ORL emission concentrates at the inner regions of the three PNe compared to the CEL emission of the same ion. This points to the ORL emission of heavy elements coming from a different plasma component than the CEL emission. We also find a significant decrease of the degree of oxygen ionization in the ORL emitting gas relative to the ``normal'' gas component, implying that the increase of the O/H abundance may be the result of the increase of the oxygen density rather than a decrease of the hydrogen density, in agreement with recent photoionization model results by \citet{gomezllanosmorisset20}. We have also constructed, for the first time in photoionized nebulae, the ADF maps of O$^+$ and O$^{2+}$. In both cases, the ADF clearly peaks centrally for the three PNe.

All these results strongly support the ``bi-abundance'' model of high-ADF PNe, which stands for the presence of two gas phases, one cold ({\te}$\sim$10$^3$\,K) and metal-rich that is responsible for the bulk of heavy element ORL emission, and a warm one ({\te}$\sim$10$^4$\,K) where the heavy element CEL emission is produced.

Under this dual gas hypothesis, we show that the main issue with the derivation of the ionic abundances of metals from either ORLs or CELs is the determination of the contribution of each gas component to the observed \hi\ emission. We therefore present a method to estimate the relative contribution of the two gas phases to the total ionic abundances when their physical conditions and the {\te}({\hi}) are known, allowing to compute corrections to the abundances derived from ORLs or CELs. We find that these abundance corrections strongly depend on the adopted values for the warm and cold region temperatures.

Using reasonable values for the undetermined parameters (electron temperature and density, and ionization of the cold region), we derive the O$^{2+}$/H contrast between the warm and cold regions \citep[known as the ACF, which is significantly higher than the ADF, see][]{gomezllanosmorisset20} and the oxygen mass ratios between the two regions. We conclude that the amount of oxygen in the cold region is of the same order of magnitude as in the warm region.

Finally, we conclude that the binary origin of the metal-rich phase can only be unravelled with detailed spatially-resolved kinematics of both gas components. However, we speculate with the possibility that the conservation of chemical stratification in the giant's envelope upon ejection can be a reasonable explanation for the link between extreme ADFs and the common envelope evolution.

\section*{DATA AVAILABILITY}

The raw MUSE data are available from the ESO archive facility at \url{http://archive.eso.org/}.
The data products used in this paper: reduced data cubes, extracted emission line maps, analysis pipeline and its outputs, are available from the authors under reasonable request. The emission line maps and the {\sc python} scripts used for the analysis and to produce the tables and figures presented in this paper are available at \url{https://github.com/Morisset/MUSE_PNe}.

\section*{Acknowledgements}

This paper is based on observations made with ESO Telescopes at the Paranal Observatory under program IDs 081.D--0857 and 097.D--0241.  We want to thank the referee, Prof. R. B. C. Henry, for a prompt evaluation of the paper. JG-R and CM would like to thank Gra\.zyna Stasi\'nska and Ana Monreal-Ibero for valuable discussions. We also thank \'Angel R. L\'opez-S\'anchez for his help in the preparation of Figure~\ref{fig:extraction}. JG-R acknowledges support from the Severo Ochoa excellence programs SEV--2015--0548 and CEX2019--000920--S, from the State Research Agency (AEI) of the Spanish Ministry of Science, Innovation and Universities (MCIU) and the European Regional Development Fund (FEDER) under grant AYA2017--83383--P and from the Canarian Agency for Research, Innovation and Information Society (ACIISI), of the Canary Islands Government, and the European Regional Development Fund (ERDF), under grant with reference ProID2021010074. JG-R, DJ and RC acknowledge support under grant P/308614 financed by funds transferred from the Spanish Ministry of Science, Innovation and Universities, charged to the General State Budgets and with funds transferred from the General Budgets of the Autonomous Community of the Canary Islands by the MCIU. CM acknowledges support from grant UNAM / PAPIIT -- IN101220 and from a visitor grant to the IAC under the Severo Ochoa excellence program CEX2019--000920--S. DJ acknowledges support from the Erasmus+ program of the European Union under grant number 2020--1--CZ01--KA203--078200. HM acknowledges the use of the computing facilities available at the Laboratory of Computational Astrophysics of the Universidade Federal de Itajub\'{a} (LAC--UNIFEI). The LAC--UNIFEI is maintained with grants from CAPES, CNPq and FAPEMIG. In particular, grants APQ--01305--17 and CNPq 436117/2018--5. This research made use of Astropy\footnote{http://www.astropy.org}, a community-developed core {\sc python} package for Astronomy \citep{astropy:2013, astropy:2018}.
The paper has been edited using the Overleaf facility.







\begin{thebibliography}{}
\expandafter\ifx\csname natexlab\endcsname\relax\def\natexlab#1{#1}\fi

\bibitem[{{Ali} \& {Dopita}(2017)}]{alidopita17}
{Ali}, A., \& {Dopita}, M.~A. 2017, \pasa, 34, e036

\bibitem[{{Ali} \& {Dopita}(2019)}]{alidopita19}
---. 2019, \mnras, 484, 3251

\bibitem[{{Ali} {et~al.}(2016){Ali}, {Dopita}, {Basurah}, {Amer}, {Alsulami},
  \& {Alruhaili}}]{alietal16}
{Ali}, A., {Dopita}, M.~A., {Basurah}, H.~M., {et~al.} 2016, \mnras, 462, 1393

\bibitem[{{Astropy Collaboration} {et~al.}(2013){Astropy Collaboration},
  {Robitaille}, {Tollerud}, {Greenfield}, {Droettboom}, {Bray}, {Aldcroft},
  {Davis}, {Ginsburg}, {Price-Whelan}, {Kerzendorf}, {Conley}, {Crighton},
  {Barbary}, {Muna}, {Ferguson}, {Grollier}, {Parikh}, {Nair}, {Unther},
  {Deil}, {Woillez}, {Conseil}, {Kramer}, {Turner}, {Singer}, {Fox}, {Weaver},
  {Zabalza}, {Edwards}, {Azalee Bostroem}, {Burke}, {Casey}, {Crawford},
  {Dencheva}, {Ely}, {Jenness}, {Labrie}, {Lim}, {Pierfederici}, {Pontzen},
  {Ptak}, {Refsdal}, {Servillat}, \& {Streicher}}]{astropy:2013}
{Astropy Collaboration}, {Robitaille}, T.~P., {Tollerud}, E.~J., {et~al.} 2013,
  \aap, 558, A33

\bibitem[{{Astropy Collaboration} {et~al.}(2018){Astropy Collaboration},
  {Price-Whelan}, {Sip{\H{o}}cz}, {G{\"u}nther}, {Lim}, {Crawford}, {Conseil},
  {Shupe}, {Craig}, {Dencheva}, {Ginsburg}, {VanderPlas}, {Bradley},
  {P{\'e}rez-Su{\'a}rez}, {de Val-Borro}, {Aldcroft}, {Cruz}, {Robitaille},
  {Tollerud}, {Ardelean}, {Babej}, {Bach}, {Bachetti}, {Bakanov}, {Bamford},
  {Barentsen}, {Barmby}, {Baumbach}, {Berry}, {Biscani}, {Boquien}, {Bostroem},
  {Bouma}, {Brammer}, {Bray}, {Breytenbach}, {Buddelmeijer}, {Burke},
  {Calderone}, {Cano Rodr{\'\i}guez}, {Cara}, {Cardoso}, {Cheedella}, {Copin},
  {Corrales}, {Crichton}, {D'Avella}, {Deil}, {Depagne}, {Dietrich}, {Donath},
  {Droettboom}, {Earl}, {Erben}, {Fabbro}, {Ferreira}, {Finethy}, {Fox},
  {Garrison}, {Gibbons}, {Goldstein}, {Gommers}, {Greco}, {Greenfield},
  {Groener}, {Grollier}, {Hagen}, {Hirst}, {Homeier}, {Horton}, {Hosseinzadeh},
  {Hu}, {Hunkeler}, {Ivezi{\'c}}, {Jain}, {Jenness}, {Kanarek}, {Kendrew},
  {Kern}, {Kerzendorf}, {Khvalko}, {King}, {Kirkby}, {Kulkarni}, {Kumar},
  {Lee}, {Lenz}, {Littlefair}, {Ma}, {Macleod}, {Mastropietro}, {McCully},
  {Montagnac}, {Morris}, {Mueller}, {Mumford}, {Muna}, {Murphy}, {Nelson},
  {Nguyen}, {Ninan}, {N{\"o}the}, {Ogaz}, {Oh}, {Parejko}, {Parley}, {Pascual},
  {Patil}, {Patil}, {Plunkett}, {Prochaska}, {Rastogi}, {Reddy Janga},
  {Sabater}, {Sakurikar}, {Seifert}, {Sherbert}, {Sherwood-Taylor}, {Shih},
  {Sick}, {Silbiger}, {Singanamalla}, {Singer}, {Sladen}, {Sooley},
  {Sornarajah}, {Streicher}, {Teuben}, {Thomas}, {Tremblay}, {Turner},
  {Terr{\'o}n}, {van Kerkwijk}, {de la Vega}, {Watkins}, {Weaver}, {Whitmore},
  {Woillez}, {Zabalza}, \& {Astropy Contributors}}]{astropy:2018}
{Astropy Collaboration}, {Price-Whelan}, A.~M., {Sip{\H{o}}cz}, B.~M., {et~al.}
  2018, \aj, 156, 123

\bibitem[{{Bacon} {et~al.}(2010){Bacon}, {Accardo}, {Adjali}, {Anwand},
  {Bauer}, {Biswas}, {Blaizot}, {Boudon}, {Brau-Nogue}, {Brinchmann},
  {Caillier}, {Capoani}, {Carollo}, {Contini}, {Couderc}, {Daguis{\'e}},
  {Deiries}, {Delabre}, {Dreizler}, {Dubois}, {Dupieux}, {Dupuy}, {Emsellem},
  {Fechner}, {Fleischmann}, {Fran{\c{c}}ois}, {Gallou}, {Gharsa}, {Glindemann},
  {Gojak}, {Guiderdoni}, {Hansali}, {Hahn}, {Jarno}, {Kelz}, {Koehler},
  {Kosmalski}, {Laurent}, {Le Floch}, {Lilly}, {Lizon}, {Loupias}, {Manescau},
  {Monstein}, {Nicklas}, {Olaya}, {Pares}, {Pasquini}, {P{\'e}contal-Rousset},
  {Pell{\'o}}, {Petit}, {Popow}, {Reiss}, {Remillieux}, {Renault}, {Roth},
  {Rupprecht}, {Serre}, {Schaye}, {Soucail}, {Steinmetz}, {Streicher}, {Stuik},
  {Valentin}, {Vernet}, {Weilbacher}, {Wisotzki}, \& {Yerle}}]{baconetal10}
{Bacon}, R., {Accardo}, M., {Adjali}, L., {et~al.} 2010, in Society of
  Photo-Optical Instrumentation Engineers (SPIE) Conference Series, Vol. 7735,
  Ground-based and Airborne Instrumentation for Astronomy III, ed. I.~S.
  {McLean}, S.~K. {Ramsay}, \& H.~{Takami}, 773508

\bibitem[{{Basurah} {et~al.}(2016){Basurah}, {Ali}, {Dopita}, {Alsulami},
  {Amer}, \& {Alruhaili}}]{basurahetal16}
{Basurah}, H.~M., {Ali}, A., {Dopita}, M.~A., {et~al.} 2016, \mnras, 458, 2694

\bibitem[{{Bhatia} \& {Kastner}(1995)}]{bhatiaetal95}
{Bhatia}, A.~K., \& {Kastner}, S.~O. 1995, \apjs, 96, 325

\bibitem[{{Boffin} \& {Jones}(2019)}]{boffinjones19}
{Boffin}, H. M.~J., \& {Jones}, D. 2019, {The Importance of Binaries in the
  Formation and Evolution of Planetary Nebulae} (Springer),
  doi:10.1007/978-3-030-25059-1

\bibitem[{{Butler} \& {Zeippen}(1989)}]{butlerzeippen89}
{Butler}, K., \& {Zeippen}, C.~J. 1989, \aap, 208, 337

\bibitem[{Chen \& Guestrin(2016)}]{chenguestrin16}
Chen, T., \& Guestrin, C. 2016, in Proceedings of the 22nd ACM SIGKDD
  International Conference on Knowledge Discovery and Data Mining, KDD '16 (New
  York, NY, USA: ACM), 785--794

\bibitem[{{Climent}(2016)}]{climent16}
{Climent}, J.~B. 2016, Master's thesis, La Laguna University

\bibitem[{{Corradi} {et~al.}(2015){Corradi}, {Garc{\'\i}a-Rojas}, {Jones}, \&
  {Rodr{\'\i}guez-Gil}}]{corradietal15}
{Corradi}, R. L.~M., {Garc{\'\i}a-Rojas}, J., {Jones}, D., \&
  {Rodr{\'\i}guez-Gil}, P. 2015, ApJ, 803, 99

\bibitem[{{Danehkar} {et~al.}(2013){Danehkar}, {Parker}, \&
  {Ercolano}}]{danehkaretal13}
{Danehkar}, A., {Parker}, Q.~A., \& {Ercolano}, B. 2013, \mnras, 434, 1513

\bibitem[{{Danehkar} {et~al.}(2016){Danehkar}, {Parker}, \&
  {Steffen}}]{danehkaretal16}
{Danehkar}, A., {Parker}, Q.~A., \& {Steffen}, W. 2016, \aj, 151, 38

\bibitem[{{Danehkar} {et~al.}(2014){Danehkar}, {Todt}, {Ercolano}, \&
  {Kniazev}}]{danehkaretal14}
{Danehkar}, A., {Todt}, H., {Ercolano}, B., \& {Kniazev}, A.~Y. 2014, \mnras,
  439, 3605

\bibitem[{{Davey} {et~al.}(2000){Davey}, {Storey}, \&
  {Kisielius}}]{daveyetal00}
{Davey}, A.~R., {Storey}, P.~J., \& {Kisielius}, R. 2000, \aaps, 142, 85

\bibitem[{{Delgado-Inglada} {et~al.}(2014){Delgado-Inglada}, {Morisset}, \&
  {Stasi{\'n}ska}}]{delgadoingladaetal14}
{Delgado-Inglada}, G., {Morisset}, C., \& {Stasi{\'n}ska}, G. 2014, MNRAS, 440,
  536

\bibitem[{{Dopita} {et~al.}(2018){Dopita}, {Ali}, {Karakas}, {Goldman}, {Amer},
  \& {Sutherland}}]{dopitaetal18}
{Dopita}, M.~A., {Ali}, A., {Karakas}, A.~I., {et~al.} 2018, \mnras, 475, 424

\bibitem[{{Dopita} {et~al.}(2017){Dopita}, {Ali}, {Sutherland}, {Nicholls}, \&
  {Amer}}]{dopitaetal17}
{Dopita}, M.~A., {Ali}, A., {Sutherland}, R.~S., {Nicholls}, D.~C., \& {Amer},
  M.~A. 2017, \mnras, 470, 839

\bibitem[{{Escalante} {et~al.}(2012){Escalante}, {Morisset}, \&
  {Georgiev}}]{escalanteetal12}
{Escalante}, V., {Morisset}, C., \& {Georgiev}, L. 2012, \mnras, 426, 2318

\bibitem[{{Fang} \& {Liu}(2011)}]{fangliu11}
{Fang}, X., \& {Liu}, X.~W. 2011, MNRAS, 415, 181

\bibitem[{{Fang} {et~al.}(2011){Fang}, {Storey}, \& {Liu}}]{fangetal11}
{Fang}, X., {Storey}, P.~J., \& {Liu}, X.~W. 2011, \aap, 530, A18

\bibitem[{{Fang} {et~al.}(2013){Fang}, {Storey}, \& {Liu}}]{fangetal13}
---. 2013, \aap, 550, C2

\bibitem[{{Ferland} {et~al.}(2017){Ferland}, {Chatzikos}, {Guzm{\'a}n},
  {Lykins}, {van Hoof}, {Williams}, {Abel}, {Badnell}, {Keenan}, {Porter}, \&
  {Stancil}}]{ferlandetal17}
{Ferland}, G.~J., {Chatzikos}, M., {Guzm{\'a}n}, F., {et~al.} 2017, \rmxaa, 53,
  385

\bibitem[{{Fitzpatrick}(1999)}]{fitzpatrick99}
{Fitzpatrick}, E.~L. 1999, \pasp, 111, 63

\bibitem[{{Froese Fischer} \& {Tachiev}(2004)}]{frosefischertachiev04}
{Froese Fischer}, C., \& {Tachiev}, G. 2004, Atomic Data and Nuclear Data
  Tables, 87, 1

\bibitem[{{Froese Fischer} {et~al.}(2006){Froese Fischer}, {Tachiev}, \&
  {Irimia}}]{froesefischeretal06}
{Froese Fischer}, C., {Tachiev}, G., \& {Irimia}, A. 2006, Atom. Data Nucl.
  Data Tables, 92, 607

\bibitem[{{Galavis} {et~al.}(1995){Galavis}, {Mendoza}, \&
  {Zeippen}}]{galavisetal95}
{Galavis}, M.~E., {Mendoza}, C., \& {Zeippen}, C.~J. 1995, \aaps, 111, 347

\bibitem[{{Garc{\'\i}a-Rojas}(2020)}]{garciarojas20}
{Garc{\'\i}a-Rojas}, J. 2020, {Physical Conditions and Chemical Abundances in
  Photoionized Nebulae from Optical Spectra} (Springer), 89--121

\bibitem[{{Garc{\'\i}a-Rojas} {et~al.}(2017){Garc{\'\i}a-Rojas}, {Corradi},
  {Boffin}, {Monteiro}, {Jones}, {Wesson}, {Cabrera-Lavers}, \&
  {Rodr{\'\i}guez-Gil}}]{garciarojasetal17}
{Garc{\'\i}a-Rojas}, J., {Corradi}, R. L.~M., {Boffin}, H. M.~J., {et~al.}
  2017, in IAU Symposium, Vol. 323, Planetary Nebulae: Multi-Wavelength Probes
  of Stellar and Galactic Evolution, ed. X.~{Liu}, L.~{Stanghellini}, \&
  A.~{Karakas}, 65--69

\bibitem[{{Garc{\'\i}a-Rojas} {et~al.}(2016){Garc{\'\i}a-Rojas}, {Corradi},
  {Monteiro}, {Jones}, {Rodr{\'\i}guez-Gil}, \&
  {Cabrera-Lavers}}]{garciarojasetal16}
{Garc{\'\i}a-Rojas}, J., {Corradi}, R. L.~M., {Monteiro}, H., {et~al.} 2016,
  ApJL, 824, L27

\bibitem[{{Garc{\'\i}a-Rojas} \& {Esteban}(2007)}]{garciarojasesteban07}
{Garc{\'\i}a-Rojas}, J., \& {Esteban}, C. 2007, ApJ, 670, 457

\bibitem[{{Garc{\'\i}a-Rojas} {et~al.}(2019){Garc{\'\i}a-Rojas}, {Wesson},
  {Boffin}, {Jones}, {Corradi}, {Esteban}, \&
  {Rodr{\'\i}guez-Gil}}]{garciarojasetal19}
{Garc{\'\i}a-Rojas}, J., {Wesson}, R., {Boffin}, H.~M.~J., {et~al.} 2019, arXiv
  e-prints, arXiv:1904.06763

\bibitem[{{Garnett} \& {Dinerstein}(2001)}]{garnettdinerstein01}
{Garnett}, D.~R., \& {Dinerstein}, H.~L. 2001, ApJ, 558, 145

\bibitem[{{Ginsburg} \& {Mirocha}(2011)}]{ginsburgmirocha11}
{Ginsburg}, A., \& {Mirocha}, J. 2011, {PySpecKit: Python Spectroscopic
  Toolkit}, , , ascl:1109.001

\bibitem[{{G{\'o}mez-Llanos} \& {Morisset}(2020)}]{gomezllanosmorisset20}
{G{\'o}mez-Llanos}, V., \& {Morisset}, C. 2020, \mnras, 497, 3363

\bibitem[{{G{\'o}mez-Llanos} {et~al.}(2020){G{\'o}mez-Llanos}, {Morisset},
  {Garc{\'\i}a-Rojas}, {Jones}, {Wesson}, {Corradi}, \&
  {Boffin}}]{gomezllanosetal20}
{G{\'o}mez-Llanos}, V., {Morisset}, C., {Garc{\'\i}a-Rojas}, J., {et~al.} 2020,
  \mnras, 498, L82

\bibitem[{{Guerrero} \& {Miranda}(2012)}]{guerreromiranda12}
{Guerrero}, M.~A., \& {Miranda}, L.~F. 2012, \aap, 539, A47

\bibitem[{{Guerrero} {et~al.}(2013){Guerrero}, {Toal{\'a}}, {Medina},
  {Luridiana}, {Miranda}, {Riera}, \& {Vel{\'a}zquez}}]{guerreroetal13}
{Guerrero}, M.~A., {Toal{\'a}}, J.~A., {Medina}, J.~J., {et~al.} 2013, \aap,
  557, A121

\bibitem[{{Hillwig} {et~al.}(2016){Hillwig}, {Bond}, {Frew}, {Schaub}, \&
  {Bodman}}]{hillwig16}
{Hillwig}, T.~C., {Bond}, H.~E., {Frew}, D.~J., {Schaub}, S.~C., \& {Bodman},
  E. H.~L. 2016, \aj, 152, 34

\bibitem[{{Igoshev} {et~al.}(2020){Igoshev}, {Perets}, \&
  {Michaely}}]{igoshev20}
{Igoshev}, A.~P., {Perets}, H.~B., \& {Michaely}, E. 2020, \mnras, 494, 1448

\bibitem[{{Jones} {et~al.}(2019){Jones}, {Boffin}, {Sowicka}, {Miszalski},
  {Rodr{\'\i}guez-Gil}, {Santander-Garc{\'\i}a}, \& {Corradi}}]{jones19}
{Jones}, D., {Boffin}, H. M.~J., {Sowicka}, P., {et~al.} 2019, \mnras, 482, L75

\bibitem[{{Jones} {et~al.}(2012){Jones}, {Mitchell}, {Lloyd}, {Pollacco},
  {O'Brien}, {Meaburn}, \& {Vaytet}}]{jones12}
{Jones}, D., {Mitchell}, D.~L., {Lloyd}, M., {et~al.} 2012, \mnras, 420, 2271

\bibitem[{{Jones} {et~al.}(2016){Jones}, {Wesson}, {Garc{\'\i}a-Rojas},
  {Corradi}, \& {Boffin}}]{jonesetal16}
{Jones}, D., {Wesson}, R., {Garc{\'\i}a-Rojas}, J., {Corradi}, R.~L.~M., \&
  {Boffin}, H.~M.~J. 2016, MNRAS, 455, 3263

\bibitem[{{Kaufman} \& {Sugar}(1986)}]{kaufmansugar86}
{Kaufman}, V., \& {Sugar}, J. 1986, Journal of Physical and Chemical Reference
  Data, 15, 321

\bibitem[{{Kingsburgh} \& {Barlow}(1994)}]{kingsburghbarlow94}
{Kingsburgh}, R.~L., \& {Barlow}, M.~J. 1994, MNRAS, 271, 257

\bibitem[{{Kisielius} {et~al.}(2009){Kisielius}, {Storey}, {Ferland}, \&
  {Keenan}}]{kisieliusetal09}
{Kisielius}, R., {Storey}, P.~J., {Ferland}, G.~J., \& {Keenan}, F.~P. 2009,
  \mnras, 397, 903

\bibitem[{{Kuruwita} {et~al.}(2016){Kuruwita}, {Staff}, \& {De
  Marco}}]{kuruwita16}
{Kuruwita}, R.~L., {Staff}, J., \& {De Marco}, O. 2016, \mnras, 461, 486

\bibitem[{{Liu} {et~al.}(2006){Liu}, {Barlow}, {Zhang}, {Bastin}, \&
  {Storey}}]{liuetal06}
{Liu}, X.~W., {Barlow}, M.~J., {Zhang}, Y., {Bastin}, R.~J., \& {Storey}, P.~J.
  2006, MNRAS, 368, 1959

\bibitem[{{Liu} {et~al.}(2001){Liu}, {Luo}, {Barlow}, {Danziger}, \&
  {Storey}}]{liuetal01}
{Liu}, X.~W., {Luo}, S.~G., {Barlow}, M.~J., {Danziger}, I.~J., \& {Storey},
  P.~J. 2001, MNRAS, 327, 141

\bibitem[{{Liu} {et~al.}(1995){Liu}, {Storey}, {Barlow}, \&
  {Clegg}}]{liuetal95}
{Liu}, X.~W., {Storey}, P.~J., {Barlow}, M.~J., \& {Clegg}, R.~E.~S. 1995,
  \mnras, 272, 369

\bibitem[{{Liu} {et~al.}(2000){Liu}, {Storey}, {Barlow}, {Danziger}, {Cohen},
  \& {Bryce}}]{liuetal00}
{Liu}, X.~W., {Storey}, P.~J., {Barlow}, M.~J., {et~al.} 2000, MNRAS, 312, 585

\bibitem[{{Lodders}(2019)}]{lodders19}
{Lodders}, K. 2019, arXiv e-prints, arXiv:1912.00844

\bibitem[{{Luridiana} {et~al.}(2015){Luridiana}, {Morisset}, \&
  {Shaw}}]{luridianaetal15}
{Luridiana}, V., {Morisset}, C., \& {Shaw}, R.~A. 2015, A\&A, 573, A42

\bibitem[{{Markwardt}(2009)}]{markwardt2009}
{Markwardt}, C.~B. 2009, in Astronomical Society of the Pacific Conference
  Series, Vol. 411, Astronomical Data Analysis Software and Systems XVIII, ed.
  D.~A. {Bohlender}, D.~{Durand}, \& P.~{Dowler}, 251

\bibitem[{{McNabb} {et~al.}(2016){McNabb}, {Fang}, \& {Liu}}]{mcnabbetal16}
{McNabb}, I.~A., {Fang}, X., \& {Liu}, X.~W. 2016, \mnras, 461, 2818

\bibitem[{{M{\'e}ndez-Delgado} {et~al.}(2021){M{\'e}ndez-Delgado}, {Esteban},
  {Garc{\'\i}a-Rojas}, {Henney}, {Mesa-Delgado}, \&
  {Arellano-C{\'o}rdova}}]{mendezdelgadoetal21}
{M{\'e}ndez-Delgado}, J.~E., {Esteban}, C., {Garc{\'\i}a-Rojas}, J., {et~al.}
  2021, \mnras, 502, 1703

\bibitem[{{Mendoza} \& {Zeippen}(1982)}]{mendozazeippen82a}
{Mendoza}, C., \& {Zeippen}, C.~J. 1982, \mnras, 199, 1025

\bibitem[{{Mesa-Delgado} {et~al.}(2009{\natexlab{a}}){Mesa-Delgado}, {Esteban},
  {Garc{\'\i}a-Rojas}, {Luridiana}, {Bautista}, {Rodr{\'\i}guez},
  {L{\'o}pez-Mart{\'\i}n}, \& {Peimbert}}]{mesadelgadoetal09a}
{Mesa-Delgado}, A., {Esteban}, C., {Garc{\'\i}a-Rojas}, J., {et~al.}
  2009{\natexlab{a}}, MNRAS, 395, 855

\bibitem[{{Mesa-Delgado} {et~al.}(2009{\natexlab{b}}){Mesa-Delgado},
  {L{\'o}pez-Mart{\'\i}n}, {Esteban}, {Garc{\'\i}a-Rojas}, \&
  {Luridiana}}]{mesadelgadoetal09b}
{Mesa-Delgado}, A., {L{\'o}pez-Mart{\'\i}n}, L., {Esteban}, C.,
  {Garc{\'\i}a-Rojas}, J., \& {Luridiana}, V. 2009{\natexlab{b}}, \mnras, 394,
  693

\bibitem[{{Mesa-Delgado} {et~al.}(2012){Mesa-Delgado},
  {N{\'u}{\~n}ez-D{\'\i}az}, {Esteban}, {Garc{\'\i}a-Rojas}, {Flores-Fajardo},
  {L{\'o}pez-Mart{\'\i}n}, {Tsamis}, \& {Henney}}]{mesadelgadoetal12}
{Mesa-Delgado}, A., {N{\'u}{\~n}ez-D{\'\i}az}, M., {Esteban}, C., {et~al.}
  2012, MNRAS, 426, 614

\bibitem[{{Monreal-Ibero} {et~al.}(2005){Monreal-Ibero}, {Roth},
  {Sch{\"o}nberner}, {Steffen}, \& {B{\"o}hm}}]{monrealiberoetal05}
{Monreal-Ibero}, A., {Roth}, M.~M., {Sch{\"o}nberner}, D., {Steffen}, M., \&
  {B{\"o}hm}, P. 2005, \apjl, 628, L139

\bibitem[{{Monreal-Ibero} \& {Walsh}(2020)}]{monrealiberowalsh20}
{Monreal-Ibero}, A., \& {Walsh}, J.~R. 2020, \aap, 634, A47

\bibitem[{{Monteiro} {et~al.}(2013){Monteiro}, {Gon{\c{c}}alves},
  {Leal-Ferreira}, \& {Corradi}}]{monteiroetal13}
{Monteiro}, H., {Gon{\c{c}}alves}, D.~R., {Leal-Ferreira}, M.~L., \& {Corradi},
  R.~L.~M. 2013, \aap, 560, A102

\bibitem[{{Morisset} {et~al.}(2015){Morisset}, {Delgado-Inglada}, \&
  {Flores-Fajardo}}]{morissetetal15}
{Morisset}, C., {Delgado-Inglada}, G., \& {Flores-Fajardo}, N. 2015, \rmxaa,
  51, 103

\bibitem[{{Munoz Burgos} {et~al.}(2009){Munoz Burgos}, {Loch}, {Ballance}, \&
  {Boivin}}]{munozburgosetal09}
{Munoz Burgos}, J.~M., {Loch}, S.~D., {Ballance}, C.~P., \& {Boivin}, R.~F.
  2009, \aap, 500, 1253

\bibitem[{{Osterbrock} \& {Ferland}(2006)}]{osterbrockferland06}
{Osterbrock}, D.~E., \& {Ferland}, G.~J. 2006, {Astrophysics of gaseous nebulae
  and active galactic nuclei}

\bibitem[{{Ott}(2012)}]{ott12}
{Ott}, T. 2012, {QFitsView: FITS file viewer}, , , ascl:1210.019

\bibitem[{Pedregosa {et~al.}(2011)Pedregosa, Varoquaux, Gramfort, Michel,
  Thirion, Grisel, Blondel, Prettenhofer, Weiss, Dubourg, Vanderplas, Passos,
  Cournapeau, Brucher, Perrot, \& Duchesnay}]{pedregosaetal11}
Pedregosa, F., Varoquaux, G., Gramfort, A., {et~al.} 2011, Journal of Machine
  Learning Research, 12, 2825

\bibitem[{{Peimbert} \& {Costero}(1969)}]{peimbertcostero69}
{Peimbert}, M., \& {Costero}, R. 1969, Boletin de los Observatorios
  Tonantzintla y Tacubaya, 5, 3

\bibitem[{{Pequignot} {et~al.}(1991){Pequignot}, {Petitjean}, \&
  {Boisson}}]{pequignotetal91}
{Pequignot}, D., {Petitjean}, P., \& {Boisson}, C. 1991, \aap, 251, 680

\bibitem[{{Porter} {et~al.}(2012){Porter}, {Ferland}, {Storey}, \&
  {Detisch}}]{porteretal12}
{Porter}, R.~L., {Ferland}, G.~J., {Storey}, P.~J., \& {Detisch}, M.~J. 2012,
  \mnras, 425, L28

\bibitem[{{Porter} {et~al.}(2013){Porter}, {Ferland}, {Storey}, \&
  {Detisch}}]{porteretal13}
---. 2013, \mnras, 433, L89

\bibitem[{{Ramsbottom} \& {Bell}(1997)}]{ramsbottombell97}
{Ramsbottom}, C.~A., \& {Bell}, K.~L. 1997, Atomic Data and Nuclear Data
  Tables, 66, 65

\bibitem[{{Reichardt} {et~al.}(2019){Reichardt}, {De Marco}, {Iaconi}, {Tout},
  \& {Price}}]{reichardt19}
{Reichardt}, T.~A., {De Marco}, O., {Iaconi}, R., {Tout}, C.~A., \& {Price},
  D.~J. 2019, \mnras, 484, 631

\bibitem[{{Richer} {et~al.}(2017){Richer}, {Su{\'a}rez}, {L{\'o}pez}, \&
  {Garc{\'\i}a D{\'\i}az}}]{richeretal17}
{Richer}, M.~G., {Su{\'a}rez}, G., {L{\'o}pez}, J.~A., \& {Garc{\'\i}a
  D{\'\i}az}, M.~T. 2017, \aj, 153, 140

\bibitem[{{Rubin}(1986)}]{rubinetal86}
{Rubin}, R.~H. 1986, \apj, 309, 334

\bibitem[{{Rynkun} {et~al.}(2019){Rynkun}, {Gaigalas}, \&
  {J{\"o}nsson}}]{rynkunetal19}
{Rynkun}, P., {Gaigalas}, G., \& {J{\"o}nsson}, P. 2019, Astron. Astrophys.,
  623, A155

\bibitem[{{Schwarz} {et~al.}(1992){Schwarz}, {Corradi}, \&
  {Melnick}}]{schwarzetal92}
{Schwarz}, H.~E., {Corradi}, R.~L.~M., \& {Melnick}, J. 1992, \aaps, 96, 23

\bibitem[{{Storey} \& {Hummer}(1995)}]{storeyhummer95}
{Storey}, P.~J., \& {Hummer}, D.~G. 1995, \mnras, 272, 41

\bibitem[{{Storey} \& {Sochi}(2014)}]{storeysochi14}
{Storey}, P.~J., \& {Sochi}, T. 2014, \mnras, 440, 2581

\bibitem[{{Storey} {et~al.}(2017){Storey}, {Sochi}, \& {Bastin}}]{storeyetal17}
{Storey}, P.~J., {Sochi}, T., \& {Bastin}, R. 2017, \mnras, 470, 379

\bibitem[{{Storey} \& {Zeippen}(2000)}]{storeyzeippen00}
{Storey}, P.~J., \& {Zeippen}, C.~J. 2000, \mnras, 312, 813

\bibitem[{{Tayal}(2011)}]{tayal11}
{Tayal}, S.~S. 2011, ApJS, 195, 12

\bibitem[{{Tayal} \& {Gupta}(1999)}]{tayalgupta99}
{Tayal}, S.~S., \& {Gupta}, G.~P. 1999, \apj, 526, 544

\bibitem[{{Tayal} \& {Zatsarinny}(2010)}]{tayal10}
{Tayal}, S.~S., \& {Zatsarinny}, O. 2010, ApJS, 188, 32

\bibitem[{{Tsamis} \& {P{\'e}quignot}(2005)}]{tsamispequignot05}
{Tsamis}, Y.~G., \& {P{\'e}quignot}, D. 2005, MNRAS, 364, 687

\bibitem[{{Tsamis} {et~al.}(2008){Tsamis}, {Walsh}, {P{\'e}quignot}, {Barlow},
  {Danziger}, \& {Liu}}]{tsamisetal08}
{Tsamis}, Y.~G., {Walsh}, J.~R., {P{\'e}quignot}, D., {et~al.} 2008, \mnras,
  386, 22

\bibitem[{{Tsamis} {et~al.}(2011){Tsamis}, {Walsh}, {V{\'\i}lchez}, \&
  {P{\'e}quignot}}]{tsamisetal11}
{Tsamis}, Y.~G., {Walsh}, J.~R., {V{\'\i}lchez}, J.~M., \& {P{\'e}quignot}, D.
  2011, MNRAS, 412, 1367

\bibitem[{{Tyndall} {et~al.}(2012){Tyndall}, {Jones}, {Lloyd}, {O'Brien}, \&
  {Pollacco}}]{tyndall12}
{Tyndall}, A.~A., {Jones}, D., {Lloyd}, M., {O'Brien}, T.~J., \& {Pollacco}, D.
  2012, \mnras, 422, 1804

\bibitem[{{Ueta} \& {Otsuka}(2021)}]{2021Ueta_arXi}
{Ueta}, T., \& {Otsuka}, M. 2021, arXiv e-prints, arXiv:2108.11007

\bibitem[{{Walsh} \& {Monreal-Ibero}(2020)}]{walshmonrealibero20}
{Walsh}, J.~R., \& {Monreal-Ibero}, A. 2020, Galaxies, 8, 31

\bibitem[{{Walsh} {et~al.}(2016){Walsh}, {Monreal-Ibero}, {Barlow}, {Ueta},
  {Wesson}, \& {Zijlstra}}]{walshetal16}
{Walsh}, J.~R., {Monreal-Ibero}, A., {Barlow}, M.~J., {et~al.} 2016, \aap, 588,
  A106

\bibitem[{{Walsh} {et~al.}(2018){Walsh}, {Monreal-Ibero}, {Barlow}, {Ueta},
  {Wesson}, {Zijlstra}, {Kimeswenger}, {Leal-Ferreira}, \&
  {Otsuka}}]{walshetal18}
---. 2018, \aap, 620, A169

\bibitem[{{Webbink}(2008)}]{webbink08}
{Webbink}, R.~F. 2008, {Common Envelope Evolution Redux}, ed. E.~F. {Milone},
  D.~A. {Leahy}, \& D.~W. {Hobill}, Vol. 352 (Springer), 233

\bibitem[{{Weilbacher} {et~al.}(2014){Weilbacher}, {Streicher}, {Urrutia},
  {P{\'e}contal-Rousset}, {Jarno}, \& {Bacon}}]{weilbacheretal14}
{Weilbacher}, P.~M., {Streicher}, O., {Urrutia}, T., {et~al.} 2014, in
  Astronomical Society of the Pacific Conference Series, Vol. 485, Astronomical
  Data Analysis Software and Systems XXIII, ed. N.~{Manset} \& P.~{Forshay},
  451

\bibitem[{{Weilbacher} {et~al.}(2020){Weilbacher}, {Palsa}, {Streicher},
  {Bacon}, {Urrutia}, {Wisotzki}, {Conseil}, {Husemann}, {Jarno}, {Kelz},
  {P{\'e}contal-Rousset}, {Richard}, {Roth}, {Selman}, \&
  {Vernet}}]{weilbacheretal20}
{Weilbacher}, P.~M., {Palsa}, R., {Streicher}, O., {et~al.} 2020, \aap, 641,
  A28

\bibitem[{{Wesson}(2016)}]{wesson16}
{Wesson}, R. 2016, \mnras, 456, 3774

\bibitem[{{Wesson} {et~al.}(2008{\natexlab{a}}){Wesson}, {Barlow}, {Liu},
  {Storey}, {Ercolano}, \& {De Marco}}]{wessonetal08}
{Wesson}, R., {Barlow}, M.~J., {Liu}, X.~W., {et~al.} 2008{\natexlab{a}},
  \mnras, 383, 1639

\bibitem[{{Wesson} {et~al.}(2018){Wesson}, {Jones}, {Garc{\'\i}a-Rojas},
  {Boffin}, \& {Corradi}}]{wessonetal18}
{Wesson}, R., {Jones}, D., {Garc{\'\i}a-Rojas}, J., {Boffin}, H.~M.~J., \&
  {Corradi}, R.~L.~M. 2018, MNRAS, 480, 4589

\bibitem[{{Wesson} {et~al.}(2003){Wesson}, {Liu}, \& {Barlow}}]{wessonetal03}
{Wesson}, R., {Liu}, X.~W., \& {Barlow}, M.~J. 2003, \mnras, 340, 253

\bibitem[{{Wesson} {et~al.}(2008{\natexlab{b}}){Wesson}, {Barlow}, {Corradi},
  {Drew}, {Groot}, {Knigge}, {Steeghs}, {Gaensicke}, {Napiwotzki},
  {Rodriguez-Gil}, {Zijlstra}, {Bode}, {Drake}, {Frew}, {Gonzalez-Solares},
  {Greimel}, {Irwin}, {Morales-Rueda}, {Nelemans}, {Parker}, {Sale},
  {Sokoloski}, {Somero}, {Uthas}, {Walton}, {Warner}, {Watson}, \&
  {Wright}}]{wessonetal08novavul}
{Wesson}, R., {Barlow}, M.~J., {Corradi}, R.~L.~M., {et~al.}
  2008{\natexlab{b}}, \apjl, 688, L21

\bibitem[{{Wiese} {et~al.}(1996){Wiese}, {Fuhr}, \& {Deters}}]{wieseetal96}
{Wiese}, W.~L., {Fuhr}, J.~R., \& {Deters}, T.~M. 1996, {Atomic transition
  probabilities of carbon, nitrogen, and oxygen : a critical data compilation}
  ({American Chemical Society})

\bibitem[{{Wyse}(1942)}]{wyse42}
{Wyse}, A.~B. 1942, ApJ, 95, 356

\bibitem[{{Yuan} {et~al.}(2011){Yuan}, {Liu}, {P{\'e}quignot}, {Rubin},
  {Ercolano}, \& {Zhang}}]{yuanetal11}
{Yuan}, H.~B., {Liu}, X.~W., {P{\'e}quignot}, D., {et~al.} 2011, MNRAS, 411,
  1035

\bibitem[{{Zhang} {et~al.}(2005){Zhang}, {Liu}, {Liu}, \&
  {Rubin}}]{zhangetal05}
{Zhang}, Y., {Liu}, X.~W., {Liu}, Y., \& {Rubin}, R.~H. 2005, \mnras, 358, 457

\end{thebibliography}



\appendix

\section{Line intensities}

\begin{table*}
\caption{Measured (F) and dereddened (I) lines fluxes$^{\rm a}$ in the integrated spectra. \label{tab:line_fluxes}}
\resizebox{\textwidth}{!}{%
\begin{tabular}{lccccccc}
\hline
Line & $\lambda_0$ (\AA) & \multicolumn{2}{c}{NGC\,6778} & \multicolumn{2}{c}{M\,1-42} &\multicolumn{2}{c}{Hf\,2-2}  \\
 &      &   I($\lambda$) &  F($\lambda$) &   I($\lambda$)   &   F($\lambda$)   &   I($\lambda$)   &   F($\lambda$)   \\
\hline
 {\oii} & 4649   &           1.71  $\pm$      0.09   &           1.86  $\pm$      0.15 &           1.75  $\pm$      0.09   &           1.95  $\pm$      0.17 &             2.5  $\pm$        0.1   &             2.7  $\pm$        0.2 \\
       & 4650   &          &          &        &         &    &   \\
{\oii}  &  4662   &           0.43  $\pm$      0.02   &           0.47  $\pm$      0.04 &           0.49  $\pm$      0.02   &           0.54  $\pm$      0.04 &           0.76  $\pm$      0.04   &           0.82  $\pm$      0.06 \\
 {\heii}  &  4686     &             4.7  $\pm$        0.2   &             5.1  $\pm$        0.4 &             8.1  $\pm$        0.4   &             8.8  $\pm$        0.7 &           1.98  $\pm$      0.10   &           2.12  $\pm$      0.17 \\
 {\fariv}  &  4711       &           0.52  $\pm$      0.03   &           0.56  $\pm$      0.05 &             0.7  $\pm$        0.7   &             0.7  $\pm$        0.8 &         0.046  $\pm$    0.004   &         0.048  $\pm$    0.005 \\
{\hei}  &  4713     &           0.93  $\pm$      0.05   &           0.99  $\pm$      0.09 &             1.2  $\pm$        0.6   &             1.3  $\pm$        0.7 &           0.66  $\pm$      0.03   &           0.70  $\pm$      0.05 \\
{\fariv}  &   4740       &           0.61  $\pm$      0.03   &           0.65  $\pm$      0.05 &           0.99  $\pm$      0.05   &           1.05  $\pm$      0.08 &           0.19  $\pm$      0.01   &           0.20  $\pm$      0.02 \\
{\hi}  &  4861       &         100.0  $\pm$        5.0   &         100.0  $\pm$        0.0 &         100.0  $\pm$        5.0   &         100.0  $\pm$        0.0 &         100.0  $\pm$        5.0   &         100.0  $\pm$        0.0 \\
{\foiii}  &  4959         &         172.8  $\pm$        8.6   &         165.8  $\pm$      11.4 &         166.3  $\pm$        8.3   &         158.5  $\pm$      11.9 &           55.2  $\pm$        2.8   &           53.1  $\pm$        3.5 \\
 {\fnitroi}  &   5197       &             3.7  $\pm$        0.2   &             3.2  $\pm$        0.2 &             4.8  $\pm$        0.2   &             4.1  $\pm$        0.3 &           1.23  $\pm$      0.06   &           1.08  $\pm$      0.07 \\
       & 5200   &          &          &        &         &    &   \\
{\cii}  &   5342   &         0.066  $\pm$    0.003   &         0.054  $\pm$    0.004 &         0.086  $\pm$    0.006   &         0.069  $\pm$    0.005 &           0.21  $\pm$      0.01   &           0.18  $\pm$      0.01 \\
{\fcliii}  &   5518       &           0.81  $\pm$      0.04   &           0.63  $\pm$      0.04 &           0.82  $\pm$      0.04   &           0.61  $\pm$      0.04 &           0.28  $\pm$      0.01   &           0.22  $\pm$      0.02 \\
{\fcliii}  &   5538       &           0.69  $\pm$      0.03   &           0.53  $\pm$      0.03 &           0.79  $\pm$      0.04   &           0.58  $\pm$      0.04 &           0.23  $\pm$      0.01   &           0.18  $\pm$      0.01 \\
{\nii}  &    5676   &         0.165  $\pm$    0.008   &         0.122  $\pm$    0.007 &   ---     &      ----     &           0.25  $\pm$      0.01   &           0.19  $\pm$      0.01 \\
{\nii}  &  5679   &           0.88  $\pm$      0.04   &           0.65  $\pm$      0.04 &           1.15  $\pm$      0.06   &           0.81  $\pm$      0.05 &           0.92  $\pm$      0.05   &           0.70  $\pm$      0.04 \\
{\fnii}  &  5755         &             4.0  $\pm$        0.2   &             2.4  $\pm$        0.2$^{\rm b}$  &             4.9  $\pm$        0.2   &             2.7  $\pm$        0.2$^{\rm b}$  &           1.10  $\pm$      0.06   &           0.29  $\pm$      0.05$^{\rm b}$  \\
{\hei}  &   5876     &           33.6  $\pm$        1.7   &           23.5  $\pm$        1.5 &           35.6  $\pm$        1.8   &           23.4  $\pm$        1.3 &           34.2  $\pm$        1.7   &           24.5  $\pm$        1.5 \\
{\foi}  &  6300        &             8.1  $\pm$        0.4   &             5.1  $\pm$        0.3 &           11.3  $\pm$        0.6   &             6.5  $\pm$        0.4 &           0.66  $\pm$      0.03   &           0.43  $\pm$      0.03 \\
 {\fsiii}  &  6312         &           1.86  $\pm$      0.09   &           1.16  $\pm$      0.07 &             2.2  $\pm$        0.1   &             1.3  $\pm$        0.1 &           0.36  $\pm$      0.02   &           0.24  $\pm$      0.02 \\
{\foi}  &   6363         &             2.7  $\pm$        0.1   &             1.7  $\pm$        0.1 &             3.7  $\pm$        0.2   &             2.1  $\pm$        0.1 &         0.199  $\pm$    0.011   &         0.127  $\pm$    0.009 \\
{\cii}  &   6462   &           0.25  $\pm$      0.01   &           0.15  $\pm$      0.01 &           0.33  $\pm$      0.02   &           0.19  $\pm$      0.01 &           0.72  $\pm$      0.04   &           0.45  $\pm$      0.03 \\
{\fnii}  &  6548         &         152.5  $\pm$        7.6   &           90.0  $\pm$        6.2 &         199.9  $\pm$      10.0   &         108.1  $\pm$        5.8 &           19.6  $\pm$        1.0   &           12.0  $\pm$        0.8 \\
{\hi}  &  6563       &         484.6  $\pm$      24.2   &         285.0  $\pm$        0.0 &         543.3  $\pm$      27.2   &         292.8  $\pm$      14.7 &         465.6  $\pm$      23.3   &         285.0  $\pm$        0.0 \\
{\fnii}  &   6584         &         458.6  $\pm$      22.9   &         268.4  $\pm$      19.3 &         596.0  $\pm$      29.8   &         319.4  $\pm$      17.4 &           60.2  $\pm$        3.0   &           36.7  $\pm$        2.8 \\
{\hei}  &   6678     &           11.3  $\pm$        0.6   &             6.5  $\pm$        0.5 &           12.6  $\pm$        0.6   &             6.6  $\pm$        0.4 &           11.2  $\pm$        0.6   &             6.7  $\pm$        0.5 \\
{\fsii}  &   6716         &           36.3  $\pm$        1.8   &           20.6  $\pm$        1.5 &           43.7  $\pm$        2.2   &           22.6  $\pm$        1.2 &             7.9  $\pm$        0.4   &             4.7  $\pm$        0.4 \\
{\fsii}  &  6731         &           39.1  $\pm$        2.0   &           22.1  $\pm$        1.7 &           50.1  $\pm$        2.5   &           25.8  $\pm$        1.5 &             6.9  $\pm$        0.3   &             4.1  $\pm$        0.3 \\
{\farv}  &   7005       &         0.019  $\pm$    0.001   &         0.010  $\pm$    0.001 &         ---   &         --- &         ---   &        --- \\
 {\hei}  &   7065     &             7.2  $\pm$        0.4   &             3.8  $\pm$        0.3 &             9.0  $\pm$        0.5   &             4.3  $\pm$        0.2 &             3.8  $\pm$        0.2   &             2.1  $\pm$        0.2 \\
 {\fariii}  &   7136       &           31.4  $\pm$        1.6   &           16.3  $\pm$        1.3 &           44.9  $\pm$        2.2   &           21.0  $\pm$        1.1 &             9.4  $\pm$        0.5   &             5.1  $\pm$        0.4 \\
{\hei}  &   7281     &           1.60  $\pm$      0.08   &           0.81  $\pm$      0.07 &           1.79  $\pm$      0.09   &           0.81  $\pm$      0.04 &           1.05  $\pm$      0.05   &           0.56  $\pm$      0.05 \\
 {\foii}  &  7318       &             4.5  $\pm$        0.2   &             1.3  $\pm$        0.2$^{\rm b}$  &             5.5  $\pm$        0.3   &             1.5  $\pm$        0.1$^{\rm b}$  &             3.7  $\pm$        0.2   &             0.5  $\pm$        0.2$^{\rm b}$  \\
       & 7319   &          &          &        &         &    &   \\
{\foii}  &  7329       &             3.6  $\pm$        0.2   &             1.0  $\pm$        0.2$^{\rm b}$  &             4.6  $\pm$        0.2   &             1.2  $\pm$        0.1$^{\rm b}$  &             3.0  $\pm$        0.2   &             0.4  $\pm$        0.2$^{\rm b}$  \\
       & 7330   &          &          &        &         &    &   \\
{\fcliv}  &   7531       &           0.24  $\pm$      0.01   &           0.11  $\pm$      0.01 &         0.311  $\pm$    0.016   &         0.134  $\pm$    0.007 &         0.050  $\pm$    0.004   &         0.025  $\pm$    0.003 \\
{\fariii}  &   7751       &             8.1  $\pm$        0.4   &             3.8  $\pm$        0.4 &           11.7  $\pm$        0.6   &             4.8  $\pm$        0.3 &             2.4  $\pm$        0.1   &             1.2  $\pm$        0.1 \\
 {\oi}  &  7772       &           0.60  $\pm$      0.03   &           0.28  $\pm$      0.03 &           0.70  $\pm$      0.09   &           0.29  $\pm$      0.04 &             2.0  $\pm$        2.0   &             1.0  $\pm$        1.0 \\
        & 7774   &          &          &        &         &    &   \\
       & 7775   &          &          &        &         &    &   \\
{\fcliv}  &   8046       &           0.53  $\pm$      0.03   &           0.23  $\pm$      0.02 &           0.74  $\pm$      0.04   &           0.29  $\pm$      0.01 &         0.086  $\pm$    0.005   &         0.041  $\pm$    0.005 \\
{\fci}  &   8728         &         0.112  $\pm$    0.006   &         0.045  $\pm$    0.005 &         0.314  $\pm$    0.017   &         0.110  $\pm$    0.006 &         0.092  $\pm$    0.006   &         0.040  $\pm$    0.005 \\
{\hi}  &    8750       &             2.5  $\pm$        0.1   &             1.0  $\pm$        0.1 &             3.2  $\pm$        0.2   &             1.1  $\pm$        0.1 &             2.0  $\pm$        0.1   &             0.9  $\pm$        0.1 \\
 {\hi}     &    8863       &             3.4  $\pm$        0.2   &             1.3  $\pm$        0.1 &             4.0  $\pm$        0.2   &             1.4  $\pm$        0.1 &  2.7  $\pm$    0.1    &    1.2  $\pm$    0.1 \\
  {\hi}     &  9015       &             4.0  $\pm$        0.2   &             1.6  $\pm$        0.2 &             5.2  $\pm$        0.3   &             1.7  $\pm$        0.1 &   2.7  $\pm$    0.1  &  1.1  $\pm$    0.1   \\
 {\fsiii}     &  9069         &           72.1  $\pm$        3.6   &           28.1  $\pm$        3.1 &         113.0  $\pm$        5.6   &           37.7  $\pm$        2.1 &           24.6  $\pm$        1.2   &           10.3  $\pm$        1.3 \\
 {\hi}     &  9229       &             6.0  $\pm$        0.3   &             2.3  $\pm$        0.3 &             7.8  $\pm$        0.4   &             2.5  $\pm$        0.1 &             5.0  $\pm$        0.3   &             2.1  $\pm$        0.3 \\

\hline
\end{tabular}
}
\begin{description}
\item $^{\rm a}$ Uncertainties in the measured fluxes (F) account for the actual measured uncertainties, while uncertainties in the dereddened fluxes account for the standard deviation of the Monte Carlo distribution, which are 0 for H$\beta$ and H$\alpha$, which are by definition set to values of 100 and 285, respectively, except in the case of M\,1-42 where extinction was computed from the average of several {\hi} lines. \\
\item $^{\rm b}$ Corrected fluxes for {\fnii} $\lambda$5755 and {\foii} $\lambda\lambda$7320+30 correspond to a recombination correction assuming {\te} = 4\,000 K.   \\
\end{description}
\end{table*}

\section{Emission line maps}

In this appendix we show complementary plots of line fluxes and abundance maps for the three PNe studied in this paper.

\begin{figure*}
\includegraphics[scale=0.35,trim={6cm 5cm 0 6cm}, clip]{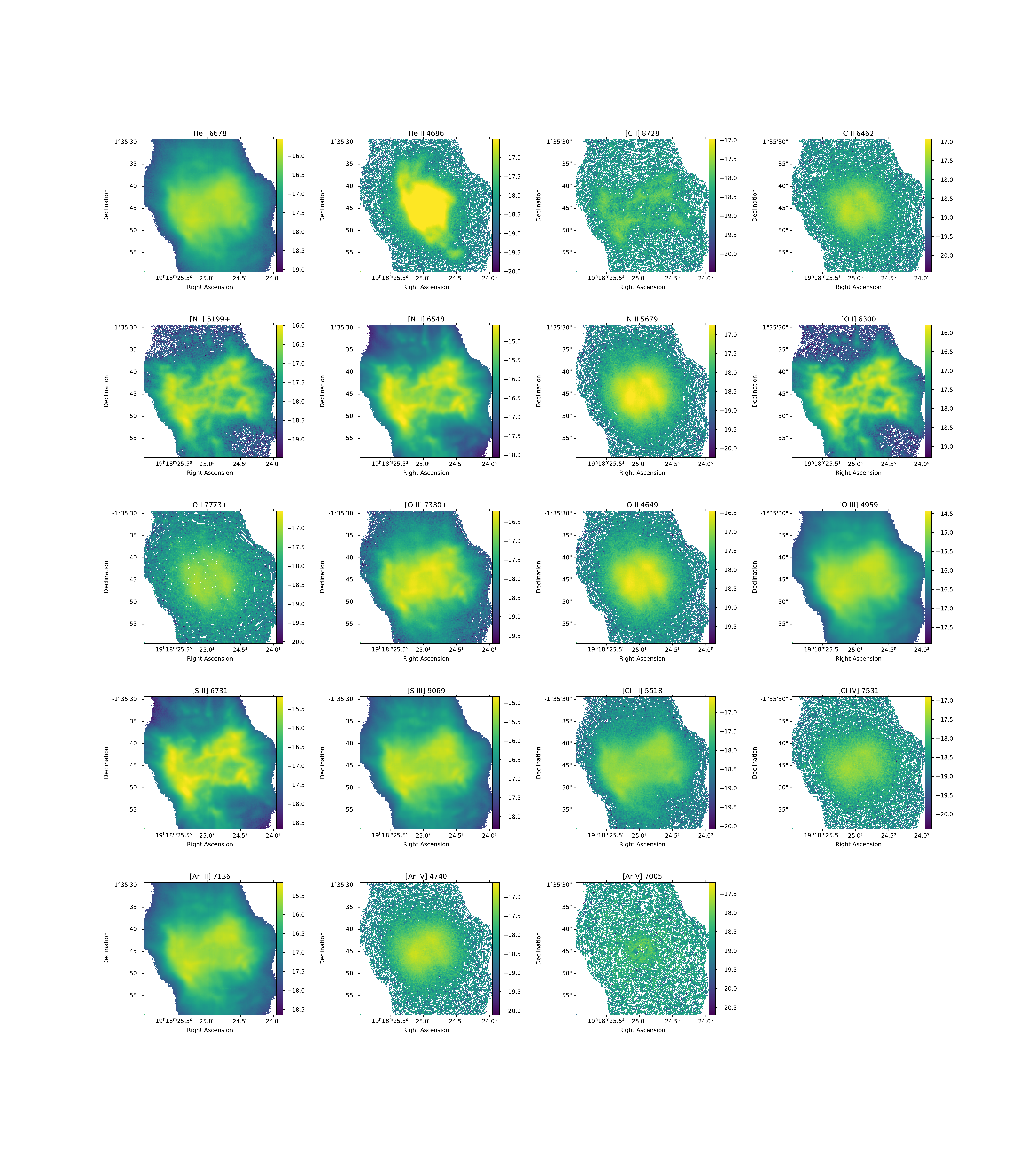}
\caption{Observed emission line flux maps for 19 lines of elements heavier than H in the MUSE datacube of NGC\,6778. The maps are ordered by atomic mass and ionization state of the selected ion. The scale is logarithmic and the flux is in units of erg cm$^{-2}$ s$^{-1}$ \AA$^{-1}$. 
\label{fig:maps_ngc6778}}
\end{figure*}

\begin{figure*}
\includegraphics[scale=0.35,trim={6cm 5cm 0 6cm}, clip]{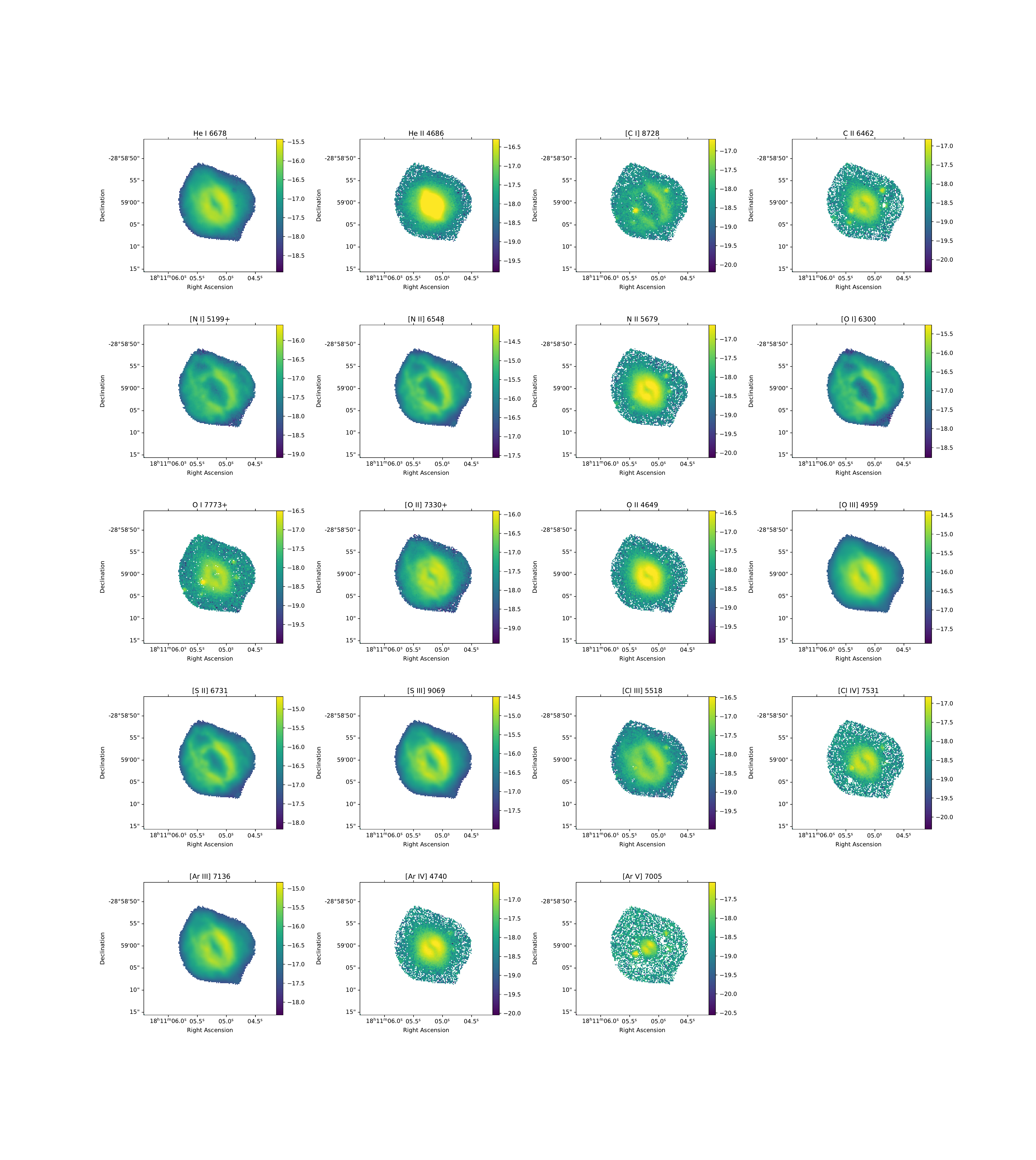}
\caption{Same figure than Fig~\ref{fig:maps_ngc6778} for M\,1-42.
\label{fig:maps_m142}}
\end{figure*}

\begin{figure*}
\includegraphics[scale=0.35,trim={6cm 5cm 0 6cm}, clip]{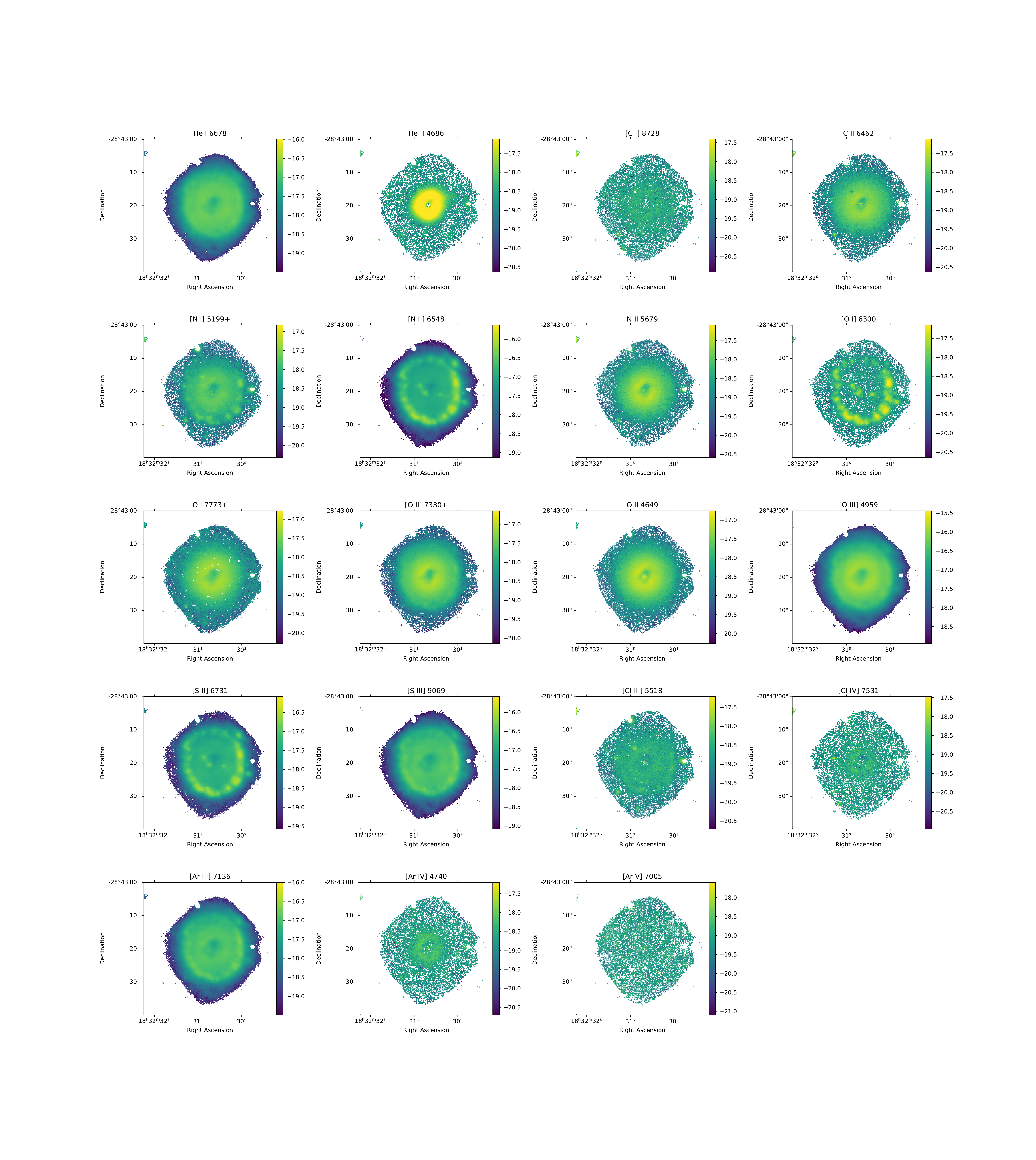}
\caption{Same figure than Fig~\ref{fig:maps_ngc6778} for Hf\,2-2
\label{fig:maps_hf22}}
\end{figure*}

\begin{figure*}
\includegraphics[scale=0.45,trim={3cm 5cm 0 5cm}, clip]{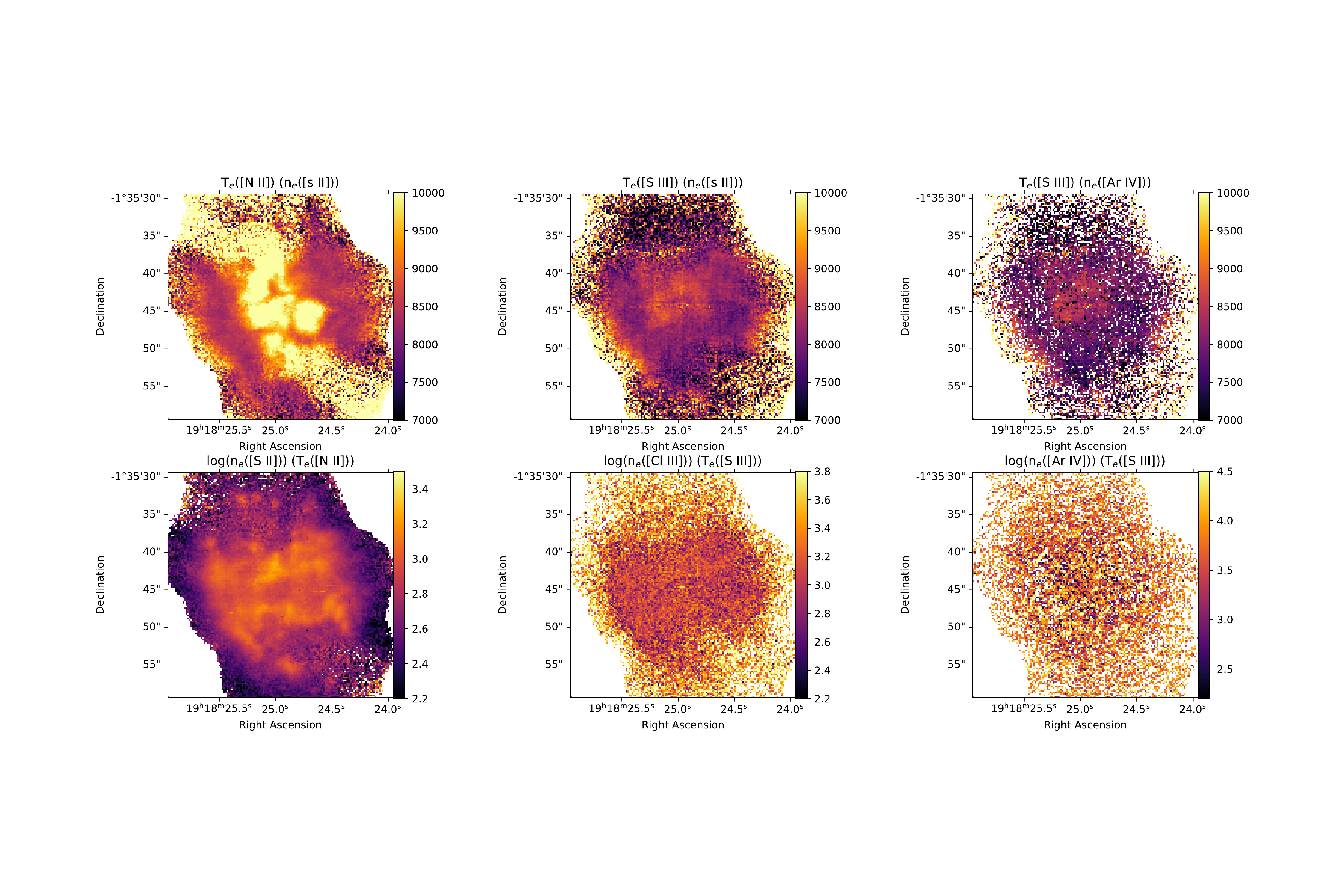}
\caption{Electron temperature and density maps obtained for NGC\,6778 from the combination of different temperature and density diagnostics. No recombination contribution to {\fnii} $\lambda$5755 is considered.
\label{fig:NGC6778_TeNe_none}}
\end{figure*}

\begin{figure*}
\includegraphics[scale=0.45,trim={3cm 5cm 0 5cm}, clip]{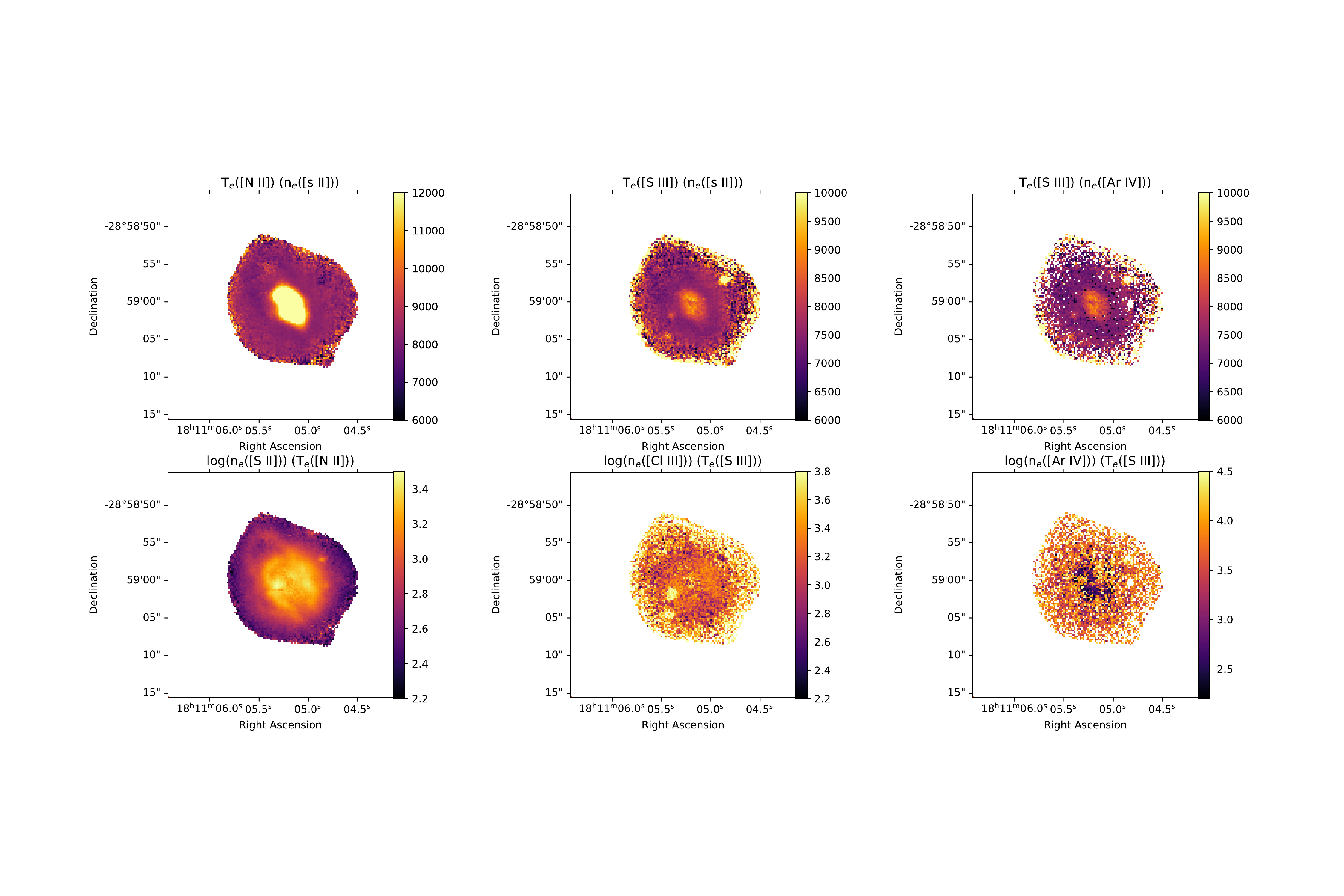}
\caption{Same as Fig.~\ref{fig:NGC6778_TeNe_none} for M\,1-42.
\label{fig:M142_TeNe_none}}
\end{figure*}

\begin{figure*}
\includegraphics[scale=0.45,trim={3cm 5cm 0 5cm}, clip]{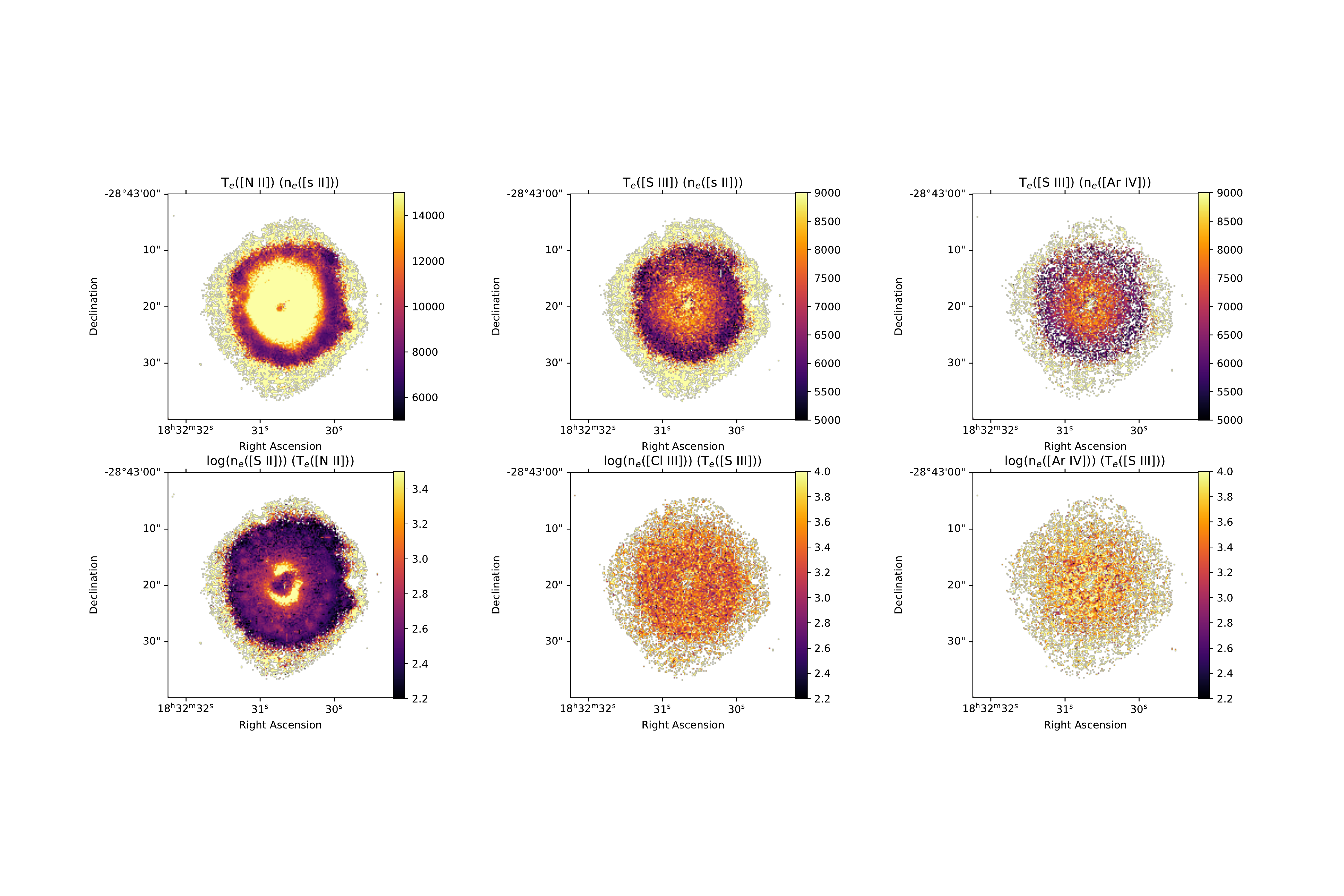}
\caption{Same as Fig.~\ref{fig:NGC6778_TeNe_none} for Hf\,2-2.
\label{fig:HF22_TeNe_none}}
\end{figure*}

\begin{figure*}
\includegraphics[scale=0.45,trim={3cm 5cm 0 5cm}, clip]{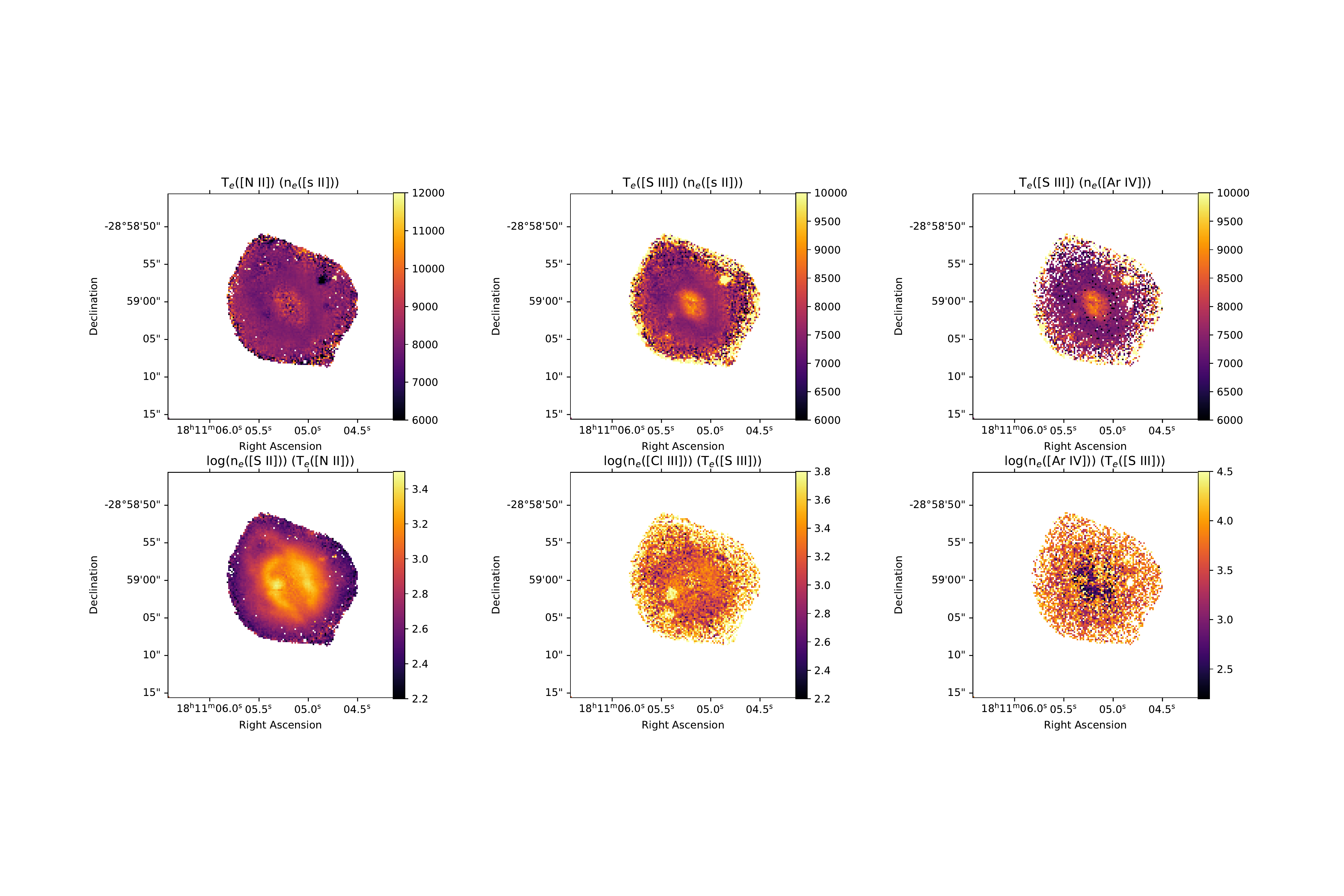}
\caption{Same as Fig.~\ref{fig:NGC6778_TeNe_4000} for M\,1-42.
\label{fig:M142_TeNe_4000}}
\end{figure*}

\begin{figure*}
\includegraphics[scale=0.45,trim={3cm 5cm 0 5cm}, clip]{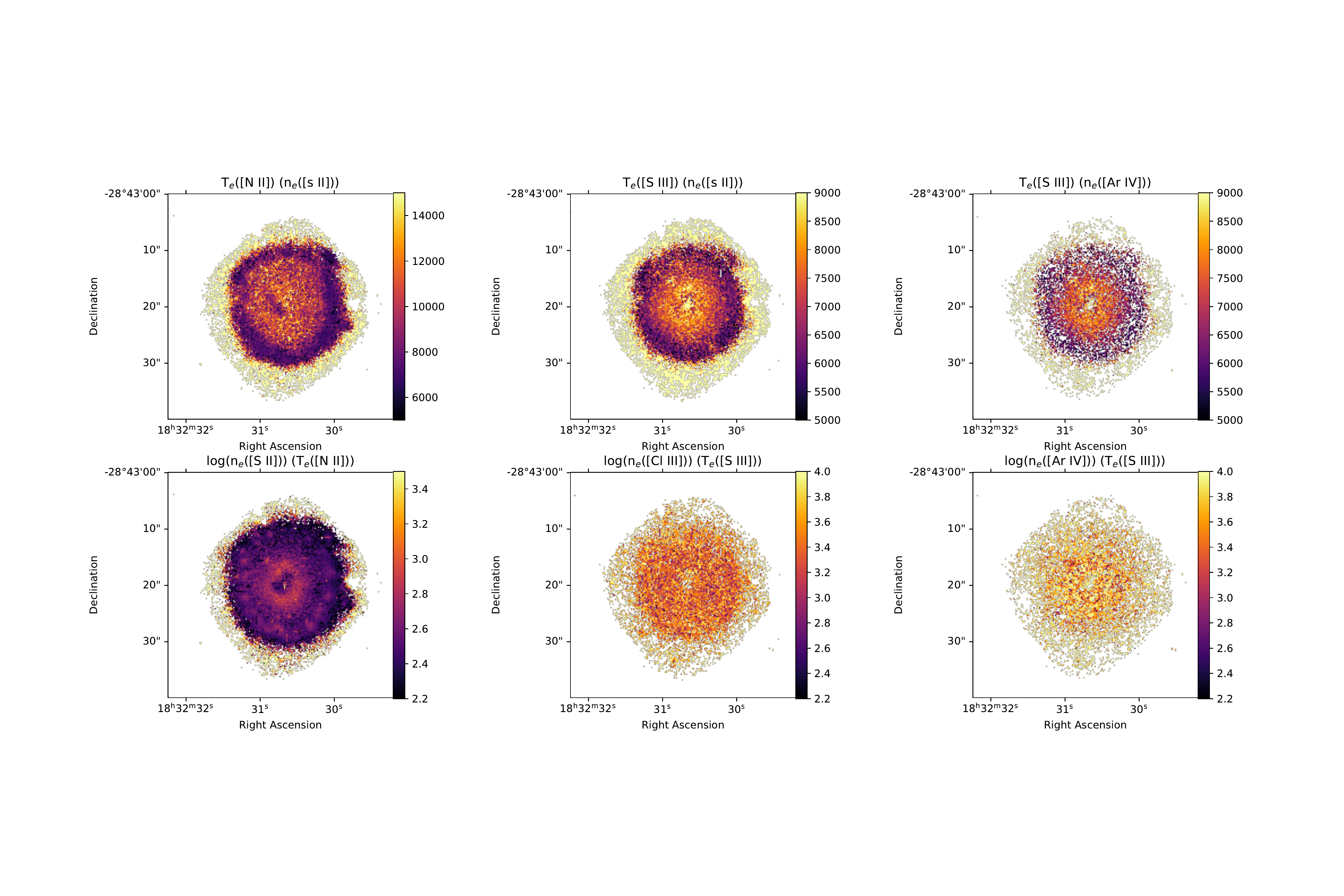}
\caption{Same as Fig.~\ref{fig:NGC6778_TeNe_4000} for Hf\,2-2.
\label{fig:HF22_TeNe_4000}}
\end{figure*}

\begin{figure*}
\includegraphics[scale=0.45,trim={3cm 5cm 0 5cm}, clip]{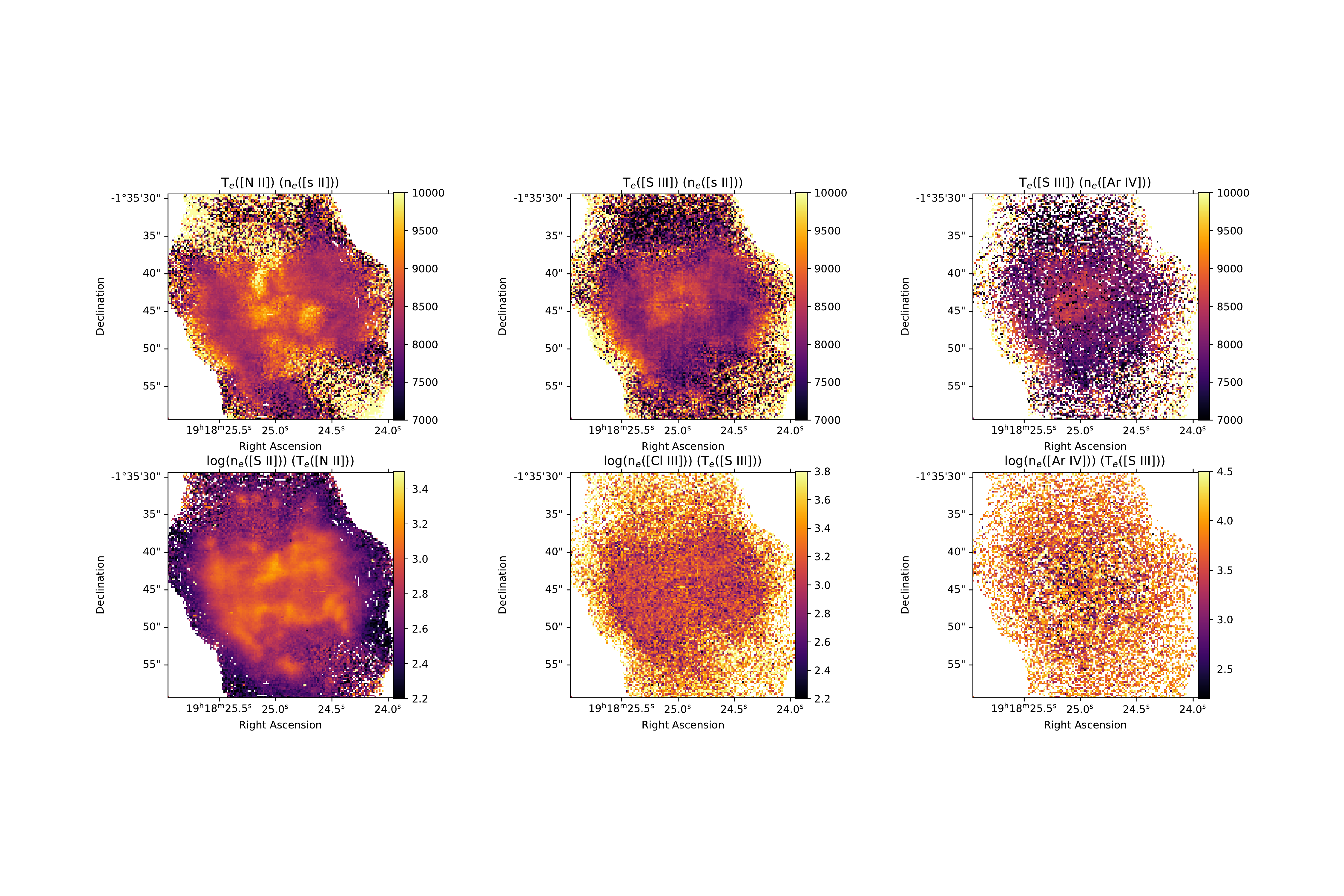}
\caption{Electron temperature and density maps obtained for NGC\,6778 from the combination of different temperature and density diagnostics. Recombination contribution to {\fnii} $\lambda$5755 assuming {\te}=1\,000\,K is considered.
\label{fig:NGC6778_TeNe_1000}}
\end{figure*}

\begin{figure*}
\includegraphics[scale=0.45,trim={3cm 5cm 0 5cm}, clip]{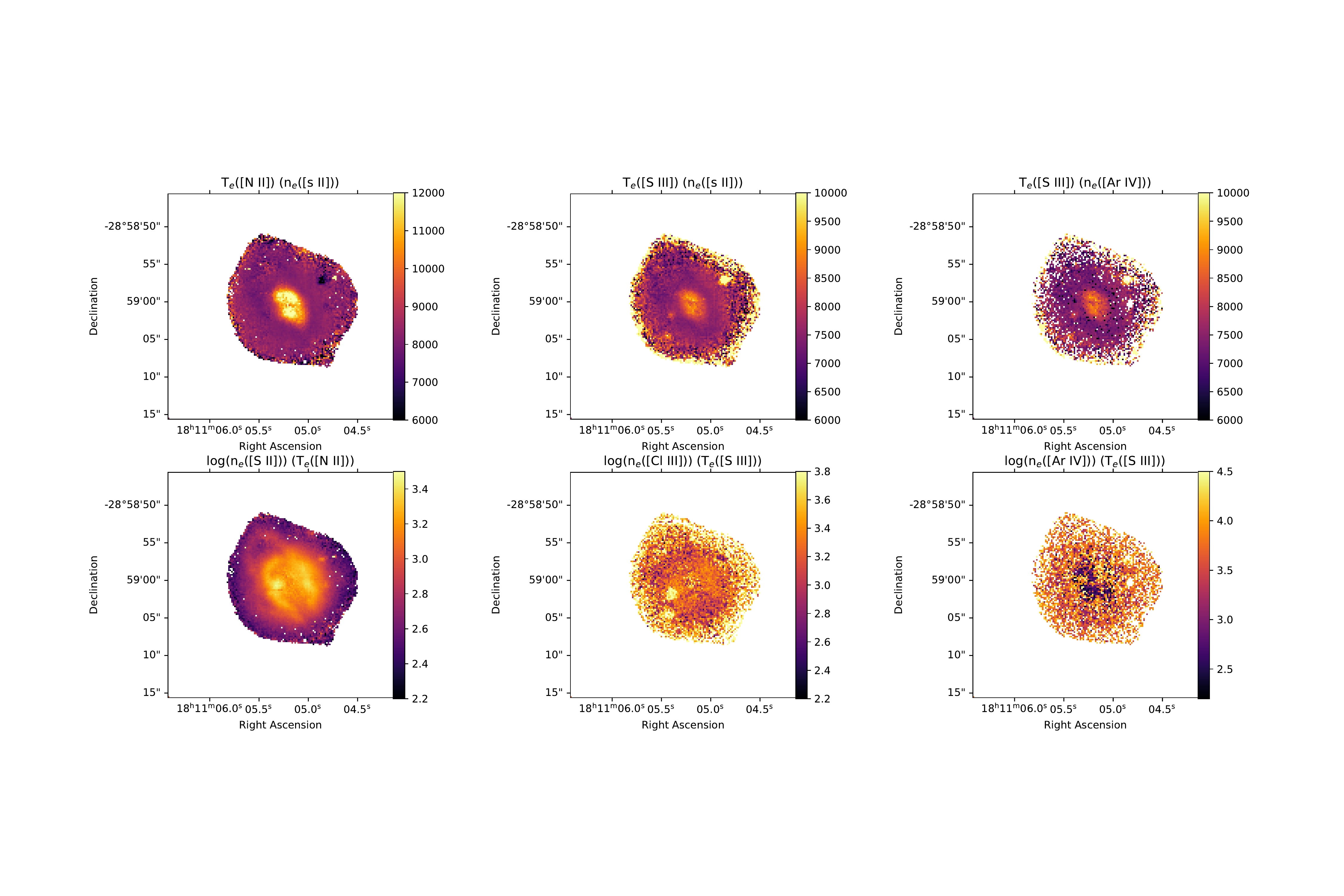}
\caption{Same as Fig.~\ref{fig:NGC6778_TeNe_1000} for M\,1-42.
\label{fig:M142_TeNe_1000}}
\end{figure*}

\begin{figure*}
\includegraphics[scale=0.45,trim={3cm 5cm 0 5cm}, clip]{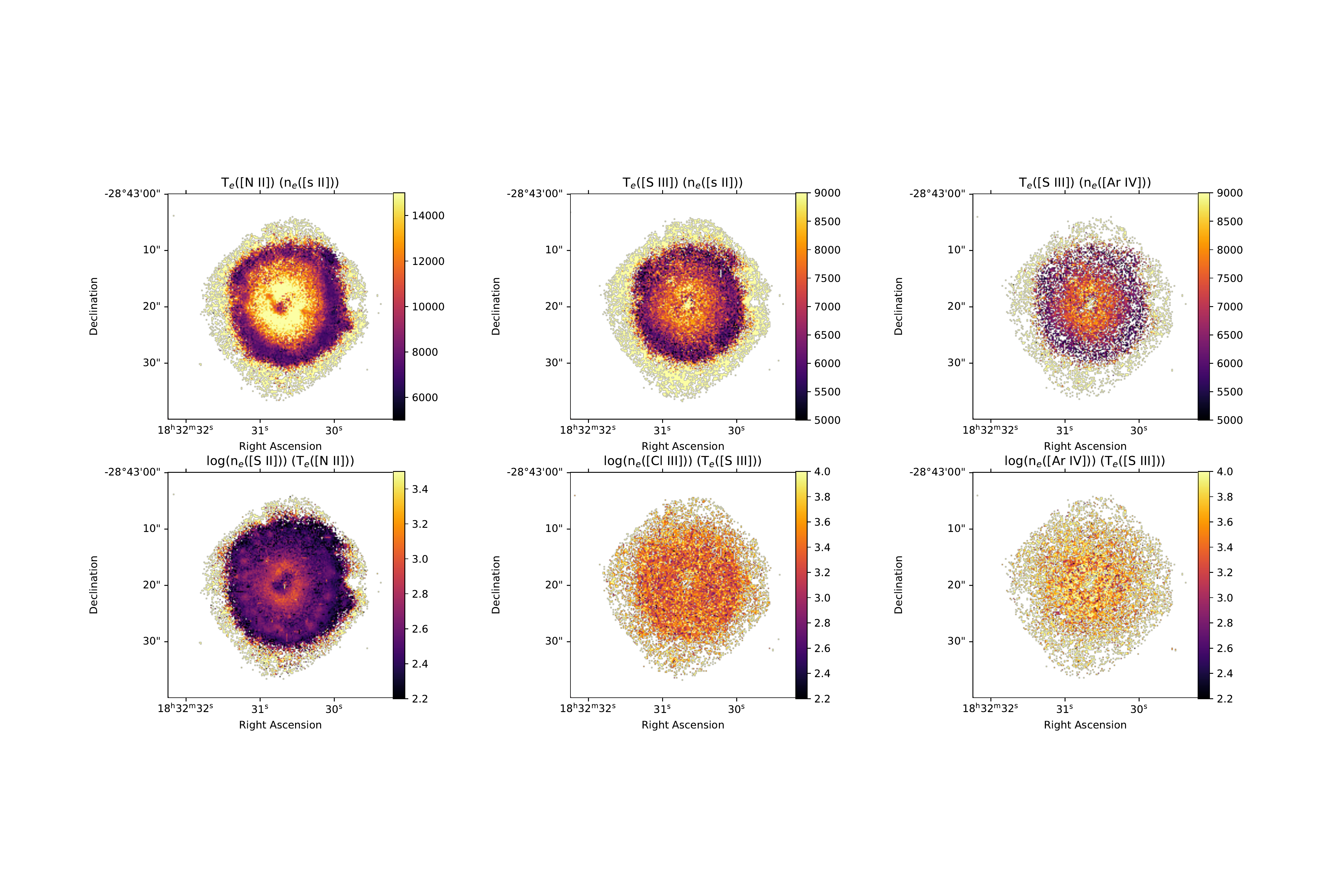}
\caption{Same as Fig.~\ref{fig:NGC6778_TeNe_1000} for Hf\,2-2.
\label{fig:HF22_TeNe_1000}}
\end{figure*}

\begin{figure*}
\includegraphics[scale=0.45,trim={3cm 5cm 0 5cm}, clip]{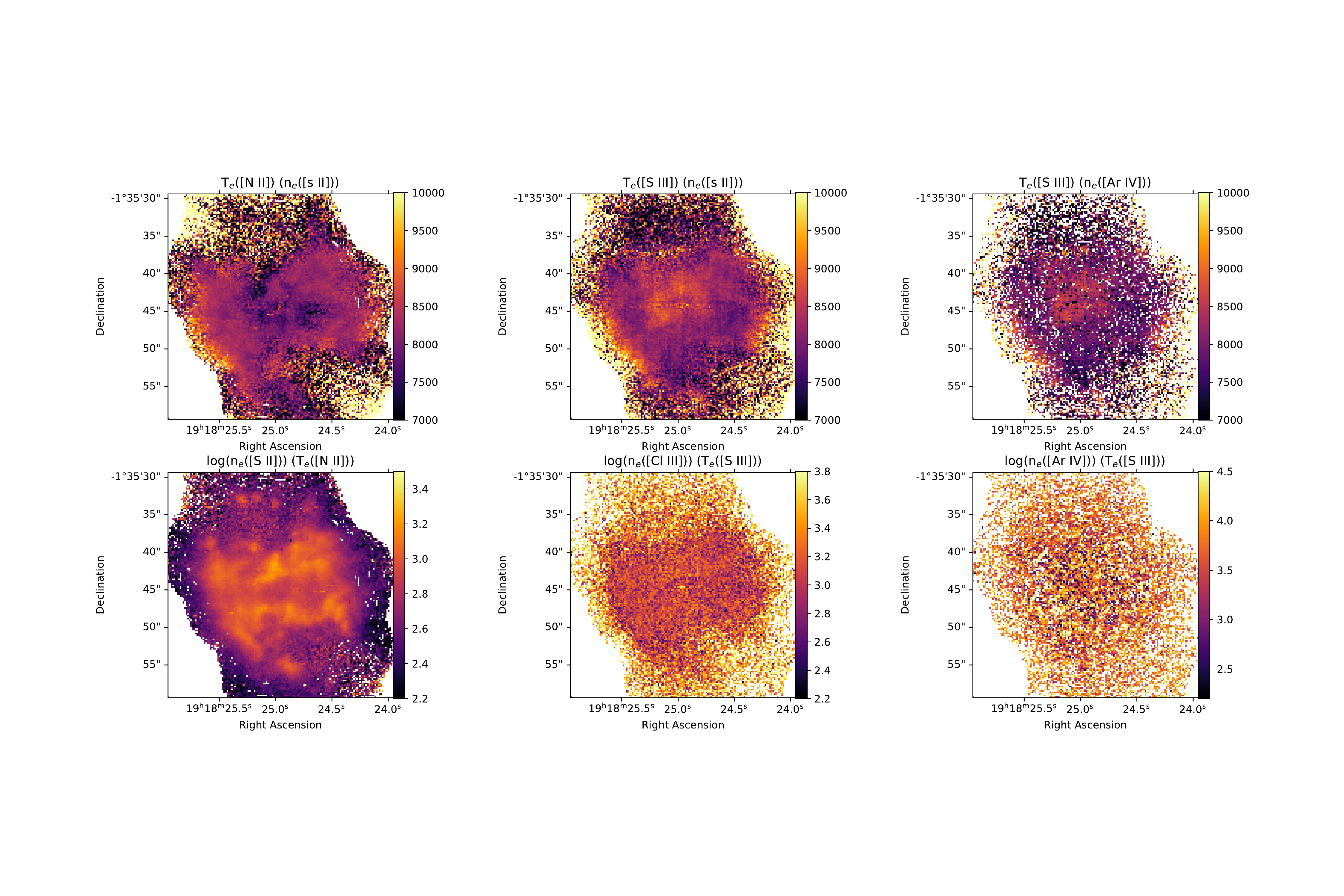}
\caption{Electron temperature and density maps obtained for NGC\,6778 from the combination of different temperature and density diagnostics. Recombination contribution to {\fnii} $\lambda$5755 assuming {\te}=8\,000\,K is considered.
\label{fig:NGC6778_TeNe_8000}}
\end{figure*}

\begin{figure*}
\includegraphics[scale=0.45,trim={3cm 5cm 0 5cm}, clip]{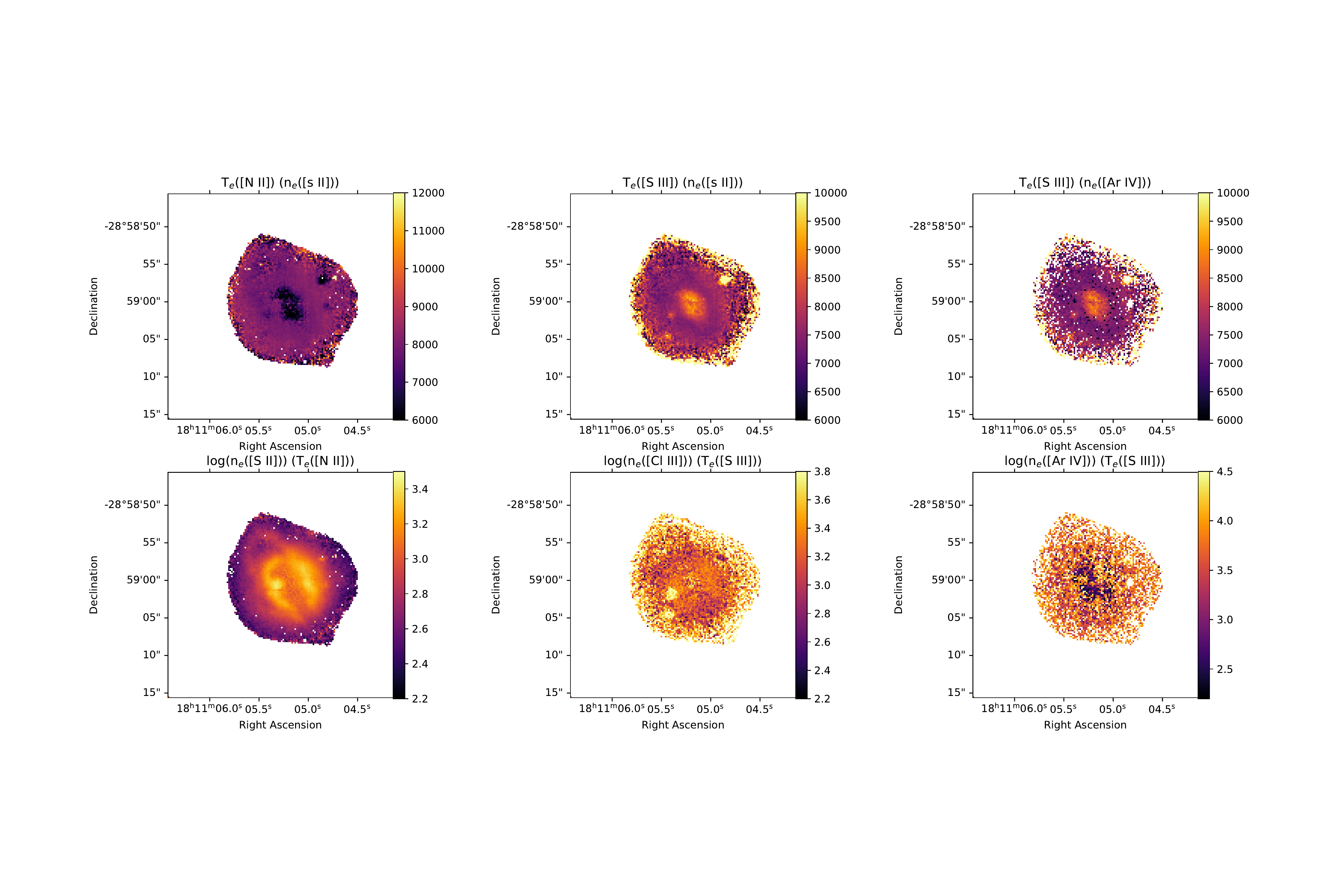}
\caption{Same as Fig.~\ref{fig:NGC6778_TeNe_1000} for M\,1-42.
\label{fig:M142_TeNe_8000}}
\end{figure*}

\begin{figure*}
\includegraphics[scale=0.45,trim={3cm 5cm 0 5cm}, clip]{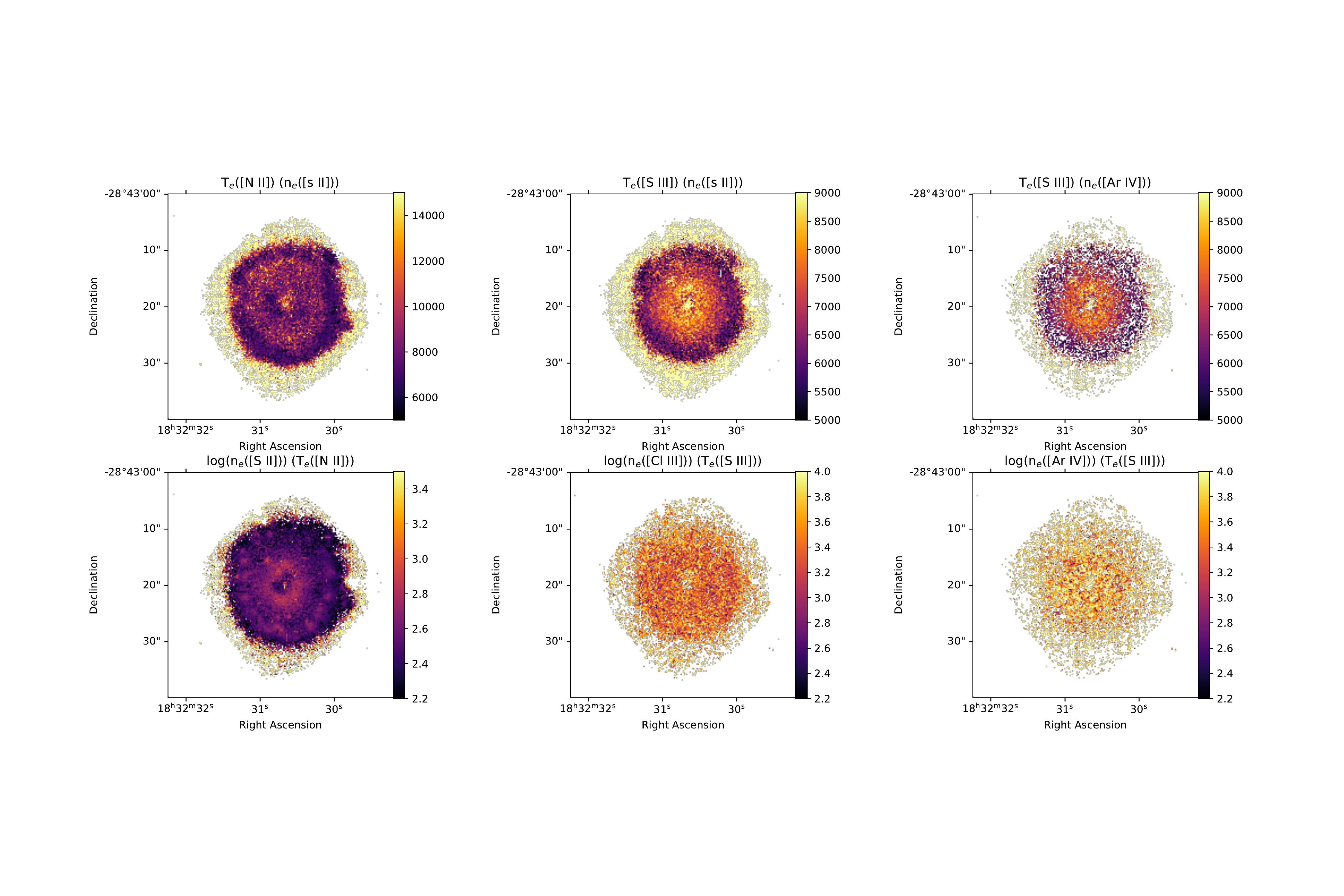}
\caption{Same as Fig.~\ref{fig:NGC6778_TeNe_8000} for Hf\,2-2.
\label{fig:HF22_TeNe_8000}}
\end{figure*}

\begin{figure*}
\includegraphics[scale=0.35,trim={6cm 5cm 0 6cm}, clip]{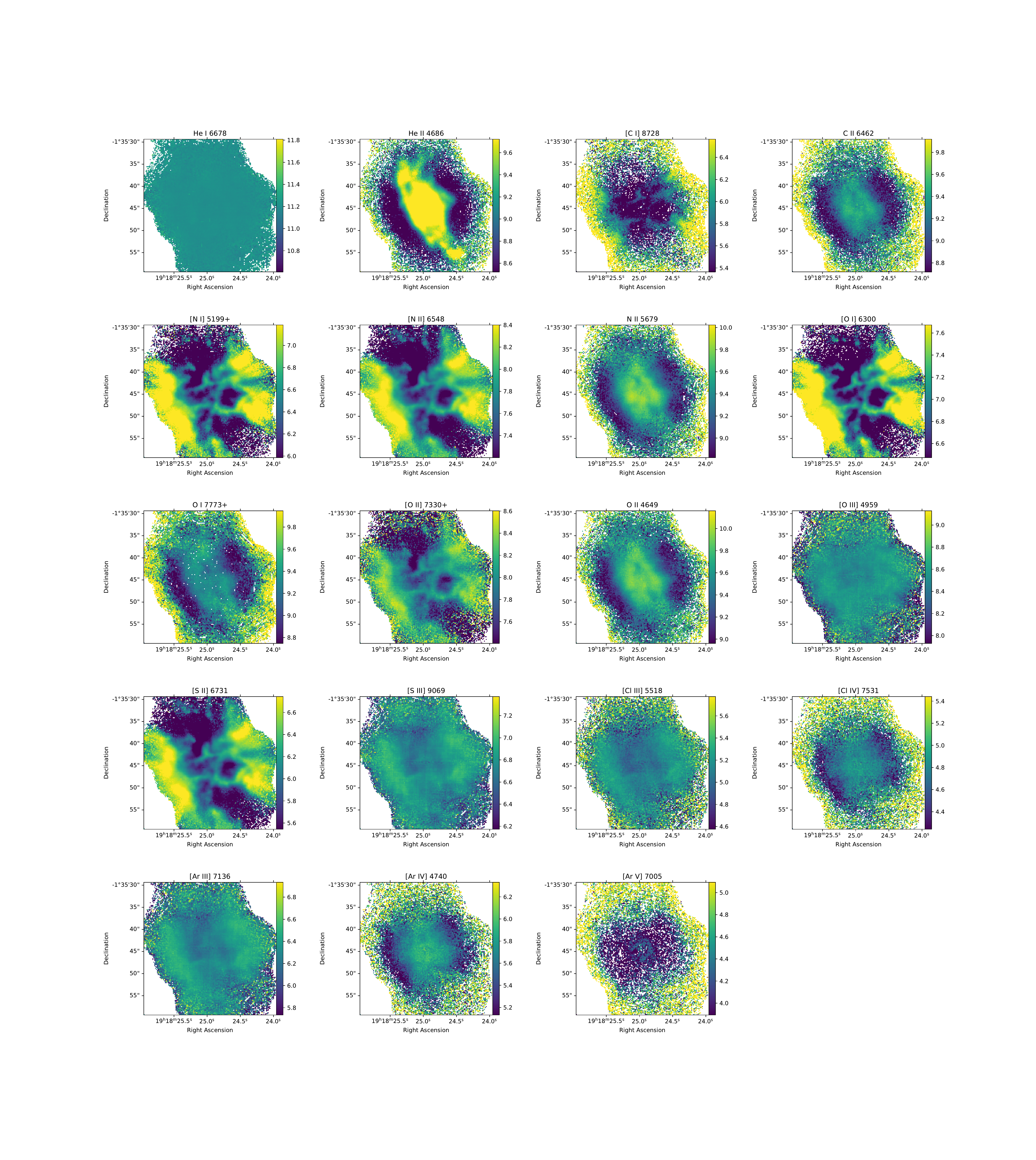}
\caption{Ionic abundance maps obtained from 19 emission lines in the MUSE datacube of NGC\,6778 when no correction to recombination contribution is made for auroral [N~{\sc ii}] and [O~{\sc ii}] lines. The maps are ordered by atomic mass and ionization state of the selected ion. The abundance scale is in unites of 12+log(X$^{i+}$/H$^+$) and the colorbar ranges from the median value of the whole FoV $\pm$ 0.6 dex. 
\label{fig:abund_ngc6778_none}}
\end{figure*}

\begin{figure*}
\includegraphics[scale=0.35,trim={6cm 5cm 0 6cm}, clip]{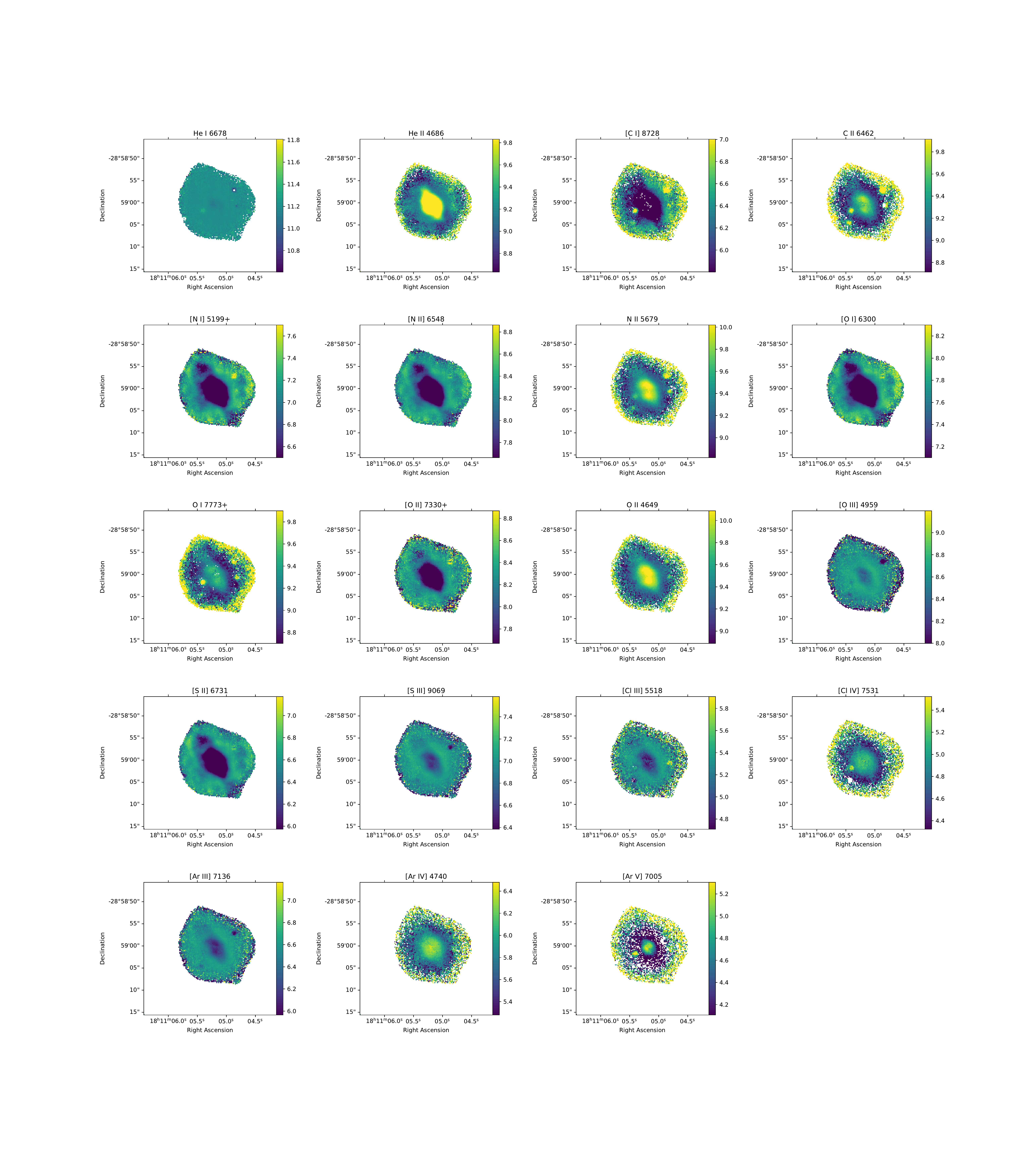}
\caption{Same figure than Fig~\ref{fig:abund_ngc6778_none} for M\,1-42.
\label{fig:abund_m142_none}}
\end{figure*}

\begin{figure*}
\includegraphics[scale=0.35,trim={6cm 5cm 0 6cm}, clip]{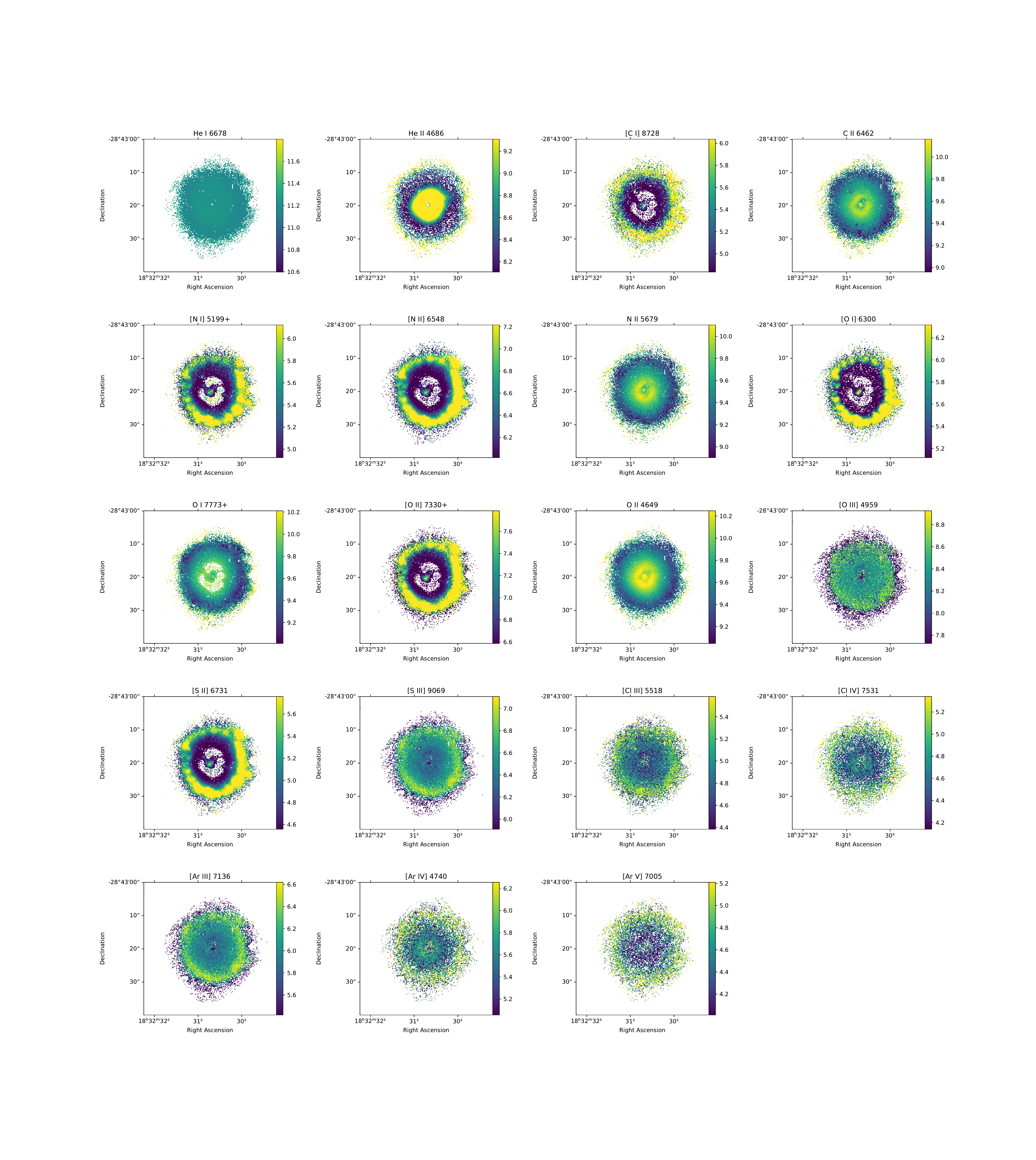}
\caption{Same figure than Fig~\ref{fig:abund_ngc6778_none} for Hf\,2-2
\label{fig:abund_hf22_none}}
\end{figure*}

\begin{figure*}
\includegraphics[scale=0.35,trim={6cm 5cm 0 6cm}, clip]{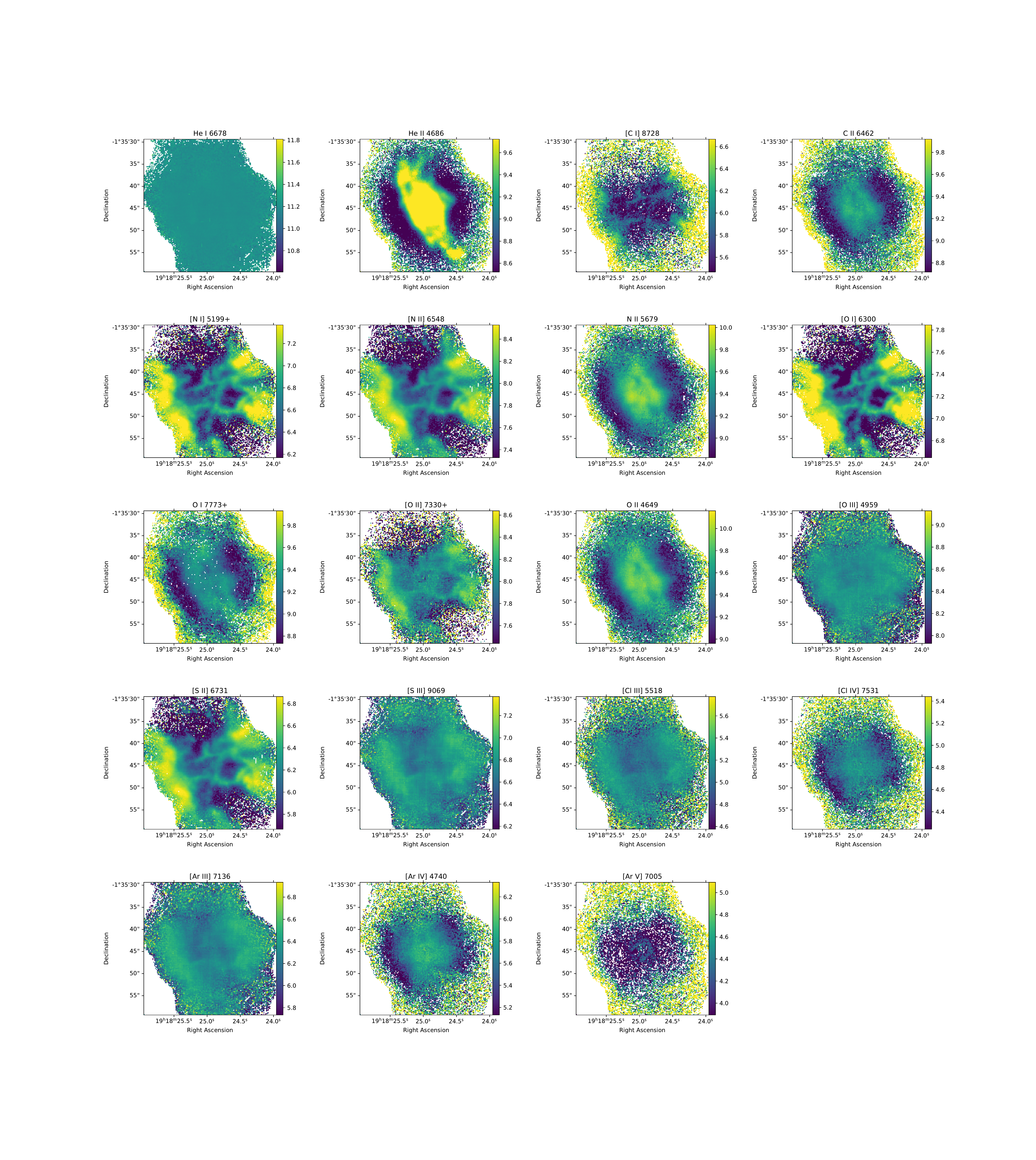}
\caption{Ionic abundance maps obtained from 19 emission lines in the MUSE datacube of NGC\,6778 when recombination contribution is made for auroral [N~{\sc ii}] and [O~{\sc ii}] lines assuming $T_e$=4\,000\,K. The maps are ordered by atomic mass and ionization state of the selected ion. The abundance scale is in unites of 12+log(X$^{i+}$/H$^+$) and the colorbar ranges from the median value of the whole FoV $\pm$ 0.6 dex. 
\label{fig:abund_ngc6778_4000}}
\end{figure*}

\begin{figure*}
\includegraphics[scale=0.35,trim={6cm 5cm 0 6cm}, clip]{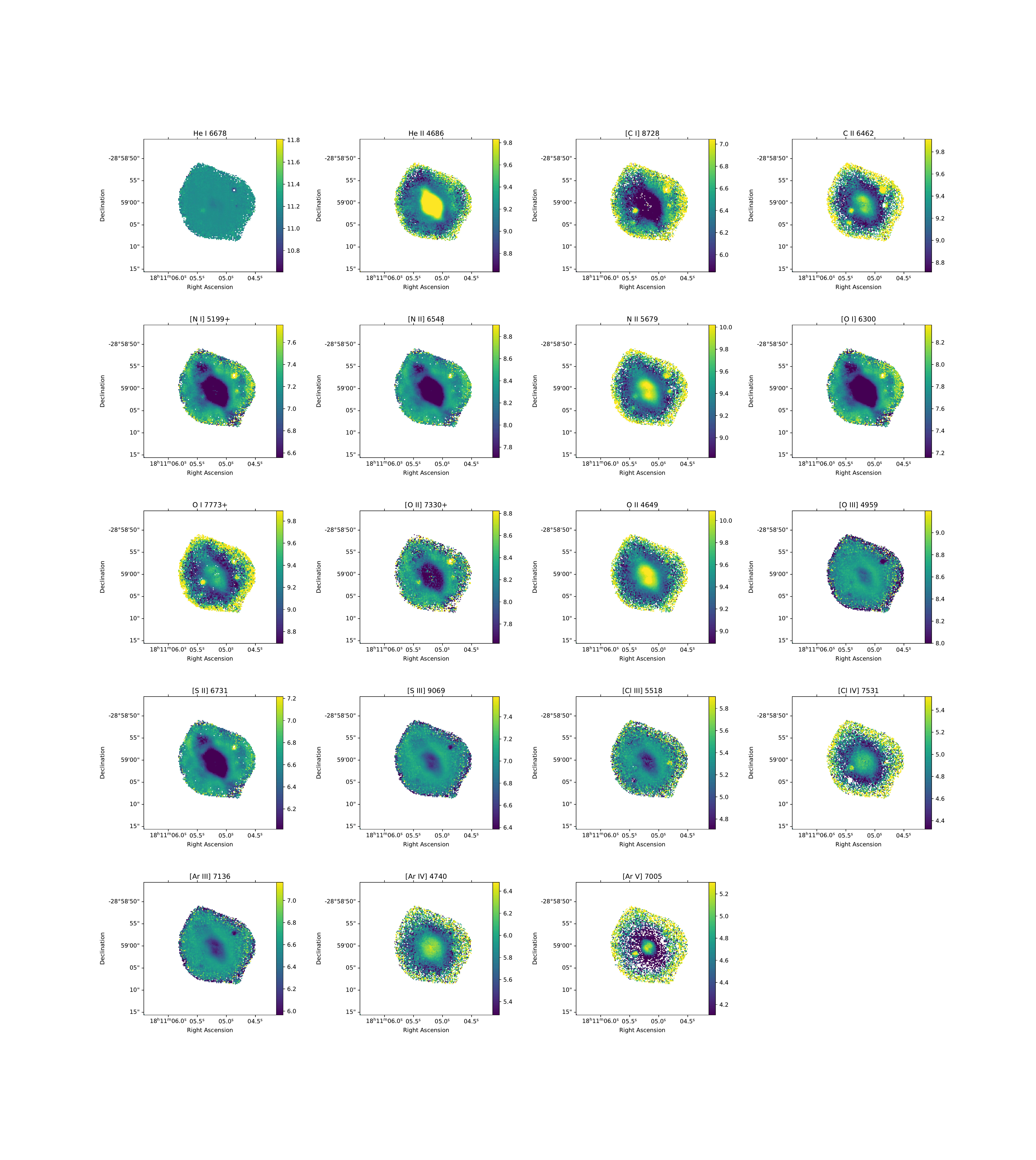}
\caption{Same figure than Fig~\ref{fig:abund_ngc6778_4000} for M\,1-42.
\label{fig:abund_m142_4000}}
\end{figure*}

\begin{figure*}
\includegraphics[scale=0.35,trim={6cm 5cm 0 6cm}, clip]{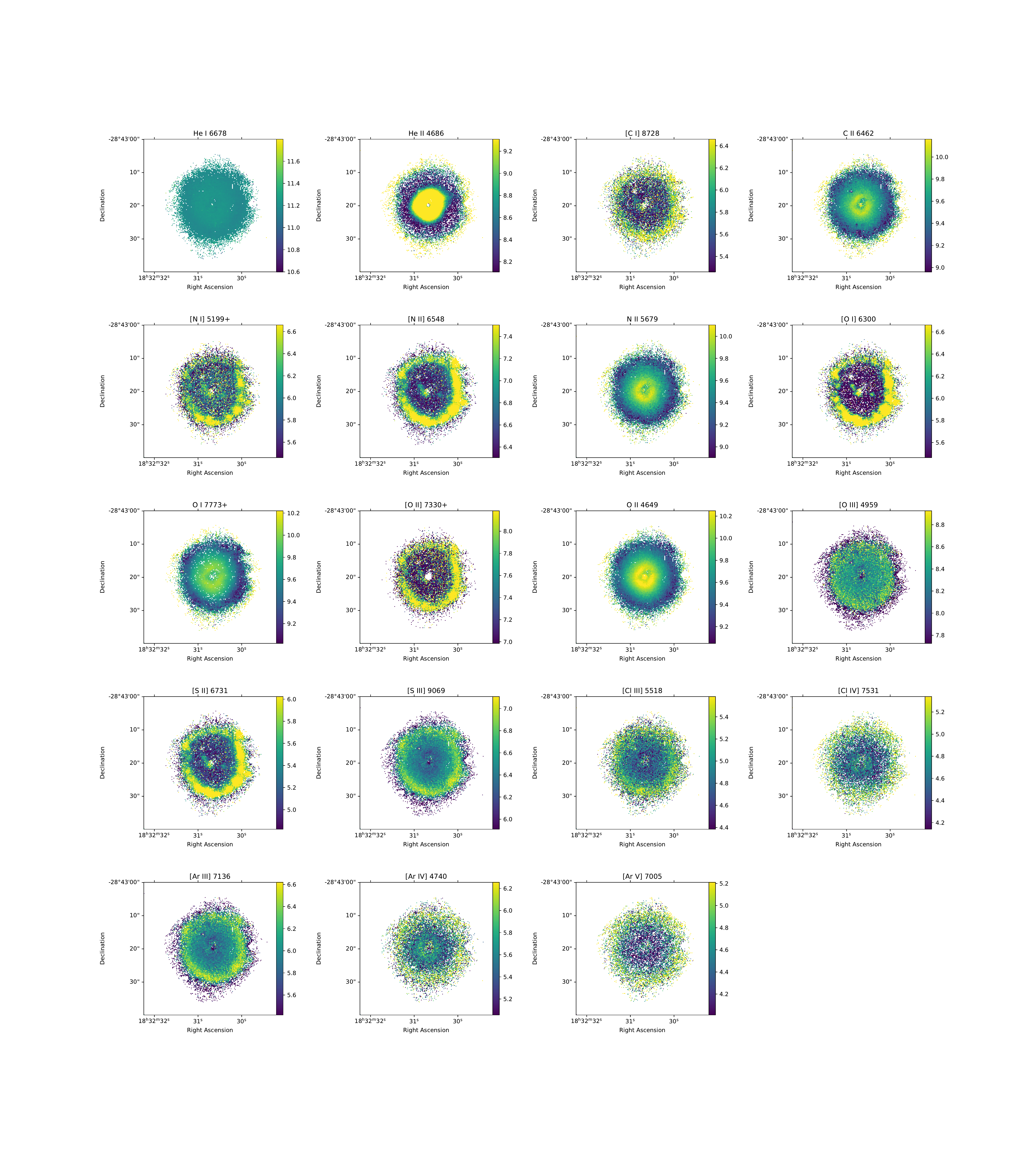}
\caption{Same figure than Fig~\ref{fig:abund_ngc6778_4000} for Hf\,2-2
\label{fig:abund_hf22_4000}}
\end{figure*}

\begin{figure}
\includegraphics[scale=0.36,trim={0.5cm 3cm 0 3cm}, clip]{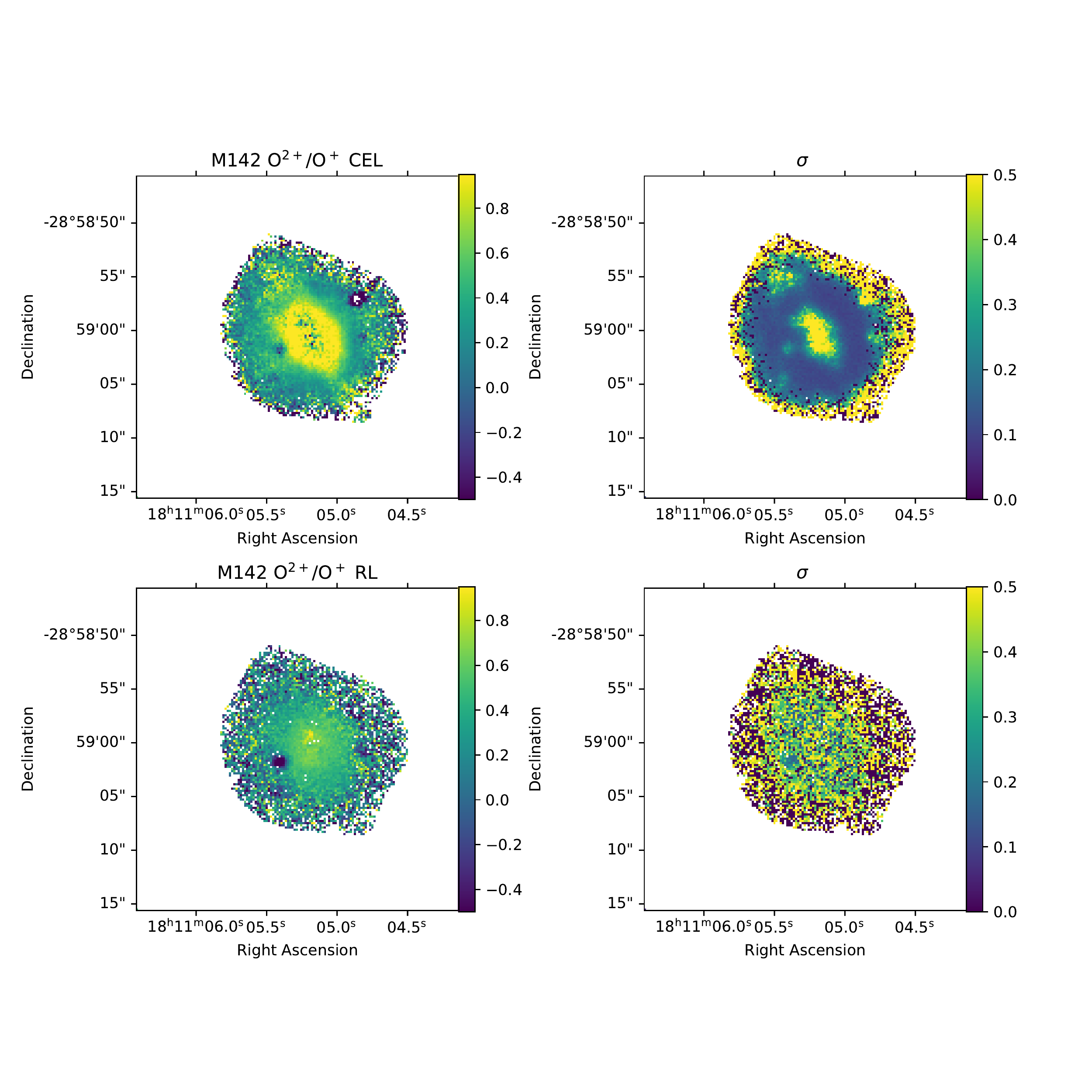}
\caption{Same as Fig.~\ref{fig:ion_degree_ngc6778} for M\,1-42.
\label{fig:ion_degree_m142}}
\end{figure}

\begin{figure}
\includegraphics[scale=0.36,trim={0.5cm 3cm 0 3cm}, clip]{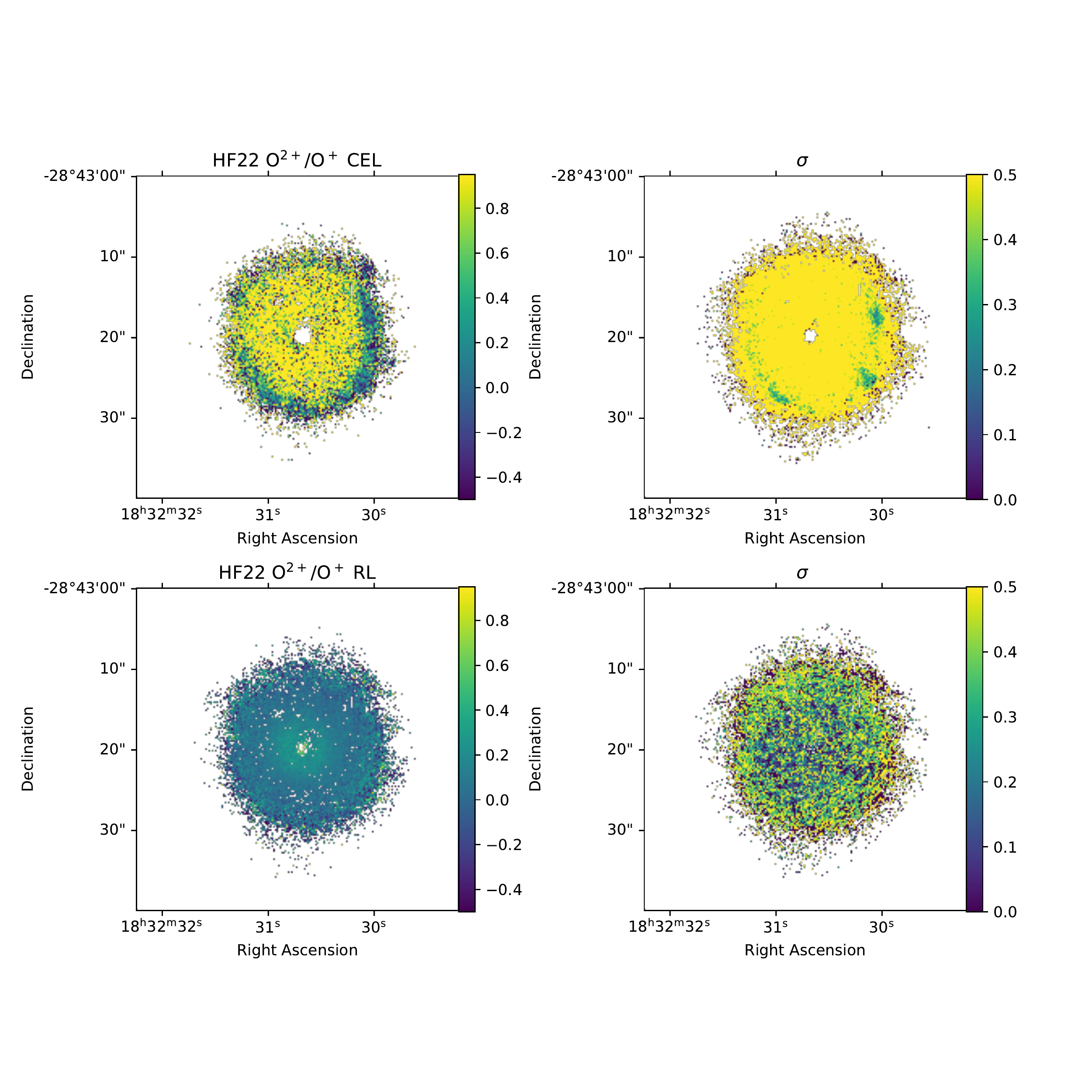}
\caption{Same as Fig.~\ref{fig:ion_degree_ngc6778} for Hf\,2-2.
\label{fig:ion_degree_hf22}}
\end{figure}


\bsp	
\label{lastpage}
\end{document}